\documentclass{emulateapj}

\usepackage{epstopdf}

\def\2pr{^{\prime \prime}}

\shorttitle{The Multi-Object Spectrographs for SDSS}

\begin{document}

\title{The Multi-Object, Fiber-Fed Spectrographs for SDSS and the\\
    Baryon Oscillation Spectroscopic Survey}

\author{
Stephen A. Smee\altaffilmark{1},
James E. Gunn\altaffilmark{2},
Alan Uomoto\altaffilmark{3},
Natalie Roe\altaffilmark{4},
David Schlegel\altaffilmark{4},
Constance M. Rockosi\altaffilmark{5},
Michael A. Carr\altaffilmark{2},
French Leger\altaffilmark{6},
Kyle S. Dawson\altaffilmark{7},
Matthew D. Olmstead\altaffilmark{7},
Jon Brinkmann\altaffilmark{8},
Russell Owen\altaffilmark{6},
Robert H. Barkhouser\altaffilmark{1},
Klaus Honscheid\altaffilmark{9},
Paul Harding\altaffilmark{10},
Dan Long\altaffilmark{8},
Robert H. Lupton\altaffilmark{2},
Craig Loomis\altaffilmark{2},
Lauren Anderson\altaffilmark{6},
James Annis\altaffilmark{11},
Mariangela Bernardi \altaffilmark{12},
Vaishali Bhardwaj\altaffilmark{6},
Dmitry Bizyaev\altaffilmark{8},
Adam S. Bolton\altaffilmark{7},
Howard Brewington\altaffilmark{8},
John W. Briggs\altaffilmark{13},
Scott Burles\altaffilmark{14},
James G. Burns\altaffilmark{9},
Francisco Javier Castander\altaffilmark{15,16},
Andrew Connolly\altaffilmark{6},
James R. A. Davenport\altaffilmark{6},
Garrett Ebelke\altaffilmark{8},
Harland Epps\altaffilmark{5},
Paul D. Feldman\altaffilmark{1},
Scott Friedman\altaffilmark{16},
Joshua Frieman\altaffilmark{11},
Timothy Heckman\altaffilmark{1}
Charles L. Hull\altaffilmark{3},
Gillian R. Knapp\altaffilmark{2},
David M. Lawrence\altaffilmark{7},
Jon Loveday\altaffilmark{17},
Edward J. Mannery\altaffilmark{6},
Elena Malanushenko\altaffilmark{8},
Viktor  Malanushenko\altaffilmark{8},
Aronne James Merrelli\altaffilmark{18},
Demitri Muna\altaffilmark{19}
Peter R. Newman\altaffilmark{8},
Robert C. Nichol\altaffilmark{20},
Daniel Oravetz\altaffilmark{8},
Kaike Pan\altaffilmark{8},
Adrian C. Pope\altaffilmark{21},
Paul G. Ricketts\altaffilmark{7},
Alaina Shelden\altaffilmark{8},
Dale Sandford\altaffilmark{5},
Walter Siegmund\altaffilmark{6},
Audrey Simmons\altaffilmark{8},
D. Shane Smith\altaffilmark{9},
Stephanie Snedden\altaffilmark{8},
Donald P. Schneider\altaffilmark{22,23},
Mark SubbaRao\altaffilmark{12},
Christy Tremonti\altaffilmark{24},
Patrick Waddell\altaffilmark{25},
Donald G. York\altaffilmark{26}
}

\altaffiltext{1}{Department of Physics and Astronomy, Johns Hopkins University, Baltimore, MD 21218, USA}
\altaffiltext{2}{Department of Astrophysical Sciences, Princeton University, Princeton, NJ 08544, USA}
\altaffiltext{3}{Observatories of the Carnegie Institution of Washington, 813 Santa Barbara Street, Pasadena, CA 91101, USA}
\altaffiltext{4}{Physics Division, Lawrence Berkeley National Laboratory, Berkeley CA 94720, USA}
\altaffiltext{5}{UC Observatories and Department of Astronomy and Astrophysics, University of California, Santa Cruz, 375 Interdisciplinary Sciences Building (ISB) Santa Cruz, CA 95064, USA}
\altaffiltext{6}{Department of Astronomy, University of Washington, Box 351580, Seattle, WA 09195, USA}
\altaffiltext{7}{Department of Physics and Astronomy, University of Utah, Salt Lake City, UT 84112, USA}
\altaffiltext{8}{Apache Point Observatory, Sunspot, NM 88349, USA}
\altaffiltext{9}{Department of Physics and Center for Cosmology and Astro-Particle Physics, Ohio State University, Columbus, OH 43210, USA}
\altaffiltext{10}{Department of Astronomy, Case Western Reserve University, Cleveland, OH 44106, USA}
\altaffiltext{11}{Fermi National Accelerator Laboratory, P.O. Box 500, Batavia, IL 60510, USA}
\altaffiltext{12}{Department of Physics and Astronomy, The University of Pennsylvania, 209 South 33rd Street, Philadelphia, PA 19104, USA}
\altaffiltext{13}{HUT Observatory, Mittelman Family Foundation, P.O. Box 5320, Eagle, CO 81631, USA}
\altaffiltext{14}{Physics Department, Massachusetts Institute of Technology, 77 Massachusetts Avenue, Cambridge, MA 02139, USA}
\altaffiltext{15}{Institut de Ci\ encies de l'Espai (IEEC-CSIC), E-08193 Ballaterra, Barcelona, Spain}
\altaffiltext{16}{Space Telescope Science Institute, Baltimore, MD 21218, USA} 
\altaffiltext{17}{Astronomy Centre, University of Sussex, Falmer, Brighton BN1 9QJ, UK}
\altaffiltext{18}{Department of Astronomy, California Institute of Technology, Pasadena, CA 91125, USA}
\altaffiltext{19}{Center for Cosmology and Particle Physics, New York University, 4 Washington Place, New York, NY 10003, USA}
\altaffiltext{20}{Institute of Cosmology and Gravitation (ICG), Dennis Sciama Building, University of Portsmouth, Portsmouth, PO1 3FX, UK}
\altaffiltext{21}{High Energy Physics Division, Argonne National Laboratory, 9700 South Cass Avenue, Lemont, IL 60439, USA}
\altaffiltext{22}{Department of Astronomy and Astrophysics, The Pennsylvania State University, University Park, PA 16802, USA}
\altaffiltext{23}{Institute for Gravitation and the Cosmos, The Pennsylvania State University, PA 16802, USA}
\altaffiltext{24}{Department of Astronomy, University of Wisconsin-Madison, Madison, WI 53703, USA}
\altaffiltext{25}{NASA Ames Research Center, Moffett Field, CA 94035, USA}
\altaffiltext{26}{Department of Astronomy and Astrophysics and the Fermi Institute, The University of Chicago, Chicago, IL 60637, USA}

\email{smee@pha.jhu.edu}

\begin{abstract}
We present the design and performance of the multi-object fiber spectrographs
for the Sloan Digital Sky Survey (SDSS) and their upgrade for the Baryon Oscillation
Spectroscopic Survey (BOSS). Originally commissioned in Fall 1999
on the 2.5-m aperture Sloan Telescope
at Apache Point Observatory, the spectrographs 
produced more than 1.5 million spectra for the SDSS and SDSS-II surveys, enabling a wide variety of Galactic
and extra-galactic science including the first observation of baryon
acoustic oscillations in 2005.  The spectrographs were upgraded in 2009 and
are currently in use for BOSS, the flagship survey of the third-generation
SDSS-III project. BOSS will measure redshifts of 1.35 million massive
galaxies to redshift 0.7 and Lyman-$\alpha$ absorption of 160,000 high
redshift quasars over 10,000 square degrees of sky, making
percent level measurements of 
the absolute cosmic distance scale of the Universe and placing tight constraints
on the equation of state of dark energy. 

The twin multi-object fiber
spectrographs utilize a simple optical layout with reflective collimators, gratings,
all-refractive
cameras, and state-of-the-art CCD detectors to produce hundreds
of spectra simultaneously in two channels over a bandpass covering
the near ultraviolet to the near infrared, with a resolving power $R = \lambda$/FWHM $\sim 2000$.
Building on proven heritage, the spectrographs
were upgraded for BOSS with volume-phase holographic gratings and
modern CCD detectors, improving the peak throughput by nearly a factor
of two, extending the bandpass to cover
360 $<$ $\lambda$ $<$ 1000 nm, and  increasing the number of fibers
from 640 to 1000 per exposure. In this paper we describe
the original SDSS spectrograph design and the upgrades implemented
for BOSS, and document the predicted and measured performances.
\end{abstract}


\section{Introduction}
\label{sec:intro}

The Sloan Digital Sky Survey \citep[SDSS;][]{york00a} project was conceived in the mid-1980s
as an ambitious endeavor to understand the large-scale structure
of the Universe. SDSS and its extension, SDSS-II, conducted a 
coordinated imaging and spectroscopic survey from 2000-2008 over approximately
10,000 deg$^{2}$ of high Galactic latitude sky.    
Now in its third phase of operation, SDSS is one of the most successful
projects in the history of astronomy. The survey has produced an enormous
catalog consisting of five-band digital images that include nearly one billion unique 
objects, and spectra of 930,000 galaxies, 120,000 quasars, and 460,000
stars, all publicly available \citep[][and references therein]{abazajian09a}.

To obtain these imaging and spectroscopic data, a dedicated 2.5m telescope \citep{gunn06a},
wide-field mosaic CCD camera \citep{gunn98a}, and twin multi-object
fiber spectrographs were constructed and installed at the Apache Point
Observatory (APO) in Sunspot, New Mexico.  
The telescope, built to accommodate the requirements for
both imaging and spectroscopy, is shared by the camera and spectrographs,
which mount at the Cassegrain focus. The imaging survey was carried out on clear,
dark nights with good seeing using the 120 mega-pixel camera, which operated in drift-scanning mode 
using a $5 \times 6$ array of $2048 \times 2048$ pixel detectors
to obtain $ugriz$ \citep{fukugita96a}, photometry.
The imaging data, once reduced
and calibrated \citep{smith02a,pier03a,ivezic04a,tucker06a,padmanabhan08a},
were used for spectroscopic target selection.
Spectroscopy was performed using the two multi-object fiber spectrographs,
collecting 640 spectra over the 3$^\circ$ diameter field in one exposure. 

In this paper, we describe the design and performance of the SDSS
spectrographs, and their recent upgrade for the Baryon Oscillation
Spectroscopic Survey \citep[BOSS][]{schlegel09a,dawson13a}. BOSS is the flagship survey in the third-generation
SDSS-III program currently underway at the 2.5m SDSS telescope \citep{eisenstein11a}. BOSS
will measure the cosmic expansion history of the universe to percent-level
precision by mapping an immense volume of sky to obtain the spatial
distributions of galaxies and quasars, and from it, the characteristic scale
imprinted by baryon acoustic oscillations (BAO) in the early universe
\citep[for a review of BAO with a respect to other cosmological probes, see][]{weinberg12a}.
A measure of the scale at low redshifts, out
to $z \sim 0.7$, will be obtained by carrying out a redshift
survey of 1.35 million massive galaxies from 10,000 deg$^{2}$ of SDSS data. BOSS will also observe Lyman-$\alpha$ absorption
in the spectra of 160,000 high-redshift quasars to measure large scale
structure at redshifts of $z \sim 2.5$.

Each SDSS spectrograph utilizes a dual-channel design with a common
reflecting collimator and a dichroic to split the beam into a blue
channel and a red channel. In each channel, just downstream of the
dichroic, a transmitting grism disperses the light, which is imaged
by an all-refractive camera onto a CCD. For BOSS the basic optical
design has been retained, with several improvements.
The ruled gratings have been replaced by volume-phase holographic
(VPH) grisms (gratings sandwiched between two prisms)
and the CCDs have been replaced with more modern devices.
These changes produce a significant improvement in throughput
and a modest extension of the wavelength range in both the blue and
red channels. Additionally, smaller diameter fibers that are better matched
to the angular scale of BOSS targets have been installed, allowing the total number
of simultaneous spectra obtained from the two spectrographs to be
increased from 640 in the original design to 1000 in the BOSS configuration.

The remainder of this paper is organized as follows. In Section 2
we begin by describing the design and construction of the original
SDSS spectrographs in some detail, published here for the first time.
This is followed in Section 3 by a discussion of the spectrograph
upgrades completed in 2009 for BOSS. The performance of both the original
SDSS spectrographs and the upgraded BOSS design is presented in Section 4. 
Finally, some highlights of the scientific
research enabled by these instruments is provided in Section 5.

\section{SDSS Spectrograph Design}

\subsection{Design Requirements}
\label{sec:SDSSreq}

The requirements for the SDSS spectrographs were set by its primary
scientific goal: the creation of a three-dimensional wide-area map of
the universe to reveal its large-scale structure.  The SDSS imaging
survey provides the two-dimensional locations of nearly one billion celestial
objects, and spectroscopy of a selected subset of targets is then used
to determine redshifts and thus distances.  The project set as a
requirement spectroscopy of one million galaxies and
100,000 quasars distributed over approximately 10,000 deg$^{2}$.

Acquisition of a large number of spectra
simultaneously over a large field of view, with moderate resolution
sufficient for accurate redshift measurements, naturally led to the
choice of a fiber-fed multi-object spectrograph.  The spectrograph
design was dictated in large part by the design of the telescope,
which was itself optimized for both wide-field, multi-band, imaging and multi-object
spectroscopy. Requirements were specified when possible; however, the instrument design was largely driven by technology available at the time.

In what follows throughout Section~\ref{sec:SDSSreq},
we summarize the requirements which dictated the design of the SDSS spectrographs. 

\subsubsection{Telescope Design}

The 2.5-m SDSS telescope is a modified distortion-free Richey-Chr\'etien
design with a 3$^\circ$ diameter field of view, and f/5 final focal
ratio, which provides a good match to fibers for spectroscopy
(180 $\mu$m diameter, $3\2pr$) and to the imaging CCDs (pixel size 24 $\mu$m, $0.4\2pr$).
The optical design incorporates two aspheric corrector lenses, a
Gascoigne-type design located near the vertex of the primary mirror, and
two interchangeable secondary correctors, one used for imaging and the
other for spectroscopy.  The imaging corrector is a thick fused
silica lens located close to the focal plane and is incorporated
into the SDSS camera, where it serves a mechanical function in addition
to providing optical correction.  The spectroscopic corrector is a
thinner lens located further from the focal plane and optimized for
chromatic focus.  The plate scale for spectroscopy is 3.627
mm/arcminute.  The spectroscopic focal surface is slightly curved, with
a maximum deviation from a plane of 2.6 mm.  One important detail of the
spectroscopic optics is that the central ray for each field point is not perpendicular to the
focal plane, necessitating a clever correction scheme for fiber
placement that will be described below.

\subsubsection{Number of Fibers}

Spectroscopy of approximately one million objects over 10,000 deg$^{2}$,
plus 10 -- 20\% additional fibers for calibration
sources and sky background measurements, implies a density of 120 deg$^{-2}$.
The 2.5-m telescope has a field of view of 7 deg$^{2}$, but each plate will view a unique area on the sky of
about 5 deg$^{2}$.  The higher density of plates is due to the need for overlap between
fields to ensure complete sky coverage without gaps, and to allow
multiple observations to cross-calibrate the entire survey.  The
required number of fibers is therefore approximately 600 per plate. 

A practical limit on the number of fibers was imposed by the detector
format, camera design, and fiber mounting scheme.
For proper spectral
sampling, the fiber images on the detector should be about 3 pixels in
diameter, with an equal space between spectra to reduce crosstalk and
allow for a measurement of the scattered light floor.  Thus, each
spectrum used six detector columns, and the $2048 \times 2048$ pixel detector
could accommodate a maximum of 341 spectra.
The actual number was
reduced to 320 spectra to avoid camera optical distortions near the
detector edges and to allow for extra gaps between groups of 20 fibers,
which was necessary for the fiber mounting scheme described in
Section \ref{sec:SDSSfiber}.  These larger gaps turned out to be quite useful
for measurements of scattered light in the wings.  The final choice of
640 fibers per plate, or 320 per spectrograph, provided some contingency
over the required 600 fibers, allowing for broken fibers, additional
calibration fibers and/or ancillary programs utilizing the extra
fibers. 

\subsubsection{Fiber Diameter}

The fiber diameter is set by the desire to maximize the signal-to-noise (S/N) ratio
for an extended source given the sky background.  For the galaxies of
interest around redshift $z= 0.1$ and the sky conditions at Apache
Point, this corresponds to a fiber size of around $3\2pr$, or a
fiber diameter of 180 $\mu$m.  Fibers of good optical quality were also
readily obtainable in this size.

 \subsubsection{Wavelength Range}

Redshifts are determined either from absorption lines or emission lines
-- in both cases only a few lines contribute most of the signal.  In
absorption, three features are dominant: the Mg b triplet at $\lambda =
5180$ \AA, Ca at $\lambda = 5270$ \AA, and the Na I doublet (D lines) at
$\lambda = 5890$ \AA.  At shorter wavelengths, the Ca II K and H lines at $\lambda =
3933, 3969$ \AA\ and the G band 4300 \AA\ may also be detected in absorption. 
In emission, H$\alpha=6353$ \AA\ is the strongest (and often the only)
line, although the [OII] doublet may also be visible at $\lambda = 3727$ \AA. 

Given the availability of these spectral features, and considering
practical limitations on UV throughput, the short wavelength cutoff was
set at 3900 \AA\ to ensure that the H and K lines of CaII are
observable even at zero redshift, while the [OII] doublet is observable
at $z > 0.05$.  Redshift determination for most nearby
galaxies could have been accomplished with a single blue arm extending
up to 6000 \AA; however, the SDSS imaging camera was designed to measure
to the detector red limit cutoff, so it was decided to take advantage of
the detector sensitivity in the spectrographs and add the red channel. 
This would enable observation of H-${\alpha}$ to a redshift of $z = 0.2$
or more, as well as the observation of quasars out to redshifts beyond $z=5$. 

Extension of the upper wavelength cutoff to 9100 \AA\
opened up a rich new vein of scientific discovery that was not
anticipated at the time of the instrument design.  In particular,
pushing the long wavelength cutoff as high as possible
extended the limit for redshift determination of luminous red galaxies
(LRG's) using the 4000 \AA\ break.  The LRG sample \citep{eisenstein01a} was used to make the
first observation of the baryon acoustic oscillation feature, which in
turn motivated the future upgrade of the spectrographs to even
longer wavelengths for BOSS, as discussed later in this paper.

\subsubsection{Resolving Power}

The spectroscopic resolution is defined as the full-width at half
maximum (FWHM) of the one-dimensional point spread function (PSF), in
wavelength units (a {\it resolution element}).  The resolving power is
the wavelength divided by this quantity, and we will often use the
phrase ``higher resolution'' to mean higher {\it resolving power}, as is
the normal usage.  Given a fixed number of pixels in the dispersion
direction and requiring proper sampling, increasing resolving power reduces the wavelength range. 
Higher resolving power also reduces the number of source photons per
pixel, increasing the exposure time required to exceed the CCD read
noise.  On the other hand, if the resolving power is too low, absorption
lines cannot be resolved and this will ultimately degrade both the
accuracy and success rate of redshift measurements. 

The resolution was therefore set by the requirement to obtain
spectroscopic redshifts of galaxies to an accuracy limited only by the
broadening due to typical velocity dispersions of 100 to 200 km/s.  This corresponds to
a resolving power of 1500-3000.  


The actual resolving power as a function of wavelength was
allowed to vary within these limits to optimize the red-blue channel
wavelength split location and the total wavelength coverage, while
maintaining well-sampled spectra with 3 pixels per resolution element on
the CCD over the full wavelength range.  These choices of spectrograph
parameters were chosen to optimize the overall redshift success rate for a 
given exposure time.

\subsubsection{Throughput and Signal to Noise Ratio}
\label{sec:SDSS_Throughput_S/N}

The requirement on throughput was set by the desire to obtain one
million spectra over 10,000 deg$^{2}$ to a limiting Petrosian
magnitude of $r = 18.15$ in five years, corresponding to roughly 100 deg$^{-2}$ galaxies.
Given the number of fibers and average weather at APO,
this implied an average exposure time of one hour.

Provided that the spectral resolution is sufficient to resolve the
absorption lines, the minimum S/N ratio needed to derive a
redshift depends mainly on the strength of the absorption lines.  For
convenience, the S/N per \AA\ of spectral continuum will
be quoted.  For an elliptical galaxy with strong absorption features, spectra obtained
in the Center for Astrophysics redshift surveys \citep{huchra83a,falco99a}
demonstrated that one can measure
a reliable redshift with S/N per \AA\ $>8$,
i.e., one needs to collect 64 object photons per \AA, assuming that
the noise is dominated by photon statistics from the source.  This
number must be increased, however, if sky background and/or readout
noise is significant.  A significant problem for some galaxies is that they have
weak absorption lines (presumably because they have a significant amount
of light from early-type stars) and yet lack strong H$\alpha$
emission.  In these cases one may need two or three times as many photons to
derive an absorption-line redshift.  We adopt as a guide the goal of
obtaining spectra with S/N of 15 per \AA.  Simulated galaxy and quasar
spectra indicated that we could in fact reach this goal with exposures of
somewhat less than one hour in typical conditions for seeing and atmospheric extinction.
The corresponding
throughput requirement, including atmospheric extinction and the telescope
throughput, varies as a function of wavelength; the maximum requirement is
about 17\% at 7000 \AA, with requirements of roughly 10\%, 15\%, and 10\% at 4000 \AA,
6000 \AA, and 8000 \AA\ respectively.  

\subsection{Fiber System Design}
\label{sec:SDSSfiber}

\subsubsection{Overview of Fiber System}

\begin{figure*}[htbp]
\begin{center}
\epsscale{0.8}
\plotone{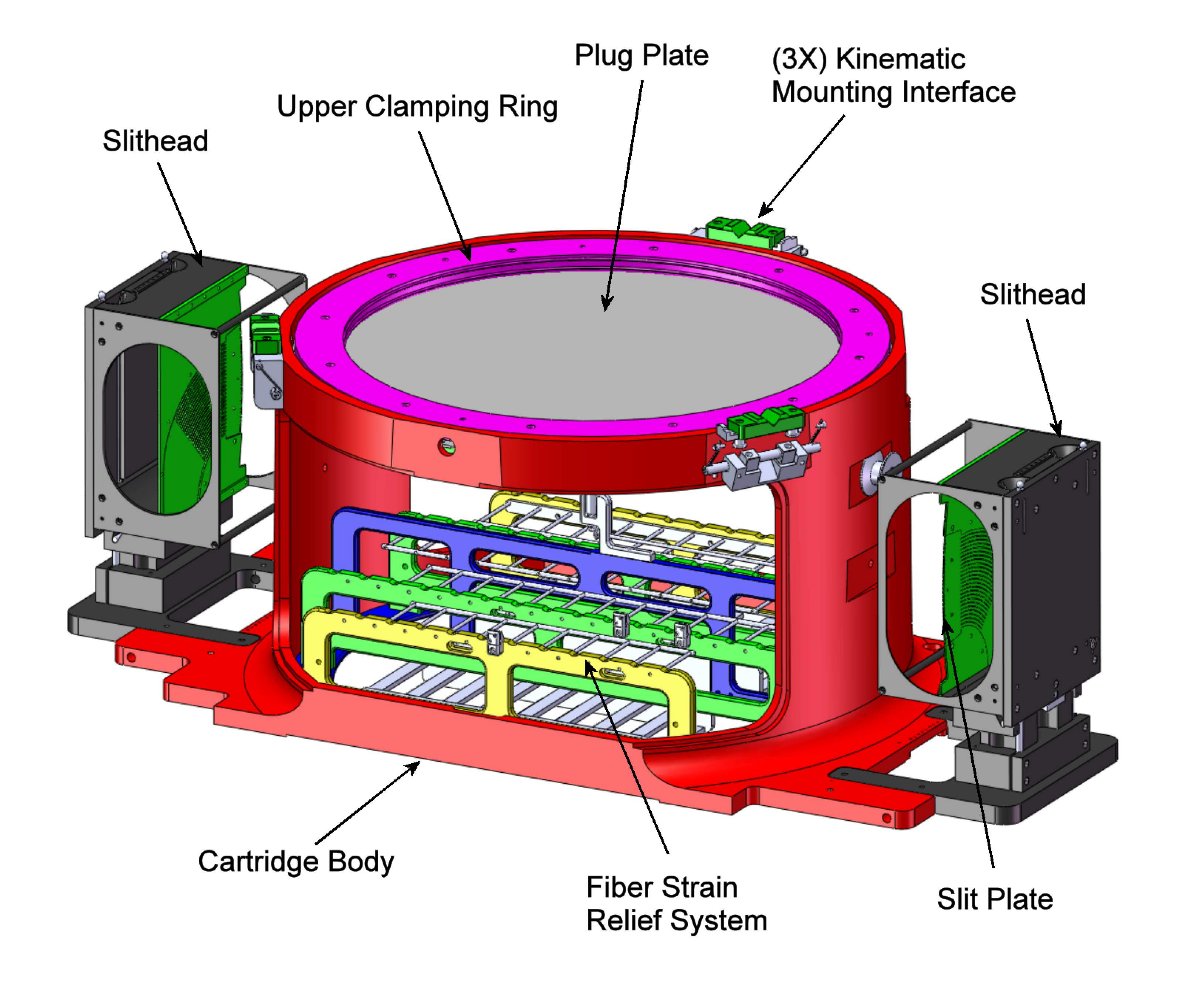}
\caption{{\bf Rendering of a fiber cartridge. The fiber cartridge consists of a cast aluminum body that supports the fiber harness, the two slitheads, and the plug-plate, which has a diameter of 800 mm. The slitheads are attached to the cartridge body with a spring-loaded seating system that provides alignment for insertion into the spectrograph bodies, but then allows the slithead to float free from the cartridge body and engage the slithead-to-spectrograph kinematic mounting system. Kinematic mounts around the periphery of the cartridge casting ensure accurate and repeatable placement of the cartridge with respect to the telescope.}}
\label{fiber_cartridge}
\end{center}
\end{figure*}

\begin{figure}[htbp]
\begin{center}
\epsscale{1.17}
\plotone{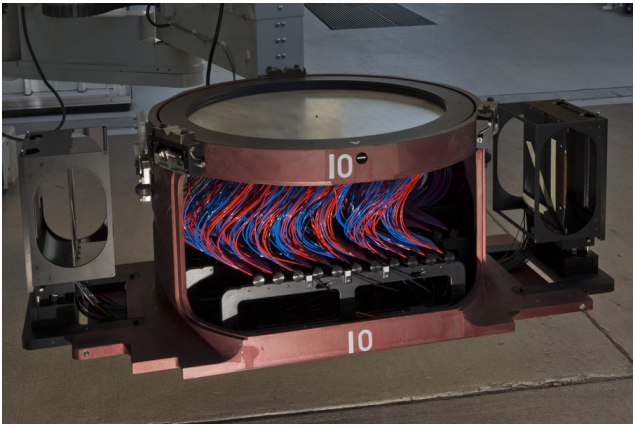}
\caption{{\bf Photograph of a BOSS fiber cartridge. Fibers plugged into the back of the plug-plate are routed in bundles to the slitheads (the two boxes standing upright at the left and right side of the cartridge).  The design shown is identical to that used for SDSS except for the number and size of the fibers. For SDSS, 320 fibers are routed to each slithead, while for BOSS each slithead carries 500 fibers.}}
\label{cartridge_photo}
\end{center}
\end{figure}

Light is transmitted from the telescope focal plane to two identical spectrographs by fiber optic strands 
180 $\mu$m in diameter ($3\2pr$ on the sky). 
Light enters the fibers at the telescope focal plane
in a cone of numerical aperture 0.1 (f/5 beam),
and the spectrograph collects light emitted from the other end of the fibers in a slightly
larger cone with numerical aperture 0.125 (f/4), due to focal
ratio degradation (FRD) that occurs as the light travels down the fiber.  
Any light emitted outside this cone will be lost, so 
a primary requirement on the fiber system is to limit FRD so as to maximize
throughput.  To this end, the spectrographs are mounted on the telescope to avoid any
relative motion between the two ends of the fibers 
and the potential stress that can result in increased FRD (an issue that was not well understood at the time).  References available in those days (early 1990s) suggested that the macrobending of fibers is benign \citep{angel77a,heacox86a,clayton89a}, however a recent study shows that the {\it repeated} bending of fibers, as would be the case for a bench-mounted spectrograph, can increase FRD over several years of operation \citep{murphy12a}. 
This scheme also maximizes throughput by keeping the fibers short,
minimizes fiber throughput variations due to physical motion and stress, and  
avoids the problems of routing and protecting long fiber runs.
The sky ends of the fibers are plugged into drilled 800 mm diameter aluminum plates called \emph{plug-plates}
that position the fibers on the spectrograph focal plane, and the other ends of the fibers are terminated in one of two slitplates. Each thin slitplate is mounted to a rigid frame with precision locating features for accurate placement in the spectrograph.
The assembly of plug-plate, fibers and slitheads is mechanically supported by a portable aluminum cartridge that can be installed on the telescope by a single operator in a few minutes.   New plug-plates are mounted on the cartridges during the day and 
plugged with fibers, then sequentially mounted on the telescope during the night. A rendering and photograph of a fiber cartridge are shown in Figures~\ref{fiber_cartridge} and~\ref{cartridge_photo}, respectively.

\begin{figure}[htbp]
\begin{center}
\epsscale{1.17}
\plotone{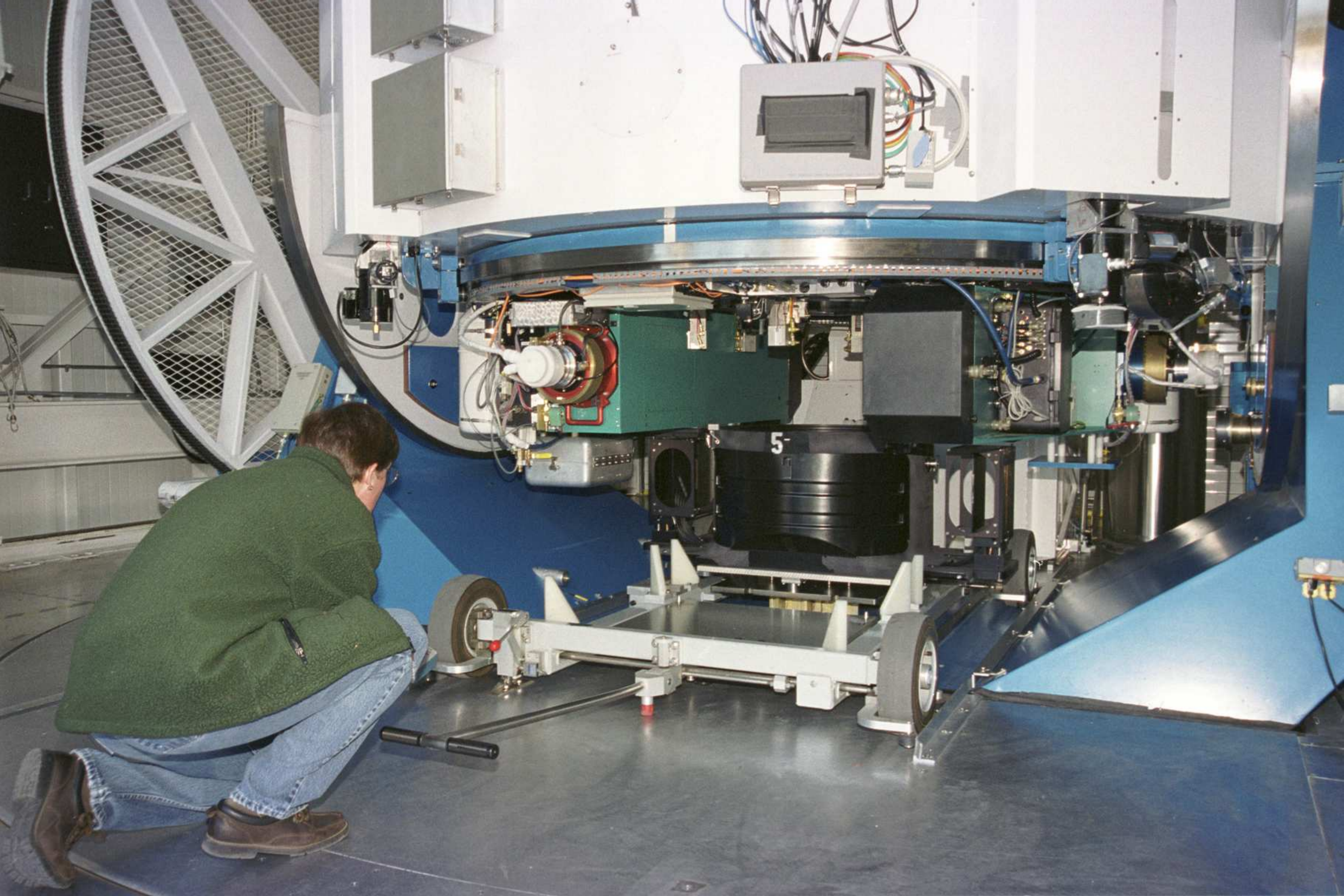}
\caption{{\bf Photograph showing a fiber cartridge being installed on the telescope.  The twin spectrographs are the green instruments on either side of the focal plane.  The cartridge is raised by a hydraulic lift in the floor below the primary cell.  When raised, the cartridge engages kinematic mounts for precise location. At the same time, the two slitheads engage the spectrographs, each of which is located by its own kinematic mounting features integral to the slithead and spectrograph optical bench.  Installation takes approximately three to five minutes.}}
\label{Linde_photo}
\end{center}
\end{figure}

For each new sky field, a cartridge is wheeled under the telescope using the Linde cart (named for its designer Carl Lindenmeyer) and attached to the telescope rotator 
using pneumatic clamps. At the same time  the attached slitheads 
enter the spectrographs through the open slithead doors and are clamped in place. A kinematic mounting interface ensures accurate, repeatable placement.
The photograph in Figure~\ref{Linde_photo} illustrates the operation.
Eight cartridges were fabricated for SDSS to provide
sufficient pre-plugged plates for an entire night of observing.

Each cartridge also has a set of coherent fibers
that are placed on pre-selected guide stars and viewed by the guider camera.
These stars are used for field rotation and translation to align the plug-plate to the field.

\subsubsection{Cartridges}
The fiber cartridge consists of a machined aluminum cast body that supports the optical fiber harnesses, spectrograph slitheads, and plug-plate. 
Assembling these components into a single robust unit protects the fragile fibers during the manipulations necessary for plugging, transport to and from the telescope, and mounting onto the instrument rotator.  The cartridges are plugged during the day, and are designed so they can be quickly installed on the telescope at night under often difficult conditions of low light and cold temperatures.

The plug-plate holder consists of two large bending rings that warp the plug-plate to match the telescope best-focus surface.  An adjustment rod centered on the back side of the plug-plate is used to fine-tune the plate curvature. 
The bending rings are mounted to the cartridge body with a set of kinematic pin mounts.  The alignment of the cartridge to the telescope is provided by another kinematic 
mount employing v-groove blocks on the cartridge that engage with v-blocks on the telescope.  This system ensures
proper and repeatable alignment between the plug-plate and telescope focal plane.

The two slitplates, each supporting 320 fibers, are mounted in their respective slitheads. These slitheads are aluminum assemblies mounted outboard of the cartridge body that support and protect the slitplates. 
The slitheads are attached to the cartridge body with a spring-loaded seating system that provides alignment for insertion into the spectrograph bodies, but then allows the slithead to float free from the cartridge body and engage the slithead-to-spectrograph kinematic mounting system. When not mounted on the telescope, these slitheads are protected by sliding covers to prevent contamination and/or mechanical contact with the delicate slitplates.

All cartridge operations occur at the same elevation, on the telescope platform and the adjacent support building. 
In the plugging lab, the exposed plug-plates from the previous night's observing are unplugged and removed
from their cartridges and new plates are installed. Once plugging and fiber mapping is completed
(a process that takes 30 -- 45 minutes), the 145 kg cartridge is stowed on a lift table installed in a
bay that provides both interior and exterior bay door access.
At night, the outside door is opened to allow the cartridges to equilibrate to the temperature of
the ambient air. 

To install a new cartridge on the telescope, an outside manipulator arm is employed to move the cartridge from the storage bay to one of \emph{two} receiving plates on the Linde cart. The cartridge is then wheeled from the storage bay to the telescope. With the telescope parked at zenith and locked into position, the Linde cart is rolled under the mounted cartridge to align the \emph{empty} receiver plate with it.  Aided by a hydraulic lift, the observer removes the exposed cartridge from the telescope. Then, maneuvering the Linde cart to align the unexposed cartridge onto the hydraulic lift, the observer mounts the new cartridge onto the telescope instrument rotator. Once the new cartridge is latched and the cart receiver plate is lowered back onto the Linde cart, the cart is rolled out from under the telescope. The telescope is now ready to move to the next field and to begin another exposure. Only three to five minutes is required to perform this cartridge change. 

\begin{figure}[htbp]
\begin{center}
\epsscale{1.1}
\vspace{2mm}
\plotone{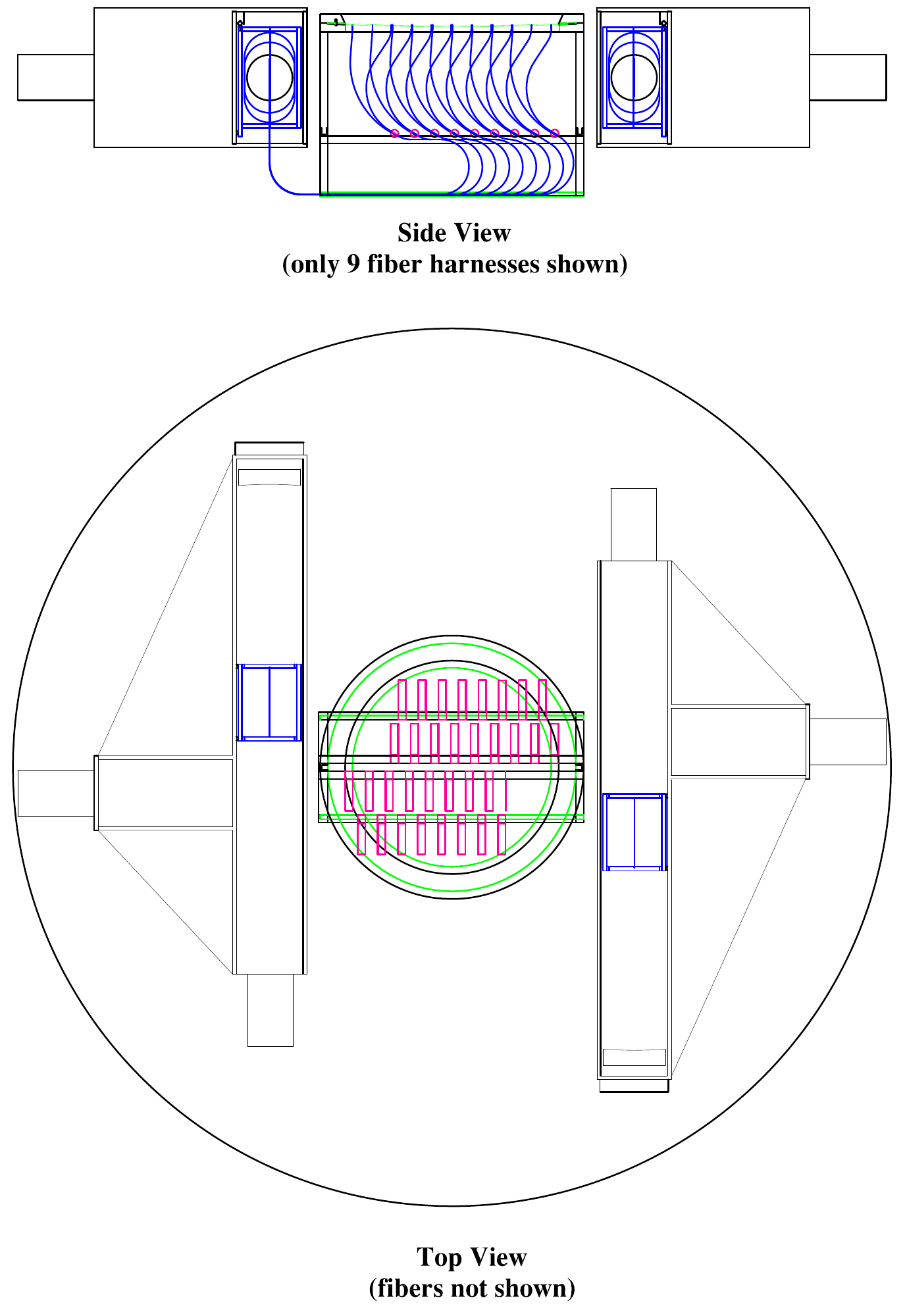}
\caption{{\bf Two schematic views of the cartridge mounted on the telescope.  Top:  a cutaway side view showing the slitheads inserted into the spectrographs (only 9 fiber harnesses are shown).  Bottom:  top view showing  the cartridge located between the two spectrographs which are mounted to the instrument rotator (depicted as the large outer circle).}}
\label{SDSScartridge}
\end{center}
\end{figure}

As the cartridge is lifted into place and clamped to the telescope, the slitheads are simultaneously
inserted into sockets in the spectrographs. The slitheads are attached to the cartridge frame by
stiff springs so that they can move slightly with respect to the rest of the cartridge.
Once the cartridge has been correctly positioned and clamped to the telescope,
each slithead is loaded against a three-point kinematic mount on the spectrograph by a single pneumatic clamp.
A flexible rubber seal between the slitheads and the spectrograph bodies prevents extraneous light
from entering  during exposures. Each slithead is coded and its identification relayed to the
observer's workstation when it is inserted. This information allows adjustments for each slithead,
e.g., image placement on the CCD and focus, to be made automatically. 
Figure~\ref{SDSScartridge} shows two schematic views of the cartridge mounted on the telescope
with the slitheads inserted into the spectrographs.

\subsubsection{Optical Fiber}
The selected optical fiber material is a silica UV-enhanced step-index fiber with a core diameter of 180 $\mu$m, a thin cladding and 
a polyimide protective layer.  The actual fiber was Polymicro Technologies, Inc.\footnote{Polymicro Technologies, Inc., http://www.polymicro.com}  FHP 180-198-218, where the
numbers refer to the diameter of the bare fiber, plus cladding and plus polyimide buffer.

\subsubsection{Fiber Harnesses}
\begin{figure}[htbp]
\begin{center}
\epsscale{0.9}
\plotone{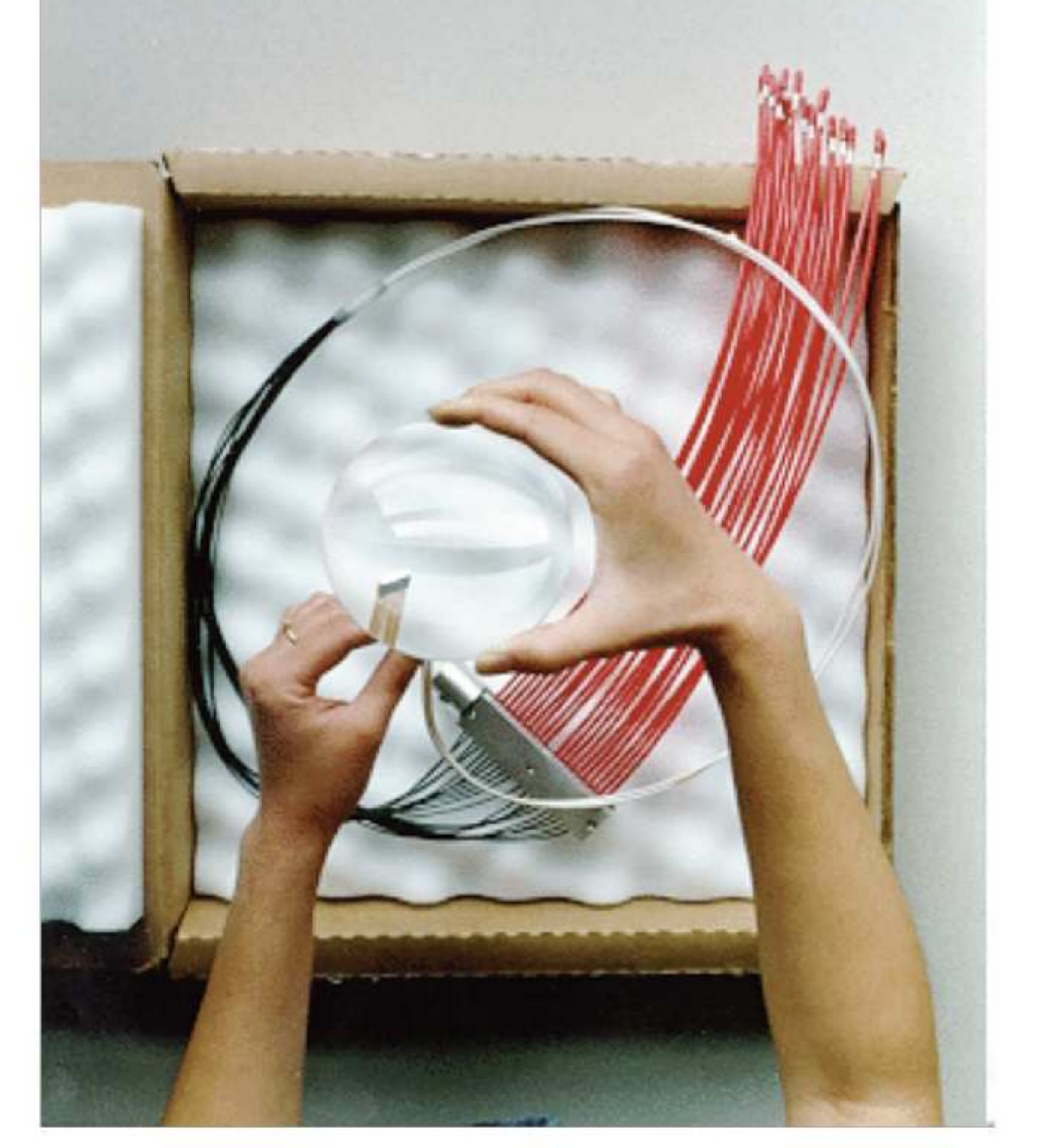}
\caption{{\bf Fiber harness for the SDSS spectrographs.  The ends, which are plugged and unplugged during operations, are protected by tough nylon tubing, which terminates in the anchor block. The lens enlarges the view of the v-groove block termination at the slit end.}}
\label{SDSSharness}
\end{center}
\end{figure}

A fiber harness consists of 20 fibers of length $1.865 \pm 0.025$ m;  each fiber cartridge contains 32 fiber harnesses.  The fiber harness
is the unit procured from C Technologies, Inc.\footnote{C Technologies, Inc., http://www.ctechnologiesinc.com} the vendor responsible for preparing the fibers by terminating both ends.   A fiber harness is shown in the
photograph in Figure~\ref{SDSSharness}.
  
The plugging end is terminated
by gluing it into a ferrule and polishing the end.  The ferrule is machined from stainless steel; the most important
tolerances on the ferrule are the centering and angle of the hole in which the fiber is inserted,  specified at 12 $\mu$m and
$1.1^{\circ}$, respectively.  The tilt of the plane of the polish on the fiber end is specified at less than $1.0^{\circ}$.  The fiber is glued into the
  ferrule using a low-shrinkage adhesive (MasterBond EP21AR\footnote{Master Bond, Inc., http://www.masterbond.com}) selected to retain flexibility for many years over a temperature range -20$^{\circ}$ C to +30$^{\circ}$ C.
To facilitate handling the fiber assemblies during plugging and unplugging while protecting the fragile fibers, a protective jacket 380 mm long is attached to the ferrule on one end and
glued into an anchor block on the other end.  The selected jacket material consists of nylon tubing 3.2  mm in diameter, allowing the fiber to move freely
inside to avoid imparting any stress.  The anchor block holds twenty fiber jackets in a row on 5.1 mm centers.   The anchor blocks are attached to the cartridge body,
providing a fixed point from which the fibers are routed to the slitplate.  The fibers are stationary from the anchor block to the slitplate, and the only portion of the fiber which moves during plugging is  protected inside the nylon  jacket.  Thin tubing, 950 mm long and 2 mm in diameter, is used to bundle the fibers into groups of 10 immediately before their termination at
the slitplate.
  
\begin{figure}[htbp]
\begin{center}
\epsscale{1.18}
\plotone{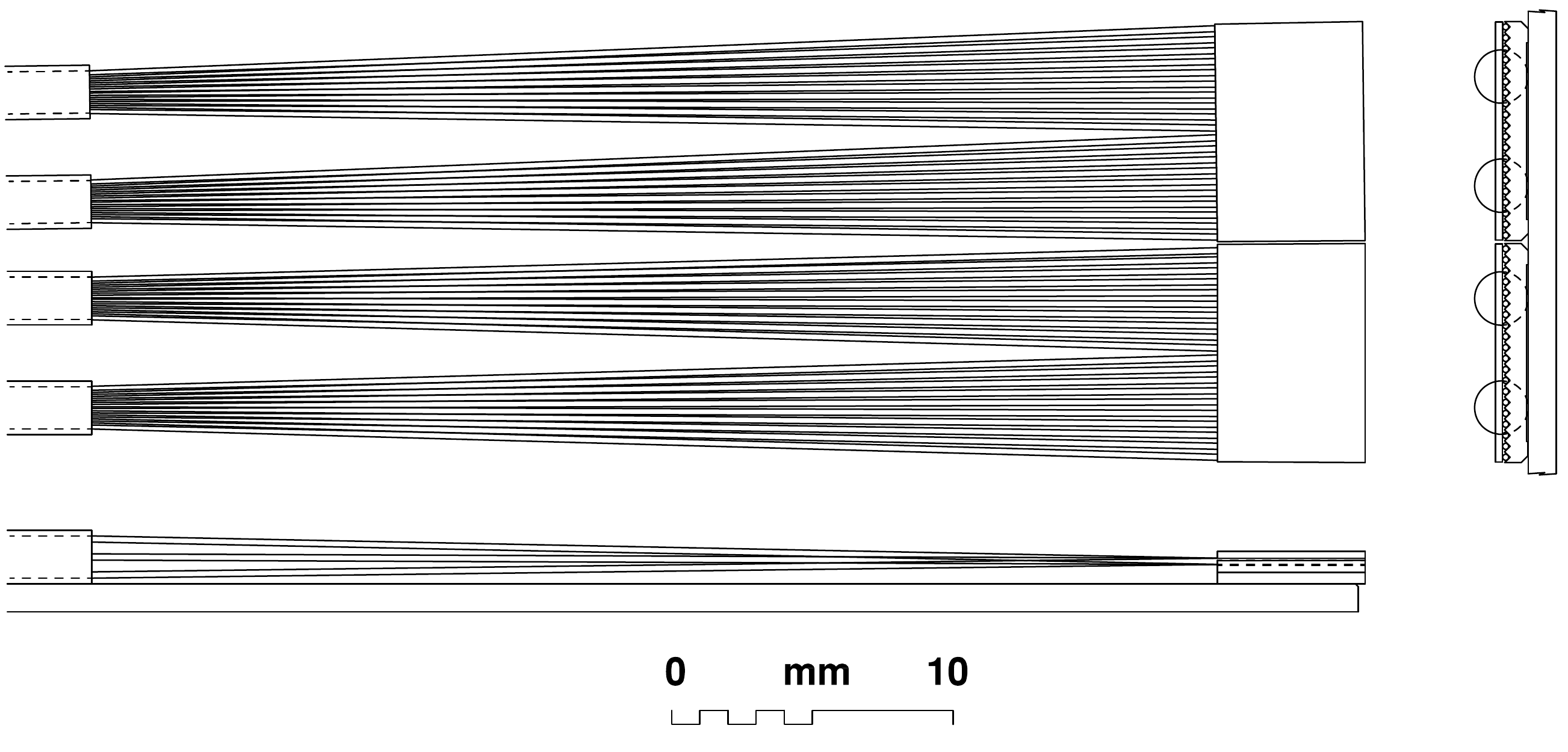}
\caption{{\bf Three views of the v-groove blocks as mounted on the slit plate. Each block contains 20 fibers, which are aligned by 20 v-shaped grooves.  The fibers are held in place by epoxy sandwiched between the fibers and a cover plate.
A hole in the slit plate under each block permits easy removal for replacement.}}
\label{SDSSvgroove}
\end{center}
\end{figure}

At the slitplate end, twenty fibers at a time are mass-terminated by gluing them (using MasterBond EP21AR) into a machined piece of stainless steel
called a v-groove block; see Figure~\ref{SDSSvgroove}.   This item is so named because it has twenty
v-shaped grooves machined on an electrical discharge machine (EDM) to precisely locate the fibers.
The grooves are not parallel, but fan out slightly so they are normal to the tangent of a circle with radius 640 mm. 
For termination, twenty fibers are arranged side by side in the v-groove block on 390 $\mu$m centers and glued into place
with a cover plate on top. The tolerance on the fiber placement is 30 $\mu$m (individual, not cumulative), and
$0.3^{\circ}$ in tilt between the groove and length of the fiber; there is an additional tolerance of $0.3^{\circ}$ on the
alignment of the grooves with respect to the body of the v-groove block.   After the fibers are aligned in the
v-groove block, covered and glued in place, the ends are polished on a flat surface, with a specified tolerance of
$1.0^{\circ}$ on the tilt of the plane of polish.
 
\subsubsection{Slitplates}
  
Each spectrograph accepts light from 320 fibers, terminated in 16 v-groove blocks that are glued to a thin, rigid aluminum slitplate with
a 640 mm radius of curvature.  The center-to-center spacing between fibers in adjacent v-groove blocks is 523 microns,
compared to 390 microns between fibers within a v-groove block. 
The v-groove block is aligned by locating the output edge against a curved alignment jig; it is fastened to the slit plate with adhesive. The adhesive is metered using an air-powered fluid dispenser to prevent contamination of critical surfaces.  The slitplates are in the beam and therefore must be as thin as possible; the total thickness
of the plate and v-groove blocks is specified to be less than 3.2 mm.

\subsubsection{Plug-Plates}
The telescope optical system is a simple, fast, large field design with a focal surface flat
to 2.6 mm but one where the principal ray deviates from the normal to the best-focus
surface by up to 37 milliradians ($2.12^\circ$).
For highest efficiency, the ends of the optical fibers should be positioned on
the best-focus surface with their axes aligned with the principal ray.
Plug-plate technology can be made to satisfy these criteria quite nicely.

The plug-plates are an aluminum alloy 2024-T3, 3.2 mm thick and 800 mm in diameter. By applying bending moments to the edge of the plate (beyond the field of view), finite element calculations show they can be deformed to match the best-focus surface to an area-weighted 62 $\mu$m RMS. The greatest departure from the best-focus surface is 200 $\mu$m and occurs at the center where the images are the best. Overall, the images are not significantly degraded from the best-focus surface.

As deformed to match the best-focus surface, the hole axes should align with the principal ray axes. This configuration is straightforward to accomplish if the plug-plate is deformed (in the opposite sense) over a properly curved mandrel, for drilling. The drilling is performed using a three-axis Computer Numerically Controlled (CNC) milling machine, as it is not necessary to tilt the drilling head or the plug-plate.  Plug-plate drilling is performed in the Physics Machine Shop at the University of Washington, Seattle WA. 

Drill test results indicate that holes can be drilled with an accuracy of 9 $\mu$m RMS in position and 4 $\mu$m RMS in diameter using short high-precision spade drill bits in a custom-made collet. In the test, four different bits were used to drill 50 holes each. The drilling time was 5.8 sec/hole. No significant degradation in drilling accuracy was observed for a range of slopes in the work-piece surface from 0 to 70 milliradians ($4.01^\circ$).

The plug-plates have a mass of 4.3 kg and are sufficiently thin so that the bending stresses, forces and material costs are reasonable. They are thick enough to provide hole depth adequate to constrain the plug angular alignment with the hole and to prevent significant gravity-induced deflections. 

Once plugged, profilometry is performed on the plug-plates in the plugging laboratory to verify that the
surface conforms to the focal plane within tolerances.  This is done using a fixture that 
mounts five digital linear indicators on a steel beam that straddles the plug-plate,
using a kinematic support system to make contact with the aluminum.
The five indicators measure the plate displacement from a calibrated zero reference along a radial
direction from the center to the edge of the plug-plate.
There are four sets of measurements taken, spaced 90$^\circ$ apart for each plug-plate.
The measurements are then recorded in a database for each plugging of each plate.\footnote{portions of the description provided in this section have been copied from http://www.astro.princeton.edu/PBOOK/welcome.htm and are reproduced here for completeness}

 \subsubsection{Fiber Tester}
Each completed fiber harness is tested by the vendor to ensure that each of the twenty fibers meets a minimum throughput requirement of 85\%.  The test device, which was supplied to the vendor, uses white light from an intensity-stabilized quartz-halogen lamp which is imaged onto the fiber under test using a source fiber and a microscope objective, which produces a uniform f/5 converging beam. A microscope eyepiece and pellicle beamsplitter provide a view of the input end of the fiber under test.

Light from the output end of the test fiber is collected by a pair of achromatic doublets focused onto a silicon photodiode. A filter between the doublets limits the source to the visible range where the optics perform well.  A calibrated aperture blocks light outside a cone of f/4. A computer-controlled translation stage accurately locates the appropriate fiber of the v-groove block in front of the aperture. This same light collection system is also used to measure light from the microscope objective, in order to make absolute throughput measurements.

An identical instrument was used to verify the manufacturer's measurements for some fraction of the fibers.\footnote{portions of the description provided in this section have been copied from http://www.astro.princeton.edu/PBOOK/welcome.htm and are reproduced here for completeness}

\subsubsection{Fiber Mapper}
\label{sec:SDSS_Mapper}

With 640 fibers on each plug-plate, it was important not to rely on manual plugging of each fiber into a
specific pre-assigned hole.  However, plugging cannot be completely random because not every fiber
can reach every position on the plate; this constraint was adopted to keep the total fiber length relatively short.  
The solution was to have hand-marked regions on each plate for specific fiber sets that can reach all
of the holes within the delineated area, and to then allow for random plugging within those regions.  After plugging is complete,
the cartridge is wheeled over to a fiber mapper system which is used to determine which fiber is in which hole. 

The fiber mapper illuminates each fiber sequentially by moving a light along the slithead;  
the illuminated fiber  appears as a bright point against the dark background of the plug-plate. 
A CCTV camera behind a narrow band filter matching the light source determines the X-Y position of each fiber on the plug-plate. 
This X-Y position is converted to sky coordinates and stored in the database, providing a map between the physical location of each
spectra on the CCD and the corresponding coordinates on the sky.  The fiber mapper is also useful for identifying broken or loose
fibers that drop down into the hole after plugging.  The entire operation takes about five minutes.

\subsection{Optical Design}
\label{sec:SDSSoptic}

\subsubsection{Optical Design Overview}

The optical design of the spectrographs was strongly influenced by a few considerations including, the use of optical fibers to feed light from the telescope to the spectrographs, available large-format science-grade CCDs, the required bandpass, and the required resolving power.  The exit face of a fiber, which is the entrance aperture into the spectrograph for one object, must be collimated, dispersed, and finally re-imaged onto a detector with adequate pixel sampling.  At the time, Tektronix/SITe 2k $\times$ 2k format CCDs having 24 $\mu$m pixels were state-of-the-art.  Given a fiber output focal ratio of f/4, the collimated beam must be re-imaged by a camera with a focal ratio of f/1.5 to provide sufficient demagnification to achieve three-pixel images of the fibers on the 24 $\mu$m pixel pitch.  At this sampling, approximately 700 resolution elements are available across the detector.  Clearly, the full 3900-9100 \AA\ bandpass will not fit within a single detector at the required resolving power of $\sim$ 2000.  Thus, two channels are needed.  Lastly, as a practical matter, it was determined that a collimated beam diameter of 160 mm would be ideal;  larger would begin to require unreasonably large optics, while a smaller beam would require higher field angles into the cameras and therefore more difficult designs.  With this beam diameter the cameras have a focal length of 240 mm, and reasonable grating designs can provide the necessary angular dispersion.

\begin{figure*}[htbp]
\begin{center}
\plotone{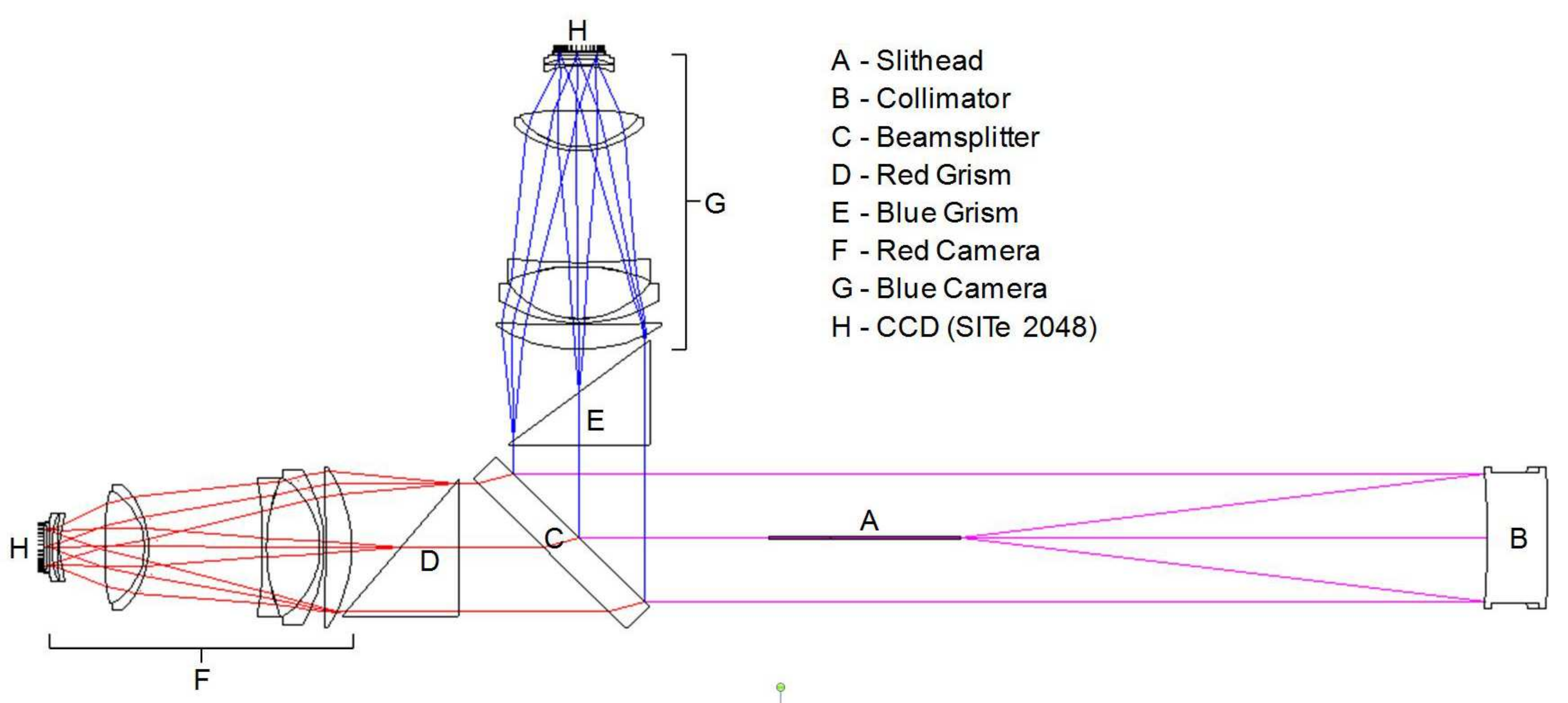}
\caption{{\bf Optical layout of the SDSS spectrographs. Light enters each spectrograph through 320, 180 $\mu$m diameter, fibers, which terminate at a curved slit plate mounted inside the slithead.  The slit plate positions the fiber ends on a radius concentric with the spherical collimating mirror, which  operates at f/4 and produces a 160 mm diameter beam.  The 45 degree dichroic beamsplitter reflects the blue portion of the bandpass ($\lambda < 6000$ \AA) and transmits the red wavelengths ($\lambda > 6000$ \AA).  Immediately after the beamsplitter in each channel is a grism, consisting of a right angle prism with a transmissive surface-relief grating replicated on the hypotenuse.  The dispersed light exits the grisms and enters all-refractive, eight-element, f/1.5 cameras.  Each camera contains a single SITe 2k $\times$ 2k CCD with 24 $\mu$m pixels. The camera demagnification from f/4 to f/1.5 produces fiber images that are just under 3 pixels in diameter, resulting in 3 pixel tall spectra on the detector.}}
\label{SDSSoptical}
\end{center}
\end{figure*}

Figure~\ref{SDSSoptical} shows the optical layout of the SDSS spectrographs.  Light enters each spectrograph through 320 fibers, which terminate at a curved slithead.  The slithead positions the fiber ends on a radius concentric with the spherical collimating mirror, which  operates at f/4 and produces a 160 mm diameter beam.  The 45 degree dichroic beamsplitter reflects the blue portion of the bandpass ($\lambda < 6000$ \AA) and transmits the red wavelengths ($\lambda > 6000$ \AA).
Located immediately behind the beamsplitter in each channel is a grism, consisting of a right angle prism with a transmissive surface-relief grating replicated on the hypotenuse.  The dispersed light exits the grisms and enters all-refractive, eight-element, f/1.5 cameras.  Each camera contains a single Tektronix/SITe 2k $\times$ 2k CCD with 24 $\mu$m pixels.  The camera demagnification from f/4 to f/1.5 produces fiber images that are just under 3 pixels in diameter.
A fiber-center separation of 390 $\mu$m at the slithead produces six pixel center-to-center spectra spacing at the detector, sufficient to avoid crosstalk between spectra.

\subsubsection{Slithead}

Light enters the spectrograph through 180 $\mu $m diameter fibers, which terminate at the curved slit plate; see Section~\ref{sec:SDSSfiber} for more details on the slithead design. The fibers are stacked vertically to form a tall, narrow, slit on a radius whose center of curvature coincides with that of the collimator. Additionally, the fibers are arranged in a fan-like pattern outward from the center of curvature toward the collimator, so that the central ray from each fiber strikes the collimator normal to the surface. Thus, the slit plate is at the focus of a one-dimensional Schmidt collimator. 

\subsubsection{Collimator}

In combination with the curved slithead, the spherical collimator mirror forms a corrector-less Schmidt collimator. It was possible to eliminate the corrector plate since some of the reduced performance could be compensated for in the spectrograph cameras; enough to satisfy the imaging requirements. Thus, the final collimator design is a single mirror with a spherical figure. The mirror itself is fabricated from a rectangular Hextek\footnote{Hextek Corp., http://www.hextek.com} gas-fusion process borosilicate blank, 175 mm wide, 419 mm tall, and 73 mm thick. The planar blank is slumped by Hextek  to near-net radius and then ground and polished to the final radius of 1264 mm. An enhanced silver coating was applied to the optical surface by Denton Vacuum.\footnote{Denton Vacuum LLC, http://www.dentonvacuum.com} The collimator forms a pupil at the center of curvature of the mirror, and this is where the gratings are located in each channel in order to minimize their required size. 

Alternative mirror technologies were investigated, including aluminum, eggcrate Zerodur, and monoliths. The gas fusion blank was determined to be optimal given weight and cost considerations. The thermal coefficient of expansion, while non-zero and much smaller than aluminum, is an acceptable compromise. Obviously an aluminum mirror would eliminate the need to refocus when the temperature changes, but is expensive and the coatings are tricky. An eggcrate Zerodur mirror is expensive and would require refocusing. Lastly, a monolith would not be much cheaper than the gas fusion mirror, and would be significantly less responsive thermally than the gas fusion mirror.

To compensate for changes in instrument temperature with time, focus adjustment is provided by the collimator mount (see Section~\ref{sec:SDSSmech}), which enables not only focus adjustment but tip/tilt adjustment as well. A pair of Hartmann doors located in front of the mirror (also described in Section~\ref{sec:SDSSmech}) allow shifts in the collimator focus to be measured rapidly. Tip/tilt adjustment allows the collimator to be precisely coaligned to the cameras, or at least to an average of the two camera axes. Thus, the center fiber can be positioned at the center of the detector in the spatial direction, and the central wavelength can be positioned at the center of the detector in the spectral direction.

\subsubsection{Beamsplitter}

A dichroic beamsplitter divides the incident collimated beam, reflecting the blue portion of the bandpass ($\lambda < 6000$ \AA) and transmitting red wavelengths ($\lambda > 6000$ \AA). It is fabricated from BK7, is 271 mm wide, 229 mm tall, and 38 mm thick, with the dichroic coating applied to the incident surface. The coating reflects the blue light very efficiently (R $>$ 98\%) and transmits the red light somewhat less efficiently (T $>$ 94\% average, including the reflection loss at the exit surface, which has a high performance broadband antireflection coating). The 10\%/90\% zone at the crossover wavelength is approximately 50 nm wide.

\subsubsection{Gratings}

The dispersing elements are grisms with zero angular deviation at 4960 \AA\ for the blue and 7400 \AA\ for the red. These are right angle BK7 prisms with a transmission grating replicated on the hypotenuse. While a reflection grating might have been used, the grism permits mounting the cameras close to the system pupil, which is located, approximately, in the middle of the grism. With a reflection grating, the cameras have to be mounted away from the grating to avoid interference with the incoming beam, making them larger and more difficult to design. A plane transmission grating does not work because the diffracted angle is large, making geometric losses high (the groove facets are foreshortened) and forcing the blaze peak outside the optical band. The configuration used here has little groove shadowing or foreshortening and results in high grating efficiency.

The ruling densities are 640 and 440 lines/mm for the blue and red grisms, respectively. Because master rulings of the necessary size and groove angle did not exist, new masters were ruled by Hyperfine, Inc.

\subsubsection{Cameras}

\begin{figure}[htbp]
\begin{center}
\epsscale{1.15}
\plotone{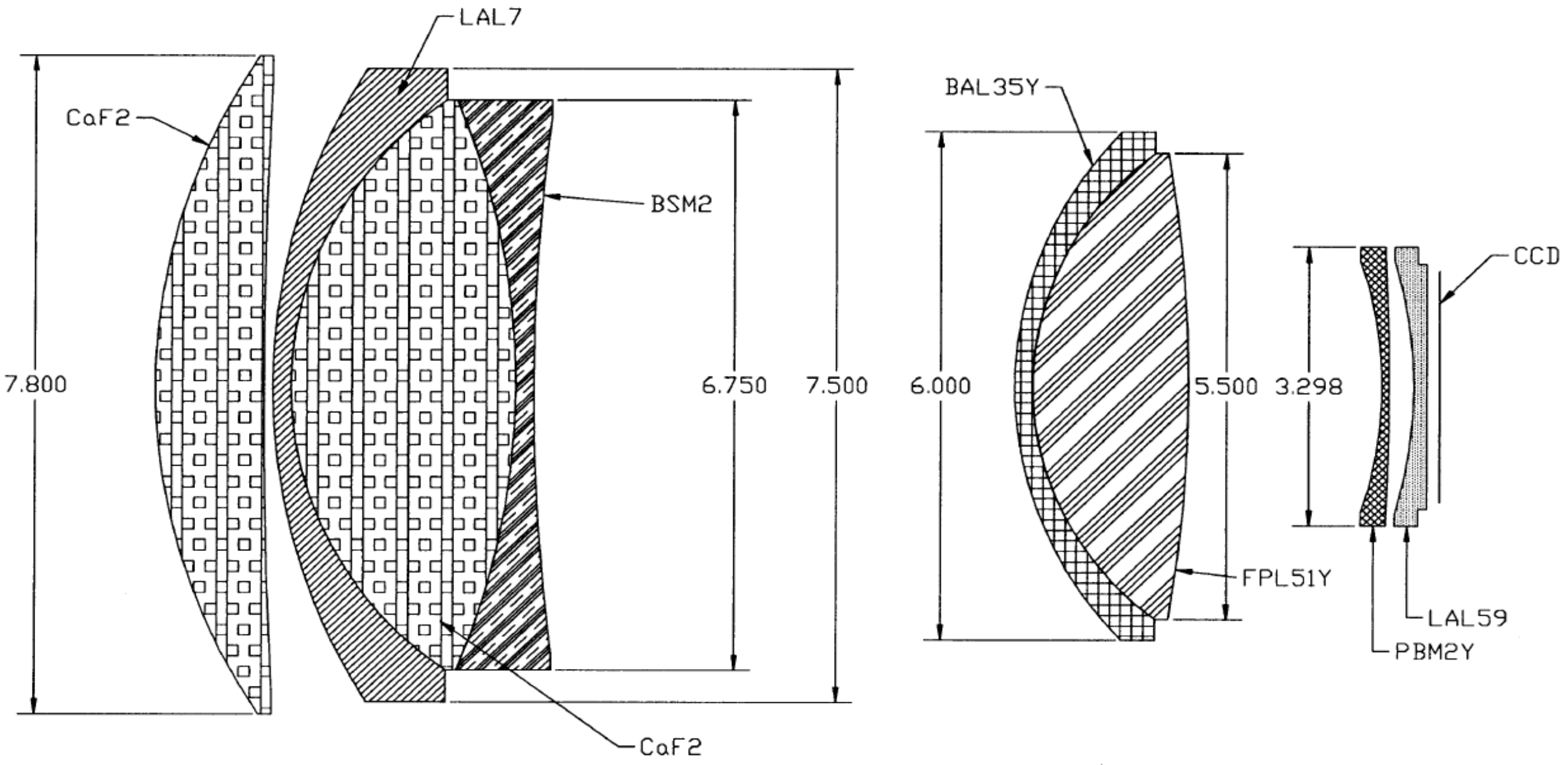}
\caption{{\bf Optical layout of the SDSS spectrograph blue channel camera designed by Harland Epps. Diameters are in inches. The red channel camera layout shares the same eight-element configuration, though the prescription for the doublet and field flatteners differ for the two camera designs. Anti-reflective coatings are optimized for the respective bandpasses of each channel.}}
\label{camera}
\end{center}
\end{figure}

The spectrograph cameras are  f/1.5 with a 240 mm focal length and 16.5 ${\deg}$ field of view; see Figure~\ref{camera} for a layout of the blue camera (the red camera layout is nearly identical).  The all-refractive design was chosen to maximize throughput since placing the detector or a secondary mirror in the unobstructed collimated beam, as would be the case for example with a Schmidt camera, would result in significant light loss. Both the red and blue cameras share the same optical prescription for the first four elements (i.e. the singlet and triplet). However, the doublets have unique prescriptions for the red and blue designs, as do the field flatteners.  Anti-reflective coatings are optimized for the respective bandpasses of each channel.

Harland Epps (University of California, Santa Cruz) was contracted to design the cameras, using the Low Resolution Imaging Spectrometer (LRIS) spectrograph camera design \citep{epps90a} as a starting point with some additional constraints.  The most important was to limit aspheric surfaces to one per camera with only mild angles (LRIS had two steep aspheric surfaces). The reason for this was to reduce schedule risk and to increase the vendor pool.  The camera employs eight lens elements arranged in five groups, including a contact triplet, a contact doublet, and three singlets. All the surfaces are spherical except for the air side surface of the second element in the doublet, which is a relatively mild asphere. Careful attention was paid to glass selection in order to maximize throughput at 390 nm, with five of the eight elements being either calcium fluoride (CaF$_{2}$) or Ohara\footnote{Ohara Corporation, http://www.oharacorp.com} i-line glasses (which have $>$ 98\% internal transmission at 365 nm). Dow Corning Q2-3067\footnote{Dow Corning, http://www.dowcorning.com} optical coupling grease was used to join the elements in the doublet and triplet. The resulting design,  shown in Figure~\ref{camera}, is similar to LRIS  but has a smaller diameter and larger field of view.

Once the fundamental camera design was in hand, it was tuned using Zemax\footnote{Radiant Zemax LLC, http://www.zemax.com} to match surfaces with existing vendor test plates, optimizing spacings to melt data received from the glass vendor, and generating optical prescriptions for the two field flattening lenses to accommodate the different CCD curvatures. The problem of CCD curvatures was particularly annoying. The CCD surfaces were not flat to start and the shape changed when the devices were cooled to operating temperature. Fortunately, they returned to the same shape every time they were cooled to operating temperature and most had a generally spherical shape that could be corrected with spherical optics. Unfortunately, they were all different so each CCD required a customized pair of field flattening lenses.

\subsubsection{Optical Prescription}

Table~\ref{OpticalPrescript} shows the spectrograph optical prescription. Some of the surface descriptions can be deciphered from this example: ``Doublet,first,back'' refers to the doublet component, first lens element, back surface (closest to the CCD). The ``Radius'' is the radius of curvature in inches, negative implying a concave left surface. ``Thickness'' is the distance from the current surface to the next; positive to the right. All materials are from Ohara Glass except for CaF$_{2}$ and the lens couplant, Dow-Corning Q2-3067. 

\begin{table*}
\begin{center}
\caption{Optical Prescription for the SDSS Spectrographs\label{OpticalPrescript}}
\begin{tabular}{lrlrlr}
\tableline\
Surface & \multicolumn{2}{c}{    Radius (mm)} & \multicolumn{2}{c}{Thickness (mm)} & Material\\
 & Blue & Red & Blue & Red & \\
\tableline
slithead                & -640.080            & -640.080              & 630.123   & 630.123   & air\\
collimator mirror       & -1263.904           & -1263.904             & -1087.831 & -1087.831 & air\\
prism                   & plano               & plano                 & 64.592    & 70.663    & BK7\\
grating                 & plano($37^{\circ}$) & plano($39.5^{\circ}$) & 52.578    & 58.420    & air\\
singlet,front           & -182.804            & -182.804              & -30.531   & -30.531   & CaF$_{2}$\\
singlet,back            & -1813.560           & -1813.560             & -2.540    & -2.540    & air\\
triplet,1st,front       & -185.522            & -185.522              & -5.156    & -5.156    & LAL7\\
triplet,1st,back        & -105.867            & -105.867              & -0.076    & -0.076    & Q2-3067\\
triplet,2nd,front       & -105.867            & -105.867              & -63.525   & -63.525   & CaF$_{2}$\\
triplet,2nd,back        & 226.136             & 226.136               & -0.076    & -0.076    & Q2-3067\\
triplet,3rd,front       & 226.136             & 226.136               & -5.105    & -5.105    & BSM2\\
triplet,3rd,back        & -661.416            & -661.416              & -136.931  & -135.941  & air\\
doublet,1st,front       & -110.490            & -109.423              & -5.080    & -5.080    & BAL35Y\\
doublet,1st,back        & -87.274             & -81.331               & -0.076    & -0.076    & Q2-3067\\
doublet,2nd,front       & -87.274             & -81.331               & -44.120   & -46.990   & FPL51Y\\
doublet,2nd,back       & 429.006\footnote{Aspheric surface coefficients: $\alpha_4 = 7.517856E-04$, $\alpha_6 = -5.900703E-05$, $\alpha_8 = 2.644114E-06$}             & 432.892\footnote{Aspheric surface coefficients: $\alpha_4 = -7.880071E-04$, $\alpha_6 = 6.800388E-05$, $\alpha_8 = -2.676036E-06$}               & -53.518   & -53.899   & air\\
flattener,1st,front     & 136.449             & 135.560               & -2.032    & -2.032    & PBM2Y\\
flattener,1st,back      & 761.975             & 608.355               & -8.128    & -6.858    & air\\
flattener,2nd,front     & 122.733             & 114.427               & -3.810    & -3.810    & LAL59\\
flattener,2nd,back      & plano               & plano                 & -3.835    & -3.683    & air\\
CCD                     & 2502.992            & -1920.037             & 0         & 0         & silicon\\
\tableline
\end{tabular}
\end{center}
\end{table*}

\subsubsection{Predicted Optical Performance}\label{subsec:sdssopticalperformance}

This section discusses the predicted optical performance of
the SDSS spectrographs, including image quality, spectral resolution,
and throughput. Measured performance is discussed in Section \ref{sec:Perf}.

\paragraph{Image Quality}

\begin{figure}[htbp]
\centering
\vspace{1mm}
\includegraphics[scale=.47]{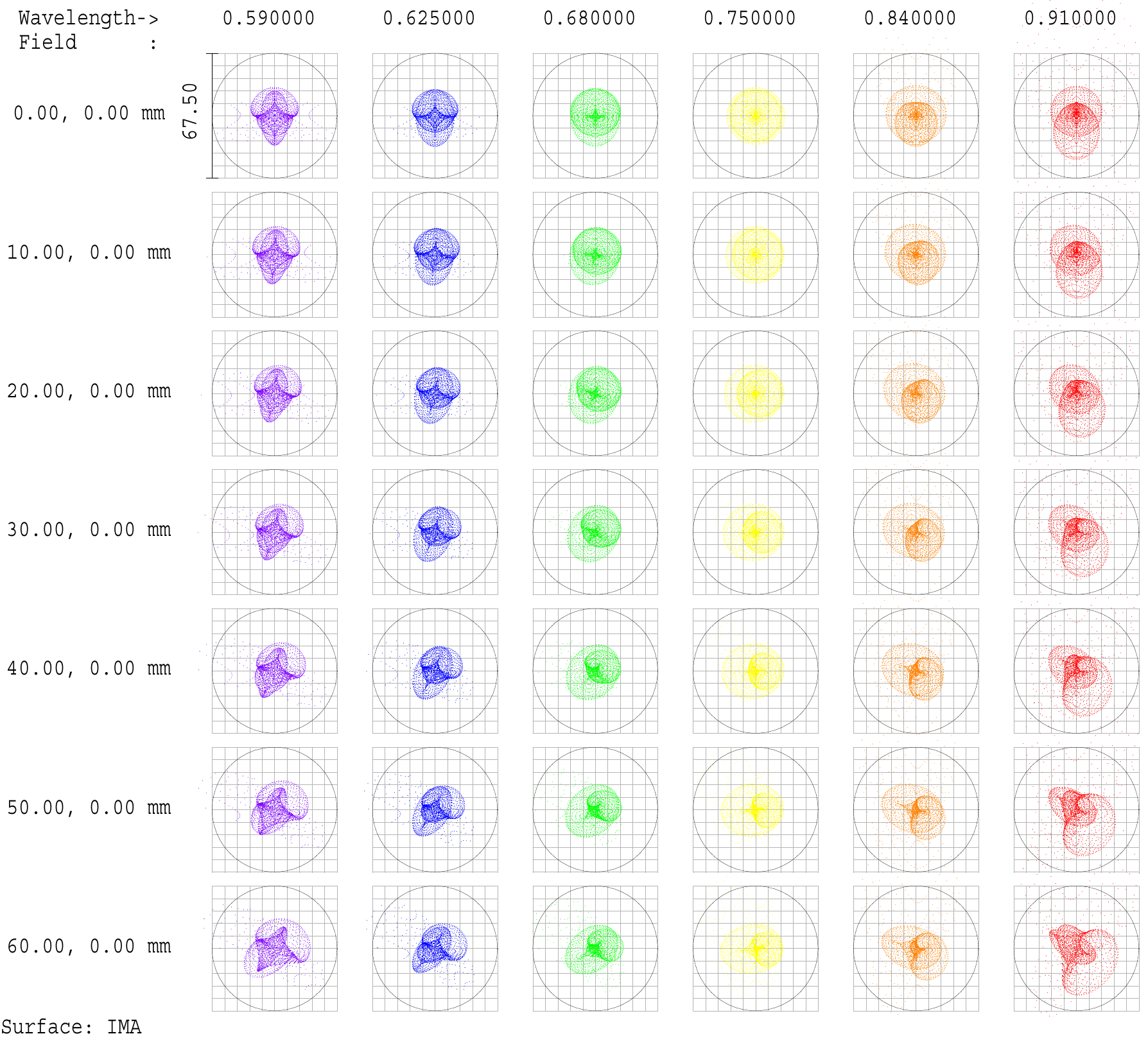}
\caption{{\bf Spot diagram for the SDSS red channel. The average RMS spot diameter is 21.5 $\mu$m and the maximum RMS diameter is 29.2 $\mu$m.}}
\label{red_spots_SDSS}
\end{figure}

\begin{figure}[htbp]
\centering
\vspace{1mm}
\includegraphics[scale=.47]{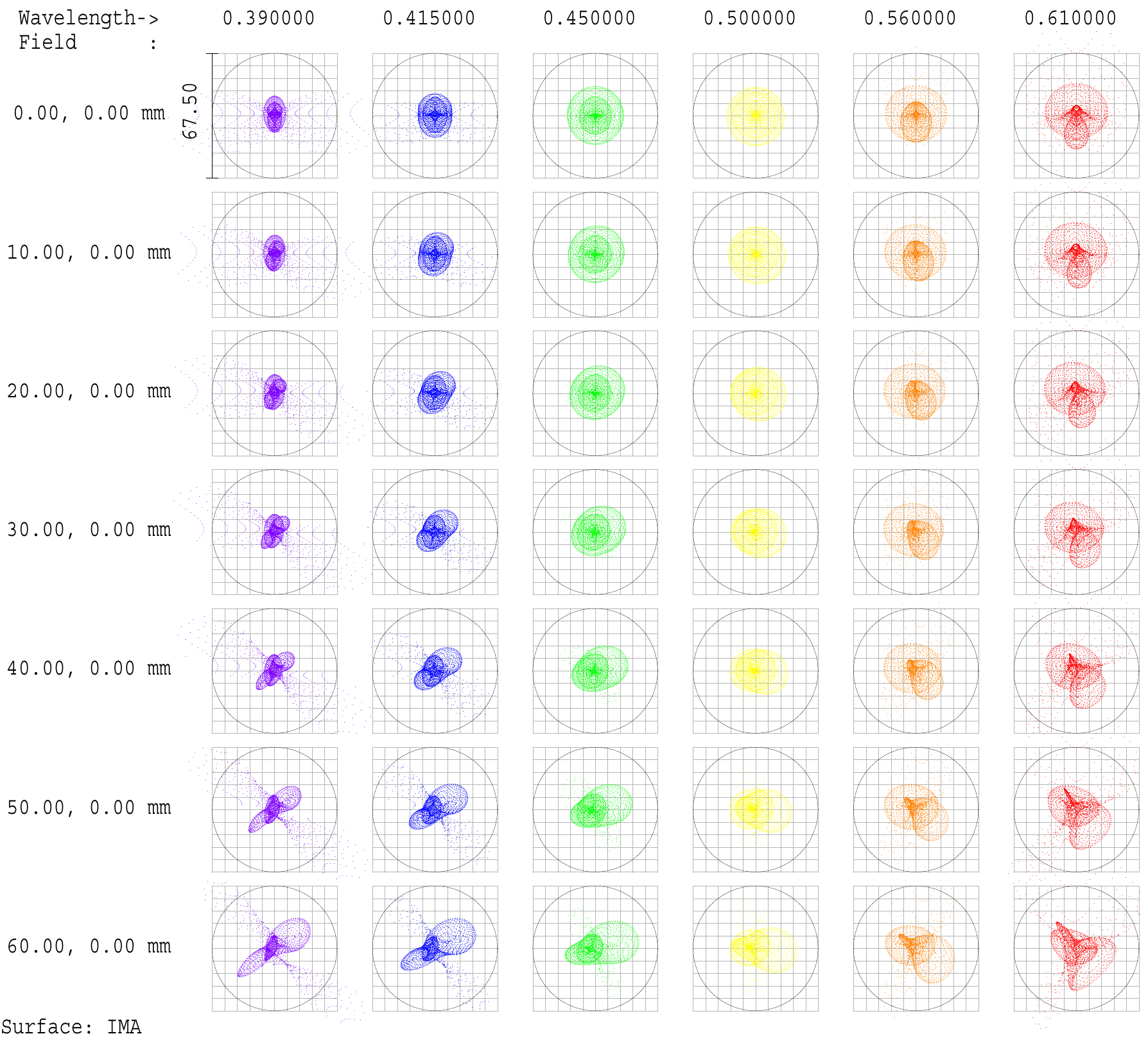}
\caption{{\bf Spot diagram for the SDSS blue channel. The average RMS diameter is 21.1 $\mu$m and the maximum RMS diameter is 27.3 $\mu$m.}}
\label{blue_spots_SDSS}
\end{figure}

Spot diagrams for the red and blue channels are shown in Figure~\ref{red_spots_SDSS}
and Figure~\ref{blue_spots_SDSS}, respectively. 
The spots are shown within a 67.5 $\mu$m diameter circle,
representing the diameter of the imaged fiber on the detector.  Each
diagram covers the full respective bandpass of the channel, and field
points cover the full length of the slit.  The average RMS spot diameter
for the red channel is 21.5 $\mu$m and the maximum RMS diameter is 29.2 $\mu$m. 
For the blue channel the average RMS diameter is 21.1 $\mu$m and the maximum
RMS diameter is 27.3 $\mu$m.  When convolved with the 67.5 $\mu$m imaged fiber
diameter, the geometric aberrations represented in these figures degrade
the image, but to a minimal degree, and the resulting image width is
well sampled by three pixels (72 $\mu$m) on the detector.

\paragraph{Spectral Resolving Power}

\begin{figure}[htbp]
\begin{center}
\epsscale{1.18}
\plotone{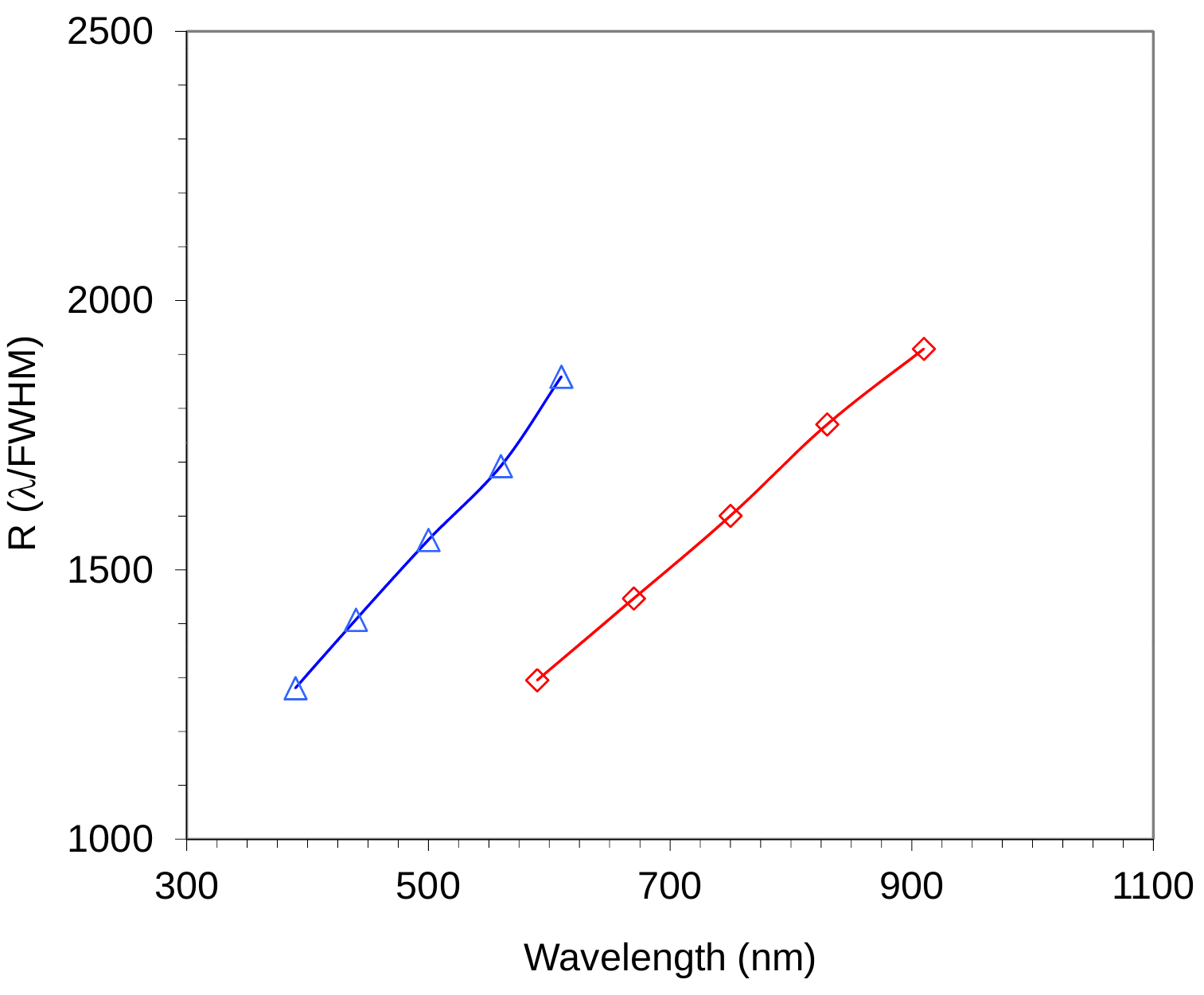}
\caption{{\bf Predicted resolving power, R,
as a function of wavelength for the SDSS spectrographs. Predictions are derived from the FWHM of simulated, uncollapsed images computed using Zemax. In reality, the resolving power of the instrument is determined using collapsed
spectra, so the prediction here is on the low side. However,
the Zemax analysis assumes perfect optics and alignment. For this
reason the predictions have been based on un-collapsed spectra, which
is the conservative approach. The expectation is that the actual resolving power
will be somewhat better than shown here.}}
\label{resolving_power_SDSS}
\end{center}
\end{figure}

To analyze the spectral resolving power, defined as 
$\lambda/\delta\lambda$, where $\delta\lambda$ is taken to be 
the spectral FWHM of the fiber
image on the detector in \AA, the following procedure was used. In Zemax,
a circular source with a diameter equal to that of the fiber was placed at the center
fiber location on the slit for the wavelength under consideration.
Many thousands of rays were launched from this circular source with
a uniform distribution, and the resulting image recorded on a simulated
detector with pixels $1/4$ the size of the 24 $\mu$m
CCD pixels (in order to better sample the image). The simulated image
data was analyzed to determine the FWHM, without collapsing the image.
The FWHM thus determined was taken to be the resolution, and the resolving 
power is plotted in Figure~\ref{resolving_power_SDSS}. 

In reality, the resolving power of the instrument is determined using collapsed
spectra, so the prediction here is on the low side. However,
the Zemax analysis assumes perfect optics and alignment. For this
reason the predictions have been based on un-collapsed spectra, which
is the conservative approach. The expectation was that the actual resolving power
would be somewhat better than shown here. 

\paragraph{Throughput}

\begin{figure*}[htbp]
\begin{center}
\epsscale{0.8}
\plotone{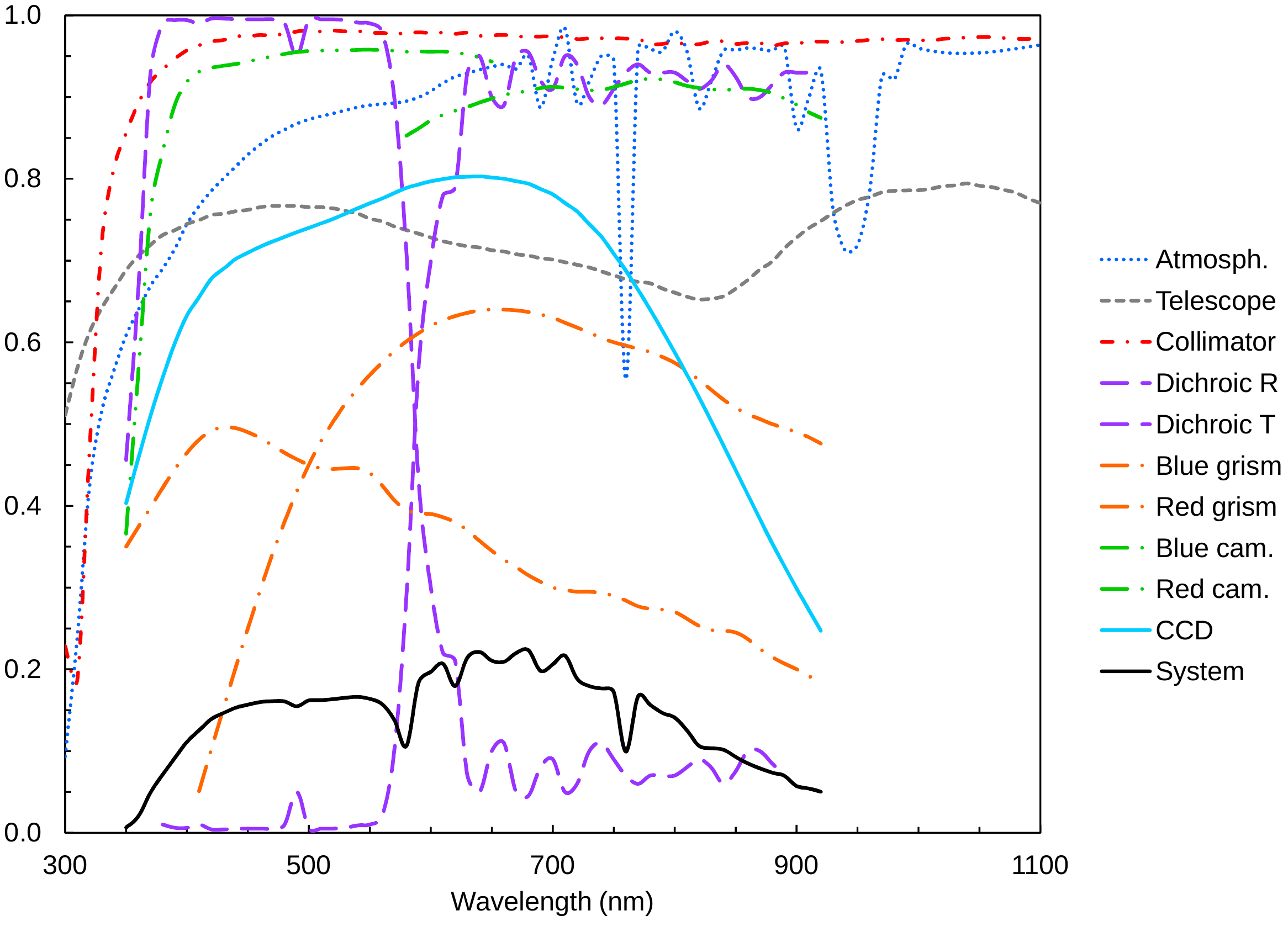}
\caption{{\bf Expected throughput for the SDSS spectrographs.  The plot shows all contributors to the throughput model having a wavelength dependence.  Not shown are those contributors with an  essentially flat response across the bandpass: fiber transmission including Fresnel losses at the two faces (0.85), focal ratio degradation overfilling the collimator stop (0.97), and ``slit'' losses for $1\2pr$ FWHM seeing conditions modeled with a double Gaussian PSF (0.94).  Not included in the model are losses due to centering and guiding errors.  Overall system throughput, shown by the solid black curve, is expected to peak at about 17\% in the blue channel and 22\% in the red channel.}}
\label{SDSS_throughput}
\end{center}
\end{figure*}

The total throughput, on-sky, for the SDSS spectrographs was predicted
from an end-to-end component model as a function of wavelength.
Included in the model were: atmospheric extinction, seeing (slit) losses, telescope, fibers, collimator,
dichroic, grism, camera, and CCD.  Atmospheric extinction was modeled at one airmass
using a Palomar curve\footnote{www.sdss.org/dr3/instruments/imager/filters/index.html}
scaled to Apache Point using
mean values for photometric nights during 1998-2008 \citep[Table 3 of][]{padmanabhan08a}.
Seeing losses were modeled using a double Gaussian PSF with a FWHM of $1\2pr$ centered
in a $3\2pr$ aperture, resulting in a 5.5\% throughput loss assumed flat across the bandpass.
The telescope efficiency is based on measured data for CO$_2$-cleaned bare aluminum mirrors
\citep{wilson99a}, along with simulated anti-reflection coating curves to match the
specifications of the two wide field corrector lenses (a small overall effect).
Measured curves were used for the dichroic, grism, camera coatings, and CCD quantum efficiency (QE).
The manufacturer's curve for Denton FSS99 silver was used for the collimator coating.
Internal transmission curves for the camera glasses were obtained from the manufacturer's data sheets.
The fiber efficiency (85\%) is based on lab measurements, representing an average value for the
fibers measured and assumed flat across the bandpass.\footnote{As per the Polymicro data sheet, the variation in transmission is only about 2\% - 3\% over the SDSS 
bandpass.}
An additional 3\% loss is included for focal ratio degradation beyond the f/4 beam
the collimator was designed to accept. 
Figure~\ref{SDSS_throughput} shows the individual component efficiencies used for
this model along with the total expected throughput.
Peak efficiencies in the blue and red channels are 17\% and 22\%, respectively.

\subsection{Mechanical Design}
\label{sec:SDSSmech}

\subsubsection{Overview}

\begin{figure}[htbp]
\begin{center}
\epsscale{1.17}
\plotone{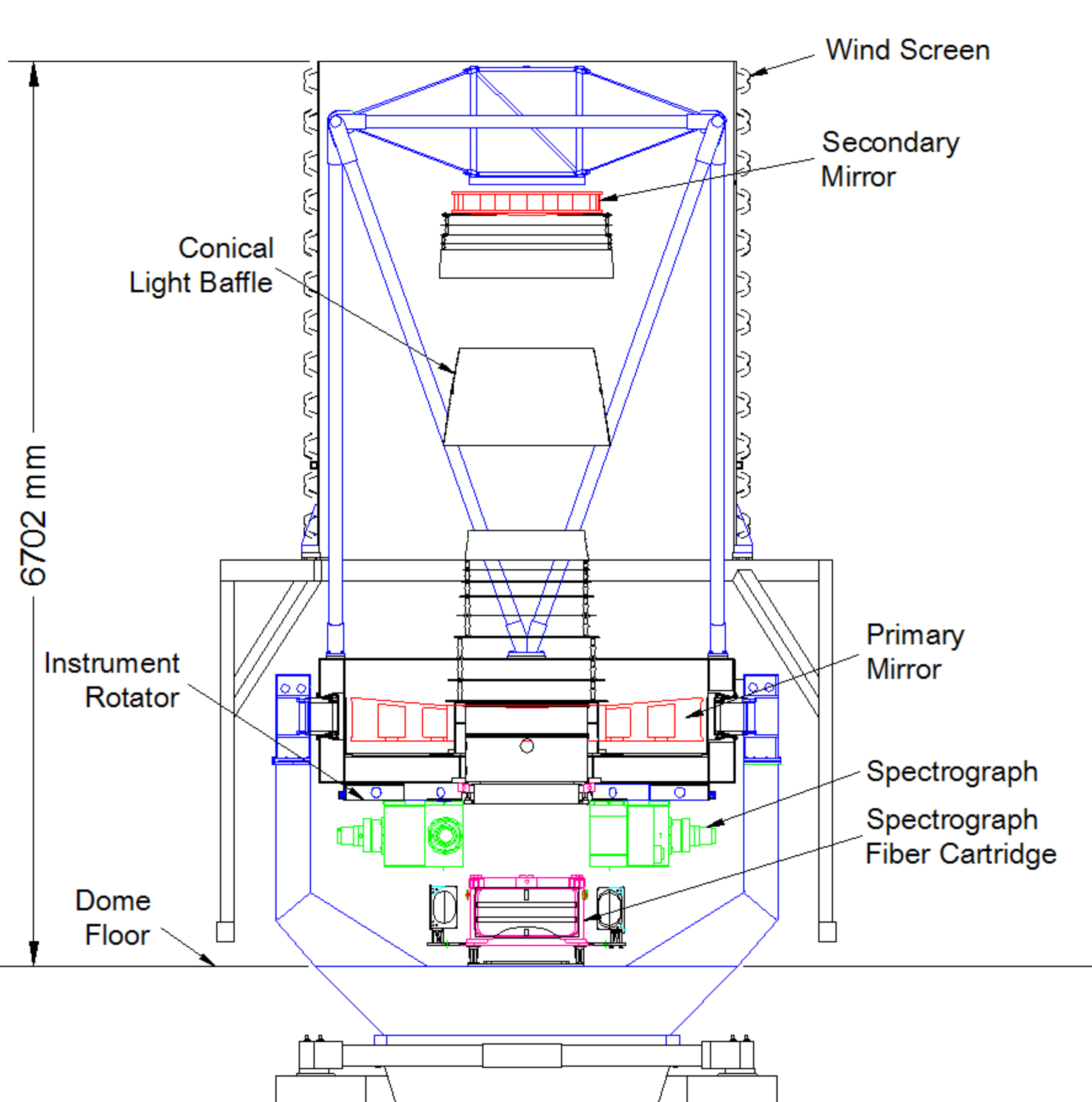}
\caption{{\bf Rear view of the SDSS telescope depicting the location of the spectrographs. Here a fiber cartridge is shown retracted from the spectrographs. The twin spectrographs, each with a mass of 320 kg, mount to the back of the Cassegrain instrument rotator adjacent to the focal plane with sufficient separation between the spectrographs to allow routine installation and removal of the imaging camera and fiber cartridges.}}
\label{SpectroMount}
\end{center}
\end{figure}

The twin spectrographs mount to the SDSS telescope Cassegrain rotator
adjacent to the focal plane with sufficient separation between the
spectrographs to allow routine installation and removal of the imaging
camera and fiber cartridges; see Figure~\ref{SpectroMount}. Compared to a bench-mounted
configuration this approach greatly reduces fiber length, and the
associated throughput loss. It also eliminates potential adverse effects resulting from the repeated bending of fibers. A consequence however, is flexure induced motion of the spectra as the telescope tracks the sky, a problem that is mitigated by the use of a stiff, enclosed optical bench. 

Each spectrograph has a mass of approximately 320 kg. The overall
length, width, and height are 2210 mm $\times$ 1000 mm $\times$ 570 mm, respectively.
The optical layout of the instrument lends itself to a simple mechanical
design, having only a few mechanisms and only one actively
controlled optic, the collimator. To ease assembly and testing, the
fiber slithead and all refractive components (i.e. camera lenses,
the dichroic, and grisms) are mounted without adjustment in logical
sub-assemblies, relying solely on machining tolerances for placement.
Shims are located at strategic locations where opto-mechanical
tolerances are tighter than could be achieved with reasonable machining
practices. This basic approach has produced a robust, reliable instrument
that has been straightforward to maintain and upgrade. In the following
sections we describe the mechanical design in more detail, with an
emphasis on the opto-mechanical sub-assemblies.

\subsubsection{Optical Bench}

\begin{figure*}[htbp]
\begin{center}
\epsscale{0.8}
\plotone{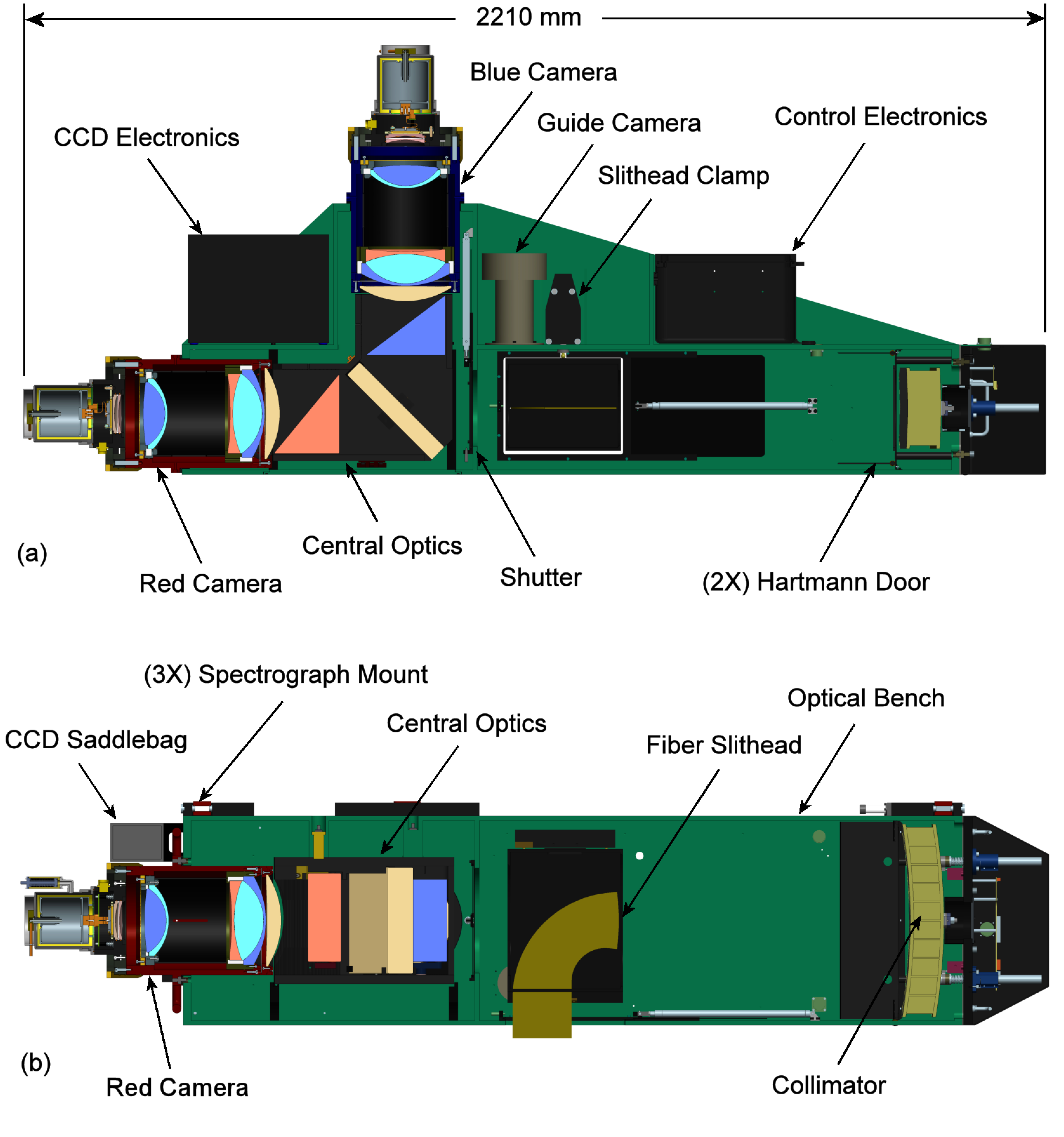}
\caption{{\bf Section views of the SDSS spectrographs. The T-shaped optical bench is an enclosed
aluminum (6061-T6) weldment with precision-machined interfaces to
locate all five opto-mechanical subassemblies: the fiber slithead,
the collimator, the central optics, and a the red and blue channel
cameras. One of the two spectrographs also supports the guide camera. Electronics control chassis mount to the external walls of the bench. Three kinematic mounts on the top of the bench interface to the back of the Cassegrain instrument rotator.}}
\label{OpticalBench}
\end{center}
\end{figure*}

Figure~\ref{OpticalBench}  shows the spectrograph optical bench and the opto-mechanical
subassemblies that interface to it. The optical bench is an enclosed
aluminum (6061-T6) weldment with precision-machined interfaces to
locate all five opto-mechanical subassemblies: the fiber slithead,
the collimator, the central optics, and the red and blue channel
cameras. The dual-channel optical layout naturally leads to the T-shaped
configuration depicted in the figure. Both the red and blue cameras,
as well as the collimator assembly, mount to exterior faces of the
bench; one at the end of each leg of the T. The fiber slithead and
the central optics assembly, which contains the dichroic and grisms,
are mounted internally through access ports. At all of these opto-mechanical
interfaces precision-machined reference datums and kinematic-locating
features are employed to ensure accurate, repeatable alignment of
the optical system. A shim at the collimator interface provides a
one-time coarse adjustment to compensate for the as-built collimator
radius of curvature. Another shim where the slithead kinematic mount
attaches to the bench provides a one-time adjustment to compensate for
machining tolerances. Instrument control electronics and the CCD controller
mount to exterior walls of the bench. The guide camera, located on
only one of the two spectrographs, mounts externally as well.

The principal considerations in designing the optical bench were ease
of assembly, precise optical alignment, and the requirement to minimize
flexure-induced motion of the spectra as the telescope tracks the
field. With a tall closed-end, box-section design, internal and external
gussets, and optimized placement of the mount locations to the rotator,
flexure-induced motion of the spectra is kept to an acceptably low
level during a typical exposure. Secondary considerations include
the desire to reduce stray light from external sources and to
keep exposed optics free of dust and debris.

\newpage

\subsubsection{Collimator Mount}

The collimator is the only actively controlled optic in the spectrograph; it is adjustable in piston, tip, and tilt.
Enabling these degrees of freedom serves multiple functions and eliminates the need for numerous mechanisms elsewhere within the spectrograph. Translating the collimator
 in piston adjusts focus in both channels, an adjustment
that occurs routinely during operation to compensate for thermal drift,
and flexure due to changes in telescope pointing and rotator angle.
A pair of Hartmann doors located in front of the collimator permits
shifts in focus to be measured rapidly. Tip/tilt adjustment allows
the collimator to be precisely coaligned to the cameras, or at least
to an average of the two camera axes. Thus, the center fiber can be
positioned at the center of the detector in the spatial direction,
and the central wavelength can be positioned at the center of the
detector in the spectral direction.  Lastly, variations in the pixel-to-pixel sensitivity in the spatial direction 
can be accurately removed by {\it dithering} (i.e. tilting) the collimator between successive flat-field exposures and processing 
the co-added image.  This is necessary because the image of the fiber on the detector has a Gaussian PSF, {\it and} the spectrograph flexes as
the telescope moves, so the fiber illumination pattern changes slowly.  This does not matter if one has a uniformly illuminated flat, but is 
disastrous if one has little or poor calibration data where now there is lots of flux because flexure has moved the fiber trace. Even if there 
is no flexure, the uniform flat provides better statistics across the fiber image. See Section \ref{sec:SDSScal} for a description of the {\it flat-field dithering} process.

\begin{figure}[htbp]
\begin{center}
\epsscale{1.1}
\plotone{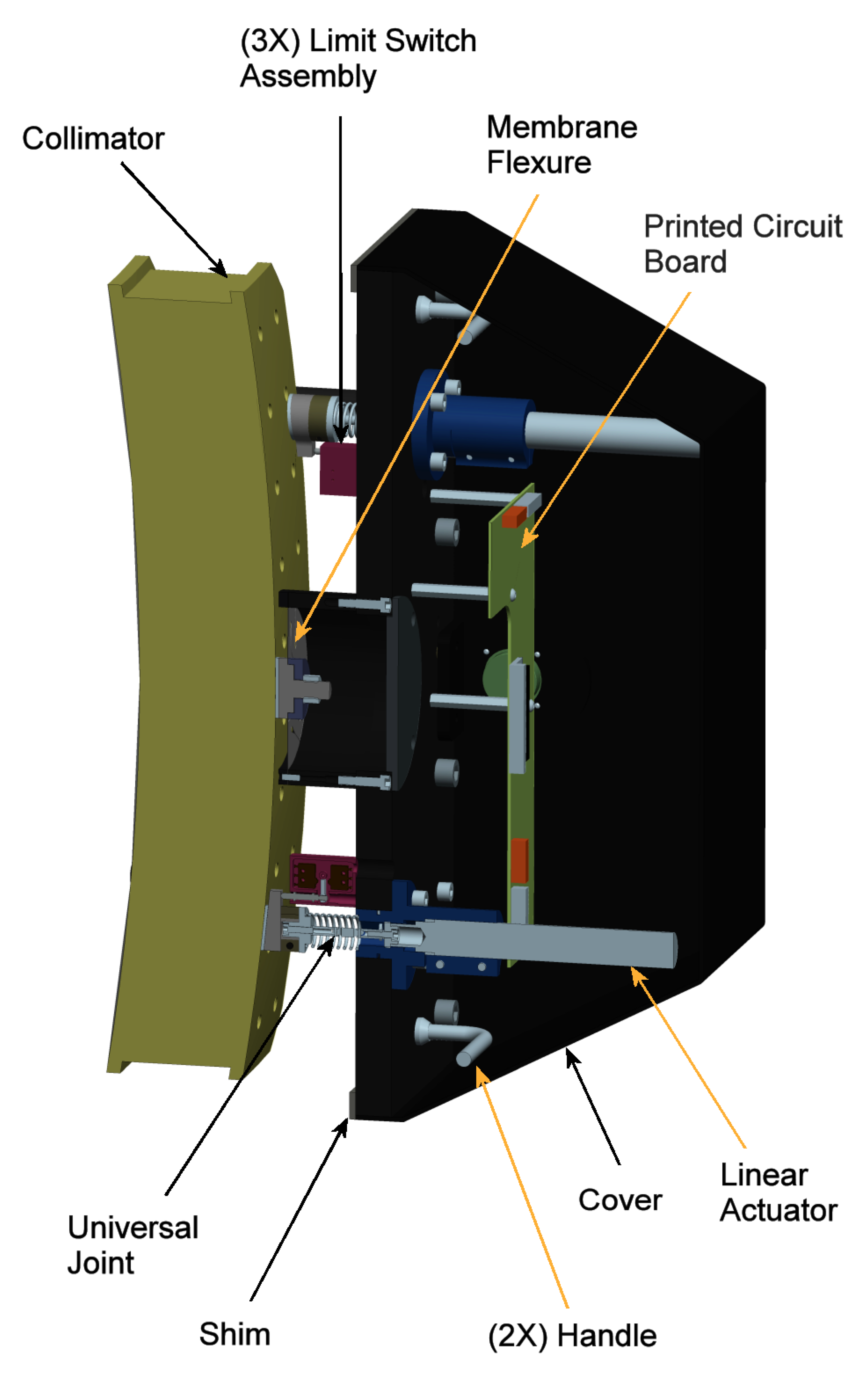}
\caption{{\bf Collimator mount for the SDSS spectrographs. The mirror
is supported from the back surface, at its center, by a circular membrane
flexure, which constrains the mirror in-plane but allows out-of-plane
compliance for tip, tilt, and piston adjustment. Motion is provided
by three linear actuators, which provide sub-micron resolution, more than sufficient for focus adjustment, adjustment of the spectra on the CCD, and flat-field dithers.}}
\label{CollimatorMount}
\end{center}
\end{figure}

Details of the collimator mount are depicted in Figure~\ref{CollimatorMount}. The mirror
is supported from the back surface, at its center, by a circular membrane
flexure, which constrains the mirror in-plane but allows out-of-plane
compliance for tip, tilt, and piston adjustment. Motion is provided
by three linear actuators (Physik Instrumente model\# M-222.20\footnote{Physik Instrumente, GmbH, http://www.physikinstrumente.com}), which
provide sub-micron resolution, more than sufficient focus adjustment, adjustment of the spectra on the CCD, and flat-field dithers. Each actuator connects
through a spring-pre-loaded universal joint to an Alloy-42 steel pad
bonded to the back of the mirror with 3M 2216\footnote{3M, http://www.3m.com} two-part epoxy. Three
limit switch assemblies behind the mirror limit travel to $\pm 3$ mm.
The linear actuators, limit switches, and flexure are attached
to a common mounting plate, which interfaces to the optical bench.
Electrical connections from the actuators and limit switches are routed
to a single connector on a custom printed circuit board mounted to
the rear of the plate. A shim between the mounting plate and the bench
facilitates one-time, gross focus adjustment. Accurate and repeatable
assembly of the subsystem is made possible by two dowel pins (one
round and one diamond shape) that engage hardened bushings pressed
into the optical bench. Safe installation is facilitated by two handles
for lifting, and four guide rods that flank the collimator ensure
the collimator clears the bench opening as it is installed. A sheet
metal cover surrounds the rear of the assembly, which protects vital
components from physical and environmental damage, and reduces stray
light. Finite element analysis conducted early in the design phase indicated
that the distortion produced by gravity and the membrane load would
be negligible. 

\subsubsection{Hartmann Doors}

The Hartmann doors are a simple bi-fold design. Each door pivots about
an upper and lower bushing pressed into the optical bench, and is
driven by a $90^\circ$ pneumatic rotary actuator (Bimba Pneu-Turn model
PT-006090-A1M\footnote{Bimba Manufacturing Company, http://www.bimba.com}) located on the top surface of the bench.
The doors are controlled by the spectrograph micro-controller and a
bank of programmable solenoid valves (Clippard model number EMC-08\footnote{Clippard Instrument Laboratory Inc., http://www.clippard.com})
located in the spectrograph electronics box. Solid-state
magnetic sensors (Bimba model number HSC-02) mounted to the actuators
report the state of the doors (i.e. open or closed). Small manual
flow control valves (Bimba model number FCP-1) at the inputs to each
actuator set the speed of rotation.

\subsubsection{Shutter}

A shutter is required to set the exposure time for wavelength calibrations,
flat fields and science exposures.
The shutter does not need to be fast or very accurate since science exposures
include calibration standards that are used to determine the zeropoint
calibration at the time of the exposure.
The minimum exposure time, 4 seconds, is set by the arc lamps used for wavelength calibration.
Flat fields require 30 second exposures.  Science exposures are fifteen minutes.

The spectrograph shutter is located just upstream of the dichroic
on a dividing wall in the optical bench residing between the central
optics and the slithead; see Figure~\ref{OpticalBench}. This location is ideal since
the entire bandpass can be blocked by a single, relatively small shutter.
The shutter is a black anodized aluminum sliding door driven by a
double-acting pneumatic cylinder (Bimba model BRM-02x-DXP). Slots
in the Delrin door frame guide the door, making a light-tight seal
when the door is closed. An oval window in the frame, slightly oversized
to the beam, serves as a light baffle. Like the Hartmann doors, the
shutter is controlled by the spectrograph micro-controller and solenoid
valves on the Clippard EMC-08 board. Solid-state magnetic sensors
(Bimba model number HSC-02) mounted to the cylinder indicate the state
of the door (i.e. open or closed). Small manual flow control valves
(Bimba model number FCP-1) at the inputs to cylinder set the speed
at which the door opens and closes.

The shutter is accurate to approximately 0.5 seconds.  Latency in the door motion due to temperature changes is responsible for most of the inaccuracy.

\subsubsection{Central Optical Assembly}

The beamsplitter and grisms are mounted in a single opto-mechanical structure,
the central optics assembly. The assembly is kinematically mounted
inside the optical bench on three gusseted posts integral to the
weldment and located on the top surface of the optical bench, thus
the central optics assembly is suspended from the top of the bench. The faces of
the three posts are precisely machined establishing a planar reference
that is square to the collimator mounting surface and the two camera
mounting surfaces. Two locating sleeves centered about two of the
three posts establish the in-plane location of the assembly. These
machined references precisely locate the assembly relative to the slithead,
collimator, and cameras. A large port in the sidewall of the optical
bench provides access for machining the post surfaces and installation.

\begin{figure}[htbp]
\begin{center}
\epsscale{1.2}
\plotone{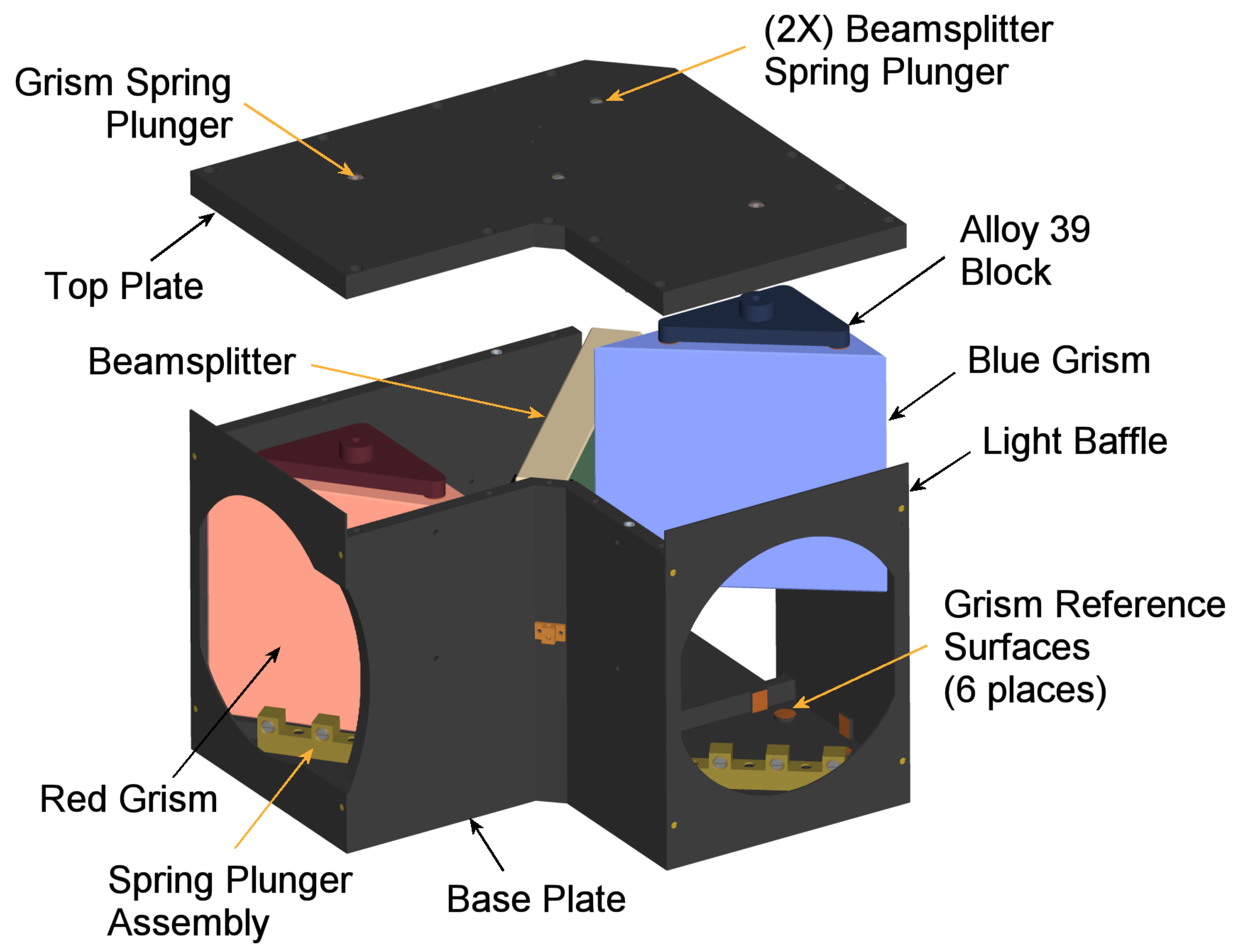}
\caption{{\bf Exploded view of the Central Optics Assembly. The dichroic
and both grisms are each located, without adjustment, by six machined
reference surfaces. Spring plungers seat the elements against these
reference surfaces and accommodate differential contraction between
the glass optics and aluminum structure. Kinematic mounting interfaces on the bottom of the base plate locate the assembly inside the optical bench.}}
\label{COA}
\end{center}
\end{figure}

Figure~\ref{COA} shows the details of the central optics assembly. The dichroic
and both grisms are each located, without adjustment, by six machined
reference surfaces (Kapton tape covers each surface to avoid metal-to-glass contact). Spring plungers seat the elements against these
surfaces accommodating differential contraction between
the glass optics and aluminum structure. A three-point-contact Alloy-39
block bonded to the top of each grism spreads the vertical load applied
by a single large plunger embedded in the top plate, and provides
a convenient lift point for installing the grism. To minimize tolerance
stack-up and improve placement accuracy, all but one of the eighteen
reference surfaces (six for each of three optics) is machined into
a single component, the base plate, which interfaces to the optical
bench. The single remaining reference surface, which controls the
tip of the beamsplitter, is located on the top plate. Black anodized surfaces
and thin light baffles at the exit faces of the assembly serve to
mitigate stray light. The entire assembly has a mass of 39 kg.

\subsubsection{Camera Opto-Mechanics}

\begin{figure}[htbp]
\begin{center}
\epsscale{1.18}
\plotone{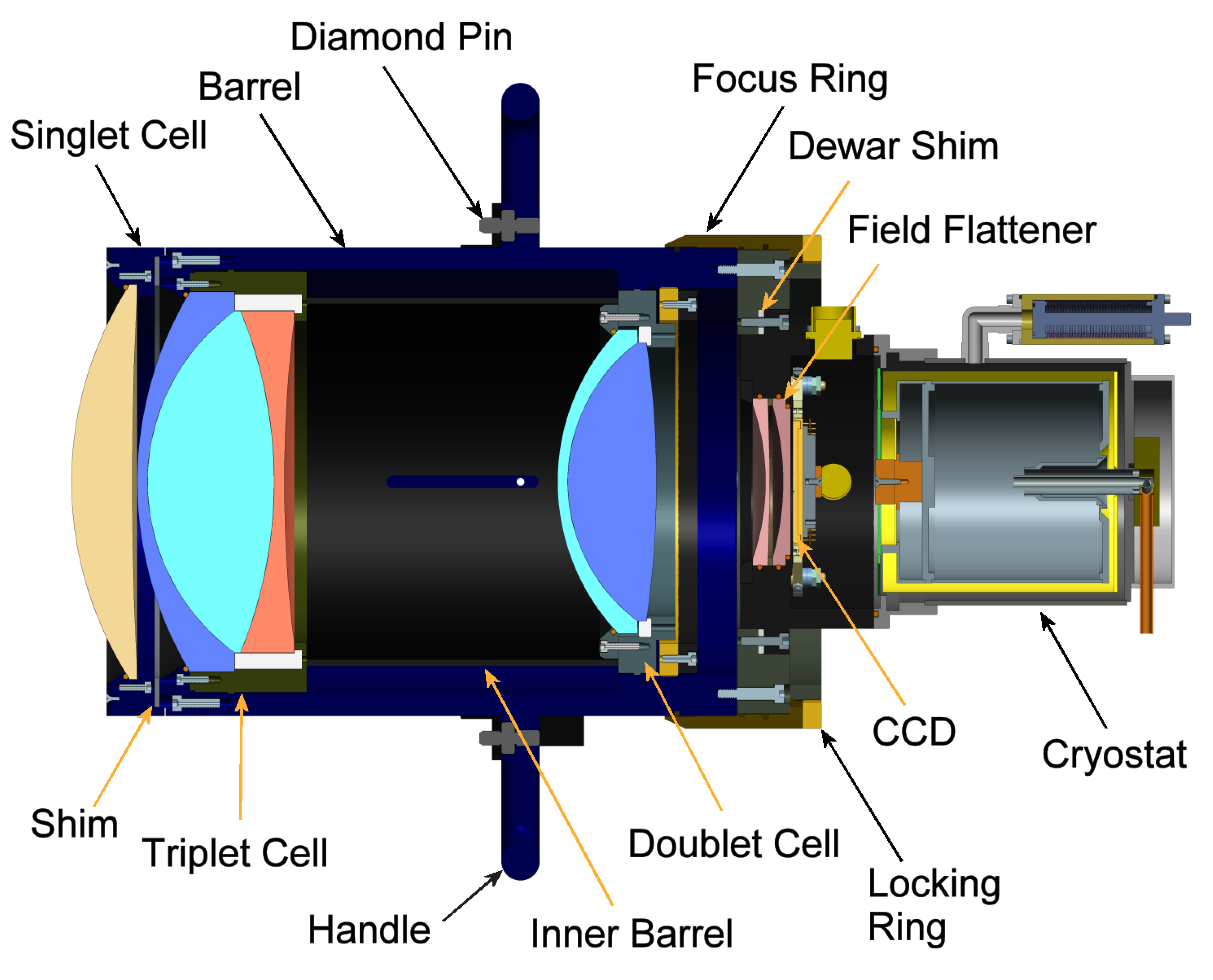}
\caption{{\bf Cross section of the SDSS blue camera highlighting the opto-mechanical design details.  The opto-mechanical details of the red camera are identical.  The design was derived largely from the Norris spectrograph camera design. The lenses and lens-groups
are mounted in athermal cells. External reference surfaces on each
cell are machined true to the opto-mechanical reference surfaces that
locate each lens, thus establishing lens concentricity from cell-to-cell
and accurate placement of the lenses along the optical axis. Manual focus adjustment is provided by a threaded focus ring at the back of the camera barrel. This adjustment is used to compensate for small focus drifts and to ensure the two channels are parfocal.}
}
\label{cam_section}
\end{center}
\end{figure}

The opto-mechanical design of the SDSS cameras, shown in Figure~\ref{cam_section},
was derived largely from the Norris spectrograph camera design \citep{cohen88a};
a logical consequence of the two cameras having very similar optical
designs, and the same optical designer, Harland Epps.  The lenses and lens-groups
are mounted in athermal cells. External reference surfaces on each
cell are machined true to the opto-mechanical reference surfaces that
locate each lens, thus establishing lens concentricity from cell-to-cell
and accurate placement of the lenses along the optical axis. The singlet
and triplet cells are bolted together in series and mounted in an
aluminum barrel along with the doublet cell. A steel inner barrel
meters the distance between the triplet and doublet, the doublet being
pressed against the inner barrel by a compression spring assembly
behind the cell. A thick shim between the singlet and triplet allows
a one-time adjustment to compensate for machining tolerance stackup.
The field flatteners are integral to the dewar, as described
in Section~\ref{sec:SDSS_Dewars}.

\begin{figure}[htbp]
\begin{center}
\epsscale{1.15}
\plotone{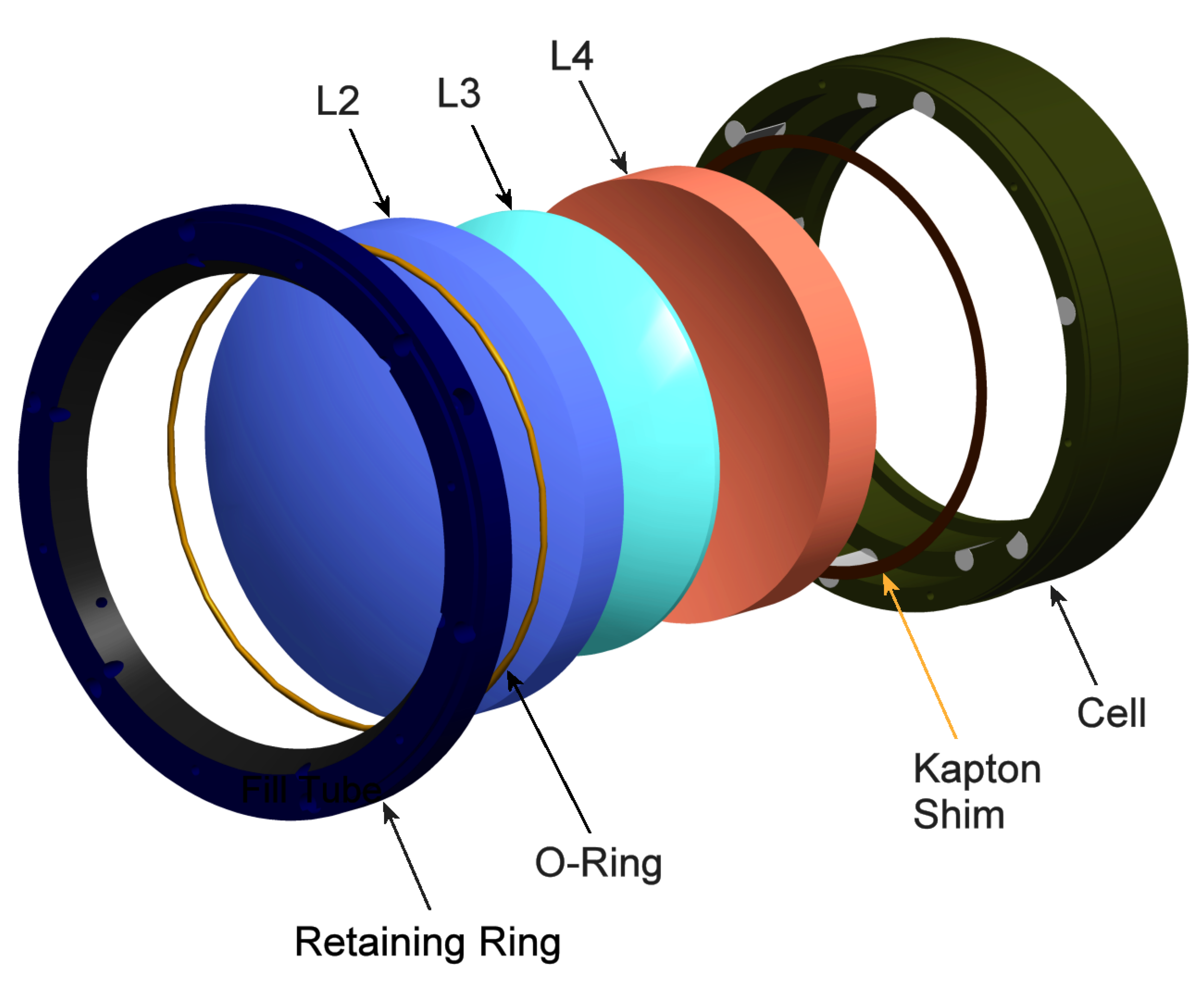}
\caption{{\bf Exploded view of the SDSS blue camera triplet. The athermal cell design consists of a type 330 stainless steel ring with six glass-filled
Teflon plugs lightly pressed into a hole pattern circumscribing
the each lens bore. Here the rear two elements are located by a single set of plugs given the small edge thickness of the middle, calcium flouride, element. These plugs are then bored on a lathe to a diameter slightly oversized (0.05 mm on diameter) to the as-built lens diameter. The net change in
the diameter of the finished bore is less than the diametrical clearance
to the lens over the temperature range of interest, given the coefficients
of expansion for the metal ring, the Teflon plug, and the glass lens.   An o-ring in the retaining ring contacts the first
lens surface, providing force to seat the lens and compliance to accommodate
differential expansion.
\\}
}
\label{triplet_section_SDSS}
\end{center}
\end{figure}

The design and construction of the athermal cell design is best described
as follows. A metal ring with an appropriate coefficient of thermal
expansion is bored oversize to the lens diameter. Six glass-filled
Teflon plugs are lightly pressed into a hole pattern circumscribing
the lens bore.  Here the rear two elements are located by a single set of plugs given the small edge thickness of the middle, calcium flouride, element. These plugs are then bored on a lathe to a diameter
slightly oversized (50 $\mu$m on diameter) to the as-built lens diameter.
In the same machining step, the remaining critical surfaces are machined,
thus achieving lens concentricity (within the limits of radial clearance),
location along the optical axis, and perpendicularity to the optical
axis. The plug diameter is calculated such that the net change in
the diameter of the finished bore is less than the diametrical clearance
to the lens over the temperature range of interest, given the coefficients
of expansion for the metal ring, the Teflon plug, and the glass lens.
Where lens groups are packaged in a single cell, multiple sets of
plugs are used. For the SDSS cameras, 6061-T6 aluminum was used for
the singlet and doublet cells, and 330 stainless steel was used for
the triplet cell. A Kapton shim rests between the lens locating face
and the metal cell. An O-ring in the retaining ring contacts the first
lens surface, providing force to seat the lens and compliance to accommodate
differential expansion. A rendered image of the triplet cell depicting
these details is shown in 
Figure~\ref{triplet_section_SDSS}.

A mounting flange on the camera barrel (located at the camera center of gravity) provides the interface to the optical bench. Two pins (one dowel
pin and one diamond pin) in the face of the flange engage two steel
bushings in the bench, ensuring repeatable placement. Handles on the
flange, and a separate, detachable guide-rail system, which bolts to
the optical bench, facilitate safe removal and installation of the
camera. The entire camera mass (including the dewar) is 40 kg.

Camera focus is achieved by translating the dewar, which houses the
detector and the two field flattening lenses, and attaches to a flange
at the rear of the camera barrel. A shim between the dewar and flange
provides a one-time coarse focus adjustment. Small focus adjustments
to compensate for seasonal temperature variations are affected by
turning the focus ring threaded onto the rear of the camera barrel.
Turning the ring translates the dewar flange, which is pre-loaded against
the ring by eight compression springs in the rear face of the camera
barrel. Two guide pins, also in the rear face of the barrel, engage
two linear bearings in the flange providing smooth axial motion with
negligible lateral motion. The large diameter of the focus ring combined
with a fine 32 thread-per-inch pitch allow focus resolution of roughly
2 $\mu$m. Marks engraved in one degree increments on the outer diameter
of the ring provide a convenient method for metering adjustments.

\subsubsection{Dewars}
\label{sec:SDSS_Dewars}

\begin{figure}[htbp]
\begin{center}
\epsscale{1.1}
\plotone{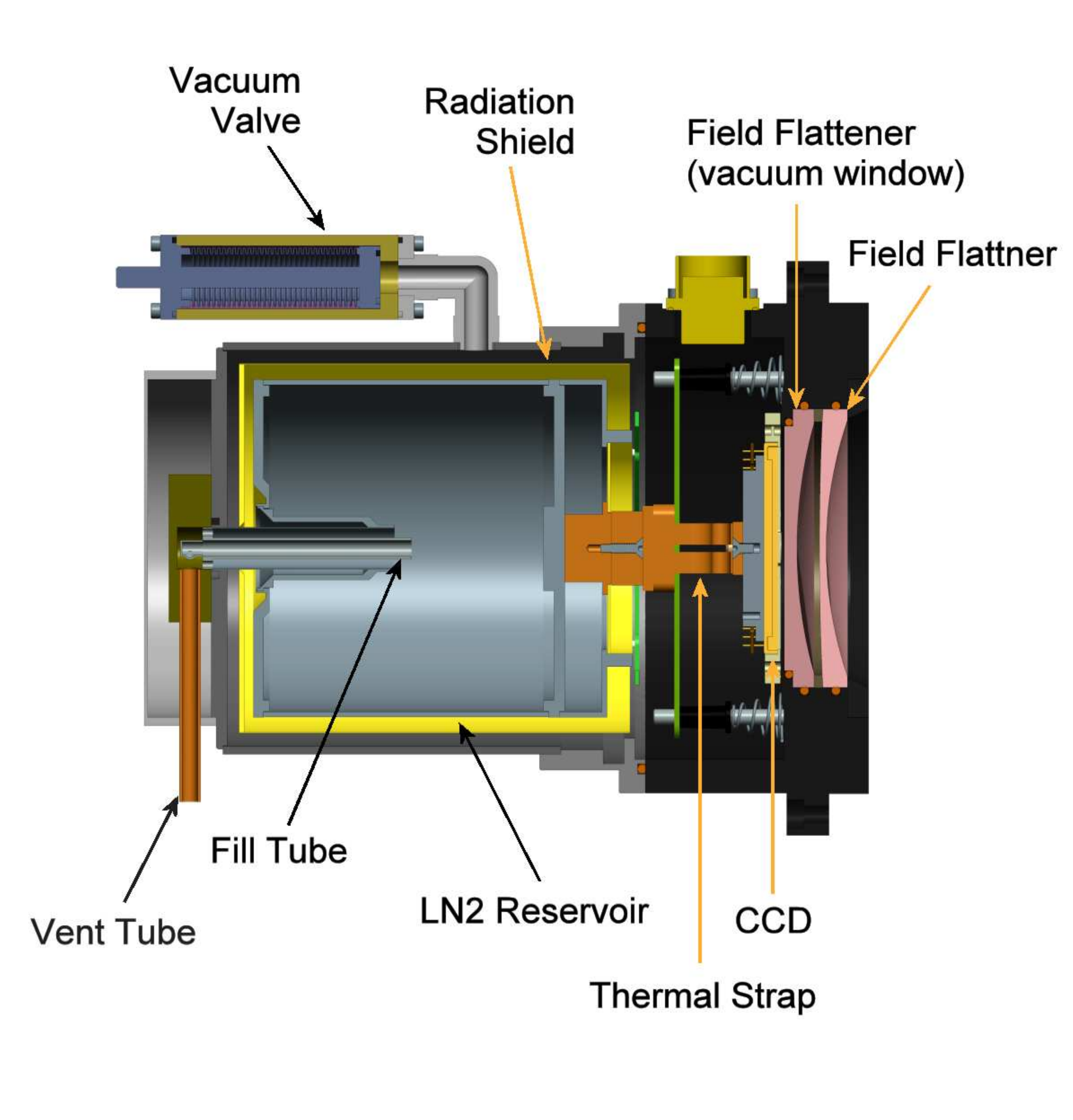}
\caption{{\bf Cross section of the CCD dewar. The liquid nitrogen dewar contains the last two elements of the camera (i.e. the field flatteners),
the detector, and the pre-amp board.  Location of the
CCD is adjustable in five degrees of freedom for centration, piston,
tip and tilt. In operation the total dewar heat load is 4 Watts. An autofill
system fills the dewar at roughly 3 hour intervals from a 10-liter
storage dewar adjacent to the spectrograph}}
\label{dewar}
\end{center}
\end{figure}

The CCD is housed in a dewar and cooled using liquid nitrogen (LN$_{2}$);
see Figure~\ref{dewar}. 

The front half of the dewar contains the last two elements of the camera (i.e. the field flatteners),
the detector, and the pre-amp board. O-rings locate the lenses on-axis
and provide radial compliance for what is a small differential thermal 
contraction over the operating temperature range. Location of the
CCD is adjustable in five degrees of freedom for centration, piston,
tip and tilt. The detector columns are aligned to the slit by rotating
the entire dewar relative to the back of the camera. 

The rear half of the dewar contains the LN$_{2}$ reservoir and is bolted to the front
half of the dewar, a logical arrangement whereby the front half containing
the optics, detector, and electronics can be assembled independent
of the rear half containing the liquid nitrogen tank. A flexible copper
braid provides the thermal connection between the reservoir and the
detector. The space surrounding the reservoir and the remainder of
the internal volume extending up to the rear of the last field flattener
is evacuated. 
Detector temperature is controlled to 180 K +/- 0.5 K using a
small heater located on the cold-finger attached to the rear of the
detector. The  LN$_{2}$ reservoir is filled from the rear of the dewar through
the center of a coaxial tube located on the reservoir axis and penetrating
to the middle of the volume; this allows half of the 0.6 liter volume
to be filled with liquid, regardless of gravity vector orientation.
Gaseous boil-off vents through the outer tube.

The thermal design of the dewar is well optimized. The liquid nitrogen
reservoir is fabricated from stainless steel and is well isolated
from its surroundings. Aside from the detector cold strap, the only
conductive paths to the reservoir are the thin-walled stainless vent
tube and the G10 spider that provides radial support at the front
end of the reservoir. Polished and gold-plated surfaces on the exterior
of the reservoir and the radiation shield minimize radiative heat
load. In operation the total dewar heat load is 4 Watts and the hold time is approximately 3 hours. 

\subsubsection{Dewar Autofill System}
\label{sec:SDSSAutofill}

Camera dewars are filled automatically using a liquid nitrogen autofill system arranged in the following manner.  A single
180-liter dewar resides on the telescope platform and is connected, during the day only, to two 10-liter intermediate dewars mounted on the instrument rotator; one next to each spectrograph.  Each 10-liter
 dewar feeds the two camera dewars on the adjacent spectrograph.  Cryogenic solenoid valves (ASCO\footnote{ASCO Valve, Inc., http://www.ascovalve.com} Red Hat series), one per dewar, control the flow. 
The intermediate dewars have sufficient capacity to operate through the night without refilling.

Dewar fill is initiated by a timer in the camera electronics that is triggered every hour.  (Filling every hour maintains a more even fill level minimizing temperature variations in the dewar.)
Termination is controlled by a simple analog circuit that uses a thermistor on the vent line to sense the liquid overflow when the dewar is full.  
With this arrangement, both spectrograph camera dewars start filling at the same time, but terminate independently depending on the amount of  LN$_{2}$ remaining at the start of the fill, and variation in parasitic losses in the plumbing. Camera dewars may be filled during observations without any impact on the science.  It typically takes six minutes to fill the dewar once the fill cycle is initiated.

For the 10-liter intermediate dewars, filling is initiated and terminated using the same analog circuit scheme described for the camera dewars.  However, two sensors are used; one located in the tank to sense when the tank is empty, the other located on the vent line to sense when it is full.  Like the camera dewars, the 10-liter dewars are filled to half capacity to avoid spillage as the telescope slews.  Again, the 10-liter intermediate dewars are only filled when the telescope is not in operation.  During operation they are disconnected from the 180-liter supply dewar. Logic in the autofill circuitry prevents the 10-liter empty sensor from triggering a fill if the lines are not connected.  Similarly, interlocks prevent the telescope axes from going live if the 180-liter supply line is connected to the intermediate dewars.

\subsection{Flexure}
\label{sec:SDSS_Flexure}

The variation in gravity load as the telescope tracks the sky causes gravity-induced deformations
within the instrument to change with time and these changes lead to
motion of the spectra on the detector during an exposure.  Motion of the spectra can
degrade resolution and increase cross-talk between the spectra.  For
this reason, the optical bench and the opto-mechanical assemblies
(slithead, collimator, central optics assembly, and cameras) were
designed to be extremely stiff. 

A goal was set to limit flexure to 0.3 pixels in one hour, or 15$^\circ$ on the sky.
This goal stems from the experience that one could reliably find the centroid of an unresolved emission line (a calibration arc line) to 1/10 the slit width, or 0.3 pixels in this case.
So assuming a set of exposures would last an hour (three fifteen minute science exposures, plus fifteen minutes for field acquisition and calibration), it would take a shift of 0.3 or more pixels before it would be noticed.  The 0.3 pixels translates to, roughly, 
0.3 \AA\ in the blue camera, or roughly 20 km/s in redshift. 
Since galaxy internal velocity dispersions are on the order of 200 km/s, a galaxy absorption line would be three or four \AA\ wide and just resolvable by the spectrograph. 
Therefore, flexing 0.3 pixels would just barely be detectable in a single galaxy absorption line.  Now it is the case that with night sky lines for reference, as well as the lines from the calibration lamps, 
this amount of flexure would not be a problem for the actual velocity measurements for high enough S/N.  However, the smearing could make sky subtraction a problem. 
Sky subtraction, at that time, was a much more difficult problem.

The 0.3 pixel budget was divided equally between each of three
subsystems: the slithead, the collimator, and the optical bench; each
being allowed to contribute no more than 0.1 pixel shift in one hour. 
It was assumed, based on the details of the design, that the cameras and
central optics assembly were rigid.  It was also assumed, based on the
designs, that the slithead and collimator would not exceed the
0.1 pixel allowance.  Hence a detailed analysis was not performed on
these two subsystems.  The collimator was analyzed for surface figure
deformation as a function of gravity for the described mounting
constraints, but these models assumed rigid actuators; something we knew
was not true but was sufficient for the purpose of quantifying the
change in surface figure. 

The optical bench, however, did receive scrutiny.  The primary concern
being the deformation in the relatively thin plate structure with
multiple heavy optical assemblies attached to it.  To assess the
stiffness of the optical bench a finite element study was conducted in
collaboration with Swales Aerospace Corporation.  The goal of the
analysis was to quantify the shift in the spectra at the detector for a
change in gravity vector orientation of 15$^\circ$. 

The results showed that flexure in the bench is influenced largely by
the placement of the three kinematic support points that attach the
instrument to the rotator, as well as the thickness of the bench top
plate.  The optimal placement of the supports was determined to be at
the edges of the bench; this set the final location for the mounts.  In
addition, the region of the bench around the mount points was stiffened
by a thick plate, which became integral to the mount itself.  The
results indicated that with this configuration, and for rotations of 15
degrees about the red camera axis, the shift in the spectra would be
less than 0.1 pixels in both channels. 

Details on flexure performance can be found in Section \ref{sec:SDSS_BOSS_Flexure}.

\subsection{Detectors, Electronics and Data Acquisition}
\label{sec:SDSSelex}

\subsubsection{Detectors}

The CCD detectors are thinned Tektronix/SITe SI424A $2048 \times 2048$ devices
with square 24 $\mu$m pixels.  The readout is done through two
amplifiers at opposite ends of a single serial register.  Readout noise
is about 5 electrons and full well is about 150,000 electrons.  A
remarkable characteristic of these devices is a high quantum efficiency in the blue
spectral region. 

\subsubsection{Detector Readout}

The readout electronics are essentially identical to those used in the
SDSS camera, which were described in detail in \citet{gunn98a}.
In brief, the electronics consist of an in-vacuum preamp/clock driver using 
Burr-Brown OPA627\footnote{Now manufactured by Texas Instruments, http://www.ti.com} FET-input low-noise operational amplifiers and a set
of solid-state switches for DC restoration and to generate the 
CCD clocks from supplied rail voltages and CMOS clock signals. This 
unit is physically and electronically identical to the ones used
in the camera. The
voltages and signals come to the dewar over a 37-wire cable, and the
video is transmitted over a 10-wire cable. These cables are routed to
a {\it saddlebag} which is mounted to the optical bench and carries three cards: 1) 
a bus receiver, which receives clock signals over an RS-485 bus serving
both dewars on a single spectrograph and produces CMOS signals for the
dewar and the other saddlebag cards; 2) a power distribution board, which
regulates and filters incoming DC power for distribution to the saddlebags
and dewar, and contains some of the temperature regulation circuitry; and 3)
a complex 8-layer signal-chain/bias board which has circuitry on
one side to generate all the needed rail voltages and on the other to
implement CDS processing of the CCD video, digitize it, and drive
an RS-485 output line to the controller. With the exception of
some component values, these cards are identical to the ones used in the camera. 

The RS-485 bus signals and the RS-485 digitized video to/from each
dewar/saddlebag are carried to a single controller chassis for each
spectrograph on two cables, the RS-485 on a standard 50-pin SCSI
cable and the video/ADC signals on a ribbon cable terminated with
DB-15s. The controller houses power supplies, using boards 
essentially identical to those used for the camera, the TDS Forth
microprocessors\footnote{Triangle Digital Systems, http://www.triangledigital.com} used for all control functions and to generate the
CCD clock waveforms, RS-485 driver/receiver circuitry, and
circuitry to drive the FOXI fiber transmitters,\footnote{FOXI transmitters were provided by Fermi National Accelerator Lab} which send the
(now multiplexed) digital data from the instrument to the data acquisition
hardware approximately a hundred meters away in the operations building. The
camera uses two micros, one for executive control functions and one
for waveform generation, but in the somewhat simpler spectrograph 
environment these functions are combined. Again, see the camera
paper for details.

\subsubsection{Data Acquisition}
\label{sec:SDSS_DAQ}

The data acquisition hardware for the spectrograph CCDs is a clone of
the photometric camera system.  The data are received on FOXI receivers,
demultiplexed and stored on a `pool' SCSI disk by a Motorola MVME167 
single-board computer. Transfers from this VME backplane to the Linux
host are handled by a custom interface designed and built at Fermilab.

On the Linux host, the observer's program (SOS) actuates the
system through remote procedure calls (RPC) over the Ethernet.  The
commands are few and simple, only those required to prep the CCD,
expose, and read out the data.  Once the data are off the CCDs and in
the data system pool, the images are downloaded over the high speed
VME link to the observer's workstation where they are written to
disk.  At the end of the night, the SDSS data was archived and sent to Fermilab.
A typical night of observing produces $\sim$ 1.5 GB of data.

The drilling database is the list of objects on a plug-plate.  It
includes the name and drilling coordinates of each object and the field
coordinates ($\alpha$, $\delta$, equinox) of the plug-plate.  This
information is needed at observing time so the telescope can be pointed
and the plug-plate ID can be incorporated into the data file.  This
database is generated at Fermilab from the photometric imaging data and
is delivered on the internet to the observer's workstation at Apache
Point well before the observations. 

As described in \ref{sec:SDSS_Mapper}, the
plugging station is equipped with a device to map the plug-plate
locations to the slithead after plugging is finished.  The plugging
database information is written directly to the observer's workstation
disk at plug time.  The name of this file is then merged with the CCD data before
it is written to disk as extended FITS keywords for each exposure.\footnote{portions of the description provided here have been copied from http://www.astro.princeton.edu/PBOOK/welcome.htm and are reproduced here for completeness} 

\subsubsection{Spectrograph Control}

Spectrograph operations are controlled from a single program (called SOP)
running on
the observer's Linux workstation.  This program, the observer's sole interface
to the spectrographic system, is a clearinghouse for observing commands
that translates observer's requests into the commands required by the
independent subsystems used for spectroscopy.  These systems include:
spectrograph microprocessor, telescope and guider, CCD data acquisition
system, drilling database, and plugging station. 
    
The spectrograph mechanical controls are handled by a Z-World Little
Giant Z180 microprocessor board.  This processor communicates with the
observer's workstation via RS-232A serial commands sent through an
Ethernet terminal server located at the telescope.  This processor
handles all motion control (shutters, Hartmann masks, collimator tip,
tilt, and focus) and also reports ambient and instrument temperatures
for focus compensation.

The observer's software communicates fully with the telescope control
system via an Ethernet telnet connection.  When a
new plug-plate is locked on the telescope, the spectrograph relays the
plug-plate ID to the observer's software, which looks up the coordinates
in the plug-plate database.  The observer's software commands the
telescope to move to the mid HA position for flat field and wavelength
calibration, then to the field for precise positioning, scaling, and
focusing.  When these operations are finished, the observer's software
starts the guider, opens the shutters, and begins observing.

\subsection{Calibration}
\label{sec:SDSScal}

Calibration for the spectrographs, both flat-fielding to tie the
responses of the fibers together for flux calibration, and the use
of spectral lamps to provide wavelength calibration, are provided as
part of the telescope structure. The screens for these functions as well
as the lamps are mounted on the windbaffle structure, which serves
as a dome/enclosure for the SDSS telescope \citep[see e.g.,][]{gunn06a}.

At the top of the windbaffle structure is a mechanism incorporating
eight lightweight aluminum-honeycomb sector-shaped panels mounted on
shafts and driven by DC gearmotors.  These panels can be closed to cover the
annular aperture of the telescope or opened to allow observing.  The
bottoms of these panels are coated with a special wide-spectrum Lambertian
paint (PolarKote\footnote{Manufactured by Light Beam Industries LLC, http://www.lightbeaminc.com}), providing a highly uniform diffuse reflective surface.

The panels are illuminated in quarters by a bank of three projectors
in each corner of the windbaffle structure just above the central
structure which supports the primary mirror and the altitude bearings.
The projectors output a quarter-annulus which fills the pupil of
the telescope and avoids illuminating the secondary structure
directly. One of these units in each of the three-projector assemblies
is illuminated by a 3100K quartz-iodine lamp,\footnote{Phillips Projection Lamp, Type 7027, model\# 409829, supplied by Specialty Bulb Co. Inc., http://www.bulbspecialists.com} and is used for flat
fields; another houses a mercury-cadmium lamp,\footnote{Osram HgCd Lamp, model\# HgCd/10, supplied by Specialty Bulb Co. Inc., http://www.bulbspecialists.com}  which has a moderately
rich emission spectrum in the blue and green, and the last houses
a neon lamp,\footnote{Osram Neon Lamp, model\# Ne/10, supplied by Specialty Bulb Co. Inc., http://www.bulbspecialists.com} which is used primarily to calibrate the red region of
the spectrum. 

The flat-field lamp is fitted with a combination of glass Schott
filters (2 mm FG3 + 2 mm BG14) 
in an attempt to mitigate the very rapid falloff of the
lamp spectrum to the blue, so that reasonable flat fields can
be obtained with good S/N in one exposure with no
danger of saturation. This approach was moderately successful, but at the expense of
producing a rather `lumpy' spectrum.

Since the spectrographs are mounted on the telescope, they flex, and
so it is necessary to be careful about flat fields. Given the very large spatial modulation of brightness on the CCD, if the image
of a fiber shifts one pixel during an exposure, the registration of
the object spectrum and the calibration flat field drift out of alignment.
Hence, very little flexure could be tolerated.  For this reason, special flat
fields are occasionally recorded by dithering the collimator mirror slightly
in angle along the spatial direction.  Doing so produces wider traces than the
normal flat field exposures, thus determining pixel-to-pixel variations on the detector, and relaxing the flexure tolerance in the spatial direction. 
One can then properly calibrate the pixel-response during science exposures
even if the spectrum moves slightly due to flexure.

In practice, this process of flat-field dithering consists of twenty
successive exposures at various tilts of the collimator mirror.
The sequence generates a map of flat-field projections offset by 0.6 pixels
between exposures. Traversing a span of 12 pixels, about four
fiber widths, these dithered flats therefore sample the detector
response over the entire spatial direction. 

\subsection{Guiding}
\label{sec:SDSSguider}

Guiding for the SDSS spectrograph is accomplished by the use of coherent
imaging fiber bundles which are connected to the plug-plates at the
locations of guide stars using ferrules which are externally identical
to the science fiber ferrules.  The imaging bundles used in SDSS-I and -II
were Sumitomo\footnote{Sumitomo Electric Lightwave Corp., http://www.sumitomoelectric.com} coherent fibers of two sizes which projected to $7\2pr$ and $11\2pr$
in the focal plane. There were nine $7\2pr$ bundles and two $11\2pr$ 
bundles. The larger ones were primarily intended for acquisition of the guide
stars, although they were also used for normal guiding as well after all the
stars were centered.

The guide fiber ferrules were modified slightly for this purpose; it is
necessary, clearly, to orient the guide fibers rotationally so the
guider will know what to do when a star is off-center.  This is
accomplished by mounting a collar with a small pin on the exterior diameter of
the ferrule.  A hole, smaller than the ferrule holes, receiving the pin
is drilled into the plate at the time it is drilled, and the plate
pluggers rotate the guide ferrule to drop the pin into the hole. The
guide fibers are {\it not} randomly plugged, but assigned to marked
holes in the plate.

The guide fibers are routed to eleven holes in a small block into
which they are cemented and polished.  That block is imaged through a
back-to-back pair of 50 mm f/1.2 Nikon camera lenses, with a 2 mm Schott BG38
filter in the parallel beam between them, onto the CCD in a Roper
Scientific (Photometrics) SenSys camera.\footnote{Roper Scientific, Inc., http://www.roperscientific.com} The CCD is a Kodak KAF0401E, which
has an $768 \times 512$ array of 
9 $\mu$m pixels. Detector readout was set in a bin-by-2 mode, which
resulted in 18 $\mu$m ($0.3\2pr$) pixels. The camera has a thermoelectric
cooler, which limits the CCD temperature to 10$^\circ$ C, but is otherwise not
temperature controlled. 
The camera came equipped with a SCSI interface,
and was connected to an Apple Macintosh computer with a commercial
SCSI-to-fiber-to-SCSI interface.

The images were buffered on the Mac and sent by fiber ethernet to
the control room where they are received and processed by the Linux
SOP computer. A PID servo was implemented in that machine to generate
correction signals to the telescope control computer. Display code
to allow the observers to monitor the performance of the guider,
either by generating basically a raw image of the fiber block or
a display in which the images of the fibers are placed approximately
where they are on the sky and oriented properly, was written. 

The performance of the system was never completely satisfactory; the CCD
was (essentially) uncooled, was noisy, had many dark defects,
and its mounting was not mechanically stable.  Halfway
through the survey new code was written that handled
centroiding properly even if the star was at the edge of the fiber
image, and explicitly used dark frames and flat fields to correct the
raw images.  This enhancement helped, but a new and better camera was clearly
needed, which did not arrive until the beginning of the BOSS survey.

\section{BOSS Upgrade Design}
\label{sec:Upgrade}

The SDSS camera and spectrographs were operated for
from 2000 -- 2005, followed by a three-year extension known as SDSS-II.
In 2005, the Astrophysical Research Consortium, which owns and operates the
Apache Point Observatory, put out a call for proposals to operate
the 2.5-m Sloan telescope and its instruments after 2008.
SEGUE-II, a one-year extension of the
Sloan Extension for Extra-Galactic Understanding and Exploration \citep[SEGUE;][]{yanny09a}
was approved for one observing season to use the SDSS spectrograph from mid-2008 to mid-2009.
In response to this announcement, the Baryon Oscillation
Spectroscopic Survey \citep[BOSS;][]{schlegel09a,dawson13a} was also
proposed and awarded five years of dark observing time beginning in 2009.
The primary scientific objective proposed by
the BOSS project was to
map the baryon acoustic peak to redshift $z=0.7$ with percent-level precision.
In addition, BOSS would pioneer a
new technique to detect large scale structure in the spectra of distant quasars by observing the
the so-called Lyman-$\alpha$ forest, a series of absorption features caused by intervening clouds of hydrogen gas.

To achieve the BOSS scientific
goals, a major upgrade of the SDSS spectrographs was required.
In particular, BOSS needed a higher multiplex factor in the number
of fibers per exposure, and higher optical throughput.  To achieve this, the BOSS proposal included plans to build a new
fiber system and to upgrade the optical system and cameras of the original SDSS spectrographs.
In this section we describe the requirements for BOSS and the hardware upgrades
to achieve the new performance.

\subsection{BOSS Requirements}
\label{sec:BOSSreq}

The BOSS science goals require spectra of approximately 1.35 million luminous galaxies as faint as $i = 19.9$, and 160,000 quasar spectra ($g < 22$), over 10,000 deg$^{2}$, in five years. 
Compared to SDSS, the BOSS spectrographs must obtain 35\% more spectra per unit time for objects that are more than a magnitude fainter.  
While most of the SDSS spectrograph is portable to the BOSS requirements, higher multi-plex factor and throughput necessitated changes to the original design.  
Here we summarize the requirements that dictated the upgrades for the BOSS spectrographs.

\subsubsection{Number of Fibers}

To achieve a percent-level precision measurement of the cosmic distance ladder,
BOSS requires redshift measurements of approximately 1.35 million luminous
galaxies to $z=0.7$ over 10,000 deg$^{2}$.
The availability of modern, large format CCDs with smaller pixels enabled
us to increase coverage of the spectrograph focal plane and an increase
in the number of fibers.
The total number of fibers for BOSS was increased to 1000 per plate,
or 500 per spectrograph.
This final number of fibers represents a practical limit given the desire
to retain most of the original SDSS spectrograph optics while minimizing
vignetting for objects imaged near the edge of the optics.

\subsubsection{Fiber Diameter} 

As discussed in Section \ref{sec:SDSSreq}, the optimal fiber diameter is set by
the desire to maximize S/N for an extended source given the sky background.
The BOSS galaxy sample extends to $z=0.7$, projecting a typical galaxy
onto the telescope focal plane at an angular scale of a few arcseconds.
The appropriate fiber size to maximize signal from the source while minimizing
sky background for most distant galaxies is $2\2pr$, corresponding to a
fiber diameter of 120 microns.
The BOSS fibers are therefore 2/3 the original SDSS fiber diameter.
The smaller diameter fiber scales with the reduction in pixel size
(24 $\mu$m for SDSS and 15 $\mu$m for BOSS) for the BOSS CCDs,
thus preserving the desired spectral sampling of three pixels.
The smaller diameter also facilitates the increase in the number of fibers
described in the previous paragraph.

\subsubsection{Wavelength Range}

BOSS requires a modest increase in wavelength range compared to SDSS.
The extension in the blue wavelength coverage is driven by the desire to
measure absorption in the Lyman-$\alpha$ forest at $z > 2.2$.
The Lyman-$\alpha$ transition occurs at a wavelength of $1216$ \AA\ in the
quasar rest frame and the Lyman-$\beta$ transition occurs at $1026$ \AA.
The minimum wavelength requirement for BOSS was set at 3560 \AA\
so as to sample roughly half of the Lyman-$\alpha$
forest illuminated by a $z = 2.2$ quasar between the Lyman-$\alpha$
and Lyman-$\beta$ transitions.
The red cutoff was also extended from  9100 \AA\ to 10,400 \AA.
The increase in coverage allows better coverage of the
continuum redward of the Ca II K and H absorption lines that are the most
common features used for redshift determination in the BOSS galaxy sample.
The additional wavelength coverage enhances our ability to reliably classify
passive galaxies up to a redshift of at least $z = 0.8$.
The BOSS wavelength coverage is larger than the original SDSS
requirements by 1600 \AA, enabled by increased detector size.

\subsubsection{Detector Read Noise and Dark Current}

To maintain Poisson-limited statistics in the $z = 2.2$ region of the Lyman-$\alpha$ forest,
the BOSS blue channel detector must have a read noise less than 3.0 $e^-$/pixel RMS.
The requirement for the red channel is less stringent because the spectra
are sky-dominated for most of the red channel wavelength range.
The read noise for the red detector is required to be less than 5.0 $e^-$/pixel RMS.
To prevent significant noise contributions during an integration, a dark current that is less than 1.0 $e^-$ for the blue CCDs and
less than 2.0 $e^-$ for the red CCDs during a 15 minute exposure is required.
All of these requirements for detector performance were enabled by
improvements in CCD technology. 

\subsubsection{Resolving Power}

The resolution requirement for SDSS was broadly defined to balance the
tradeoff between resolving absorption lines in relatively massive
galaxies and maximizing wavelength coverage.
The resolution requirement for BOSS is driven by the ability to fit the
Balmer series in calibration stars
between 3800$<\lambda < $4900 \AA.  From these fits we model surface gravity and
stellar parameters and derive a synthetic spectrum for each standard star
to calibrate spectrophotometry for the exposure.

Experience with SDSS indicates that a resolving power greater than $R=1400$ in this wavelength range is adequate for precise and reliable stellar classification.
The BOSS galaxy spectra require resolution sufficient for achieving RMS redshift error
of 300 km/s, achieved with a resolving power of $R>1000$ for
the remainder of the wavelength range.

\subsubsection{Throughput and Signal to Noise Ratio}
\label{BOSS_Throughput_S/N}

The required throughput is set by the desire to obtain spectra of
1.35 million luminous galaxies to a limiting magnitude of $i = 19.9$
in five years.
From observations of fainter galaxies in SDSS and a well-tested
data reduction pipeline, it became clear that successful redshift
classifications were possible with significantly lower S/N than
was achieved in SDSS.
We established a rough goal of S/N$>3-4$ per \AA\
for the faintest targets.  With one hour exposures and
1000 fibers, BOSS requires a factor of two improvement
in peak throughput.
The improvement is possible through the use of state-of-the-art large format
volume-phase holographic (VPH) gratings and
modern CCDs with higher quantum efficiency.

\subsection{Fiber System Upgrade}
\label{sec:BOSSfiber}

A major part of the SDSS spectrograph upgrade for BOSS was the
acquisition of nine new fiber cartridges with 1000   fibers
to replace the original SDSS cartridges which had 640 fibers. For BOSS, the cartridge design, including the castings themselves, is identical to that used for SDSS with the exception of revised slitheads to accommodate more fibers having a smaller diameter.  The fiber mapper system follows the same philosophy as that used for SDSS, and fiber performance was tested in the same manner as the original SDSS fibers.  For more detail on the cartridge design, the fiber mapper, and the method for assessing fiber performance see Section \ref{sec:SDSSfiber}.
Here we describe the BOSS fiber system upgrade, which for the most part is limited to the implementation of a larger number of smaller fibers.

\subsubsection{Fibers}

The optical fiber selected for BOSS is a step-index, all-silica, UV-enhanced fiber. 
It is similar to the fiber originally chosen for SDSS, except that it has a smaller core
diameter of 120 $\mu$m.  The core is surrounded by 25 $\mu$m of a lower-index cladding
with a cladding-to-core ratio of at least 1.4, which confines the radiation by total internal
reflection. The cladding is protected by a polyimide sheath, or buffer, bringing the outer
diameter to 190 $\mu$m.  The tolerance on the outer diameter was specified at $\pm$ 3 $\mu$m,
with a maximum decentering of the core relative to the outer diameter
of 6 $\mu$m.  These tolerances were tighter than the $\pm$ 5 $\mu$m
outer diameter that was specified for BOSS prototype fibers with
a larger outer diameter of 197 micron.
The fiber was manufactured by Polymicro Technologies, Inc. and is 
known as FBP 120-170-190.  The total length of each fiber is $1830 \pm 25$ mm.

The design of the fiber harnesses for BOSS follows that of the original SDSS fiber harnesses as described in Section 2.2, with the same overall dimensions and components.  The same manufacturer, C-Technologies, Inc, was selected during a competitive bidding process.   As in the original SDSS  design, a fiber harness consists of 20 fibers, with one end mass-terminated by gluing the fibers 
into a metal v-groove block, while the free ends of the fibers are individually terminated in stainless steel ferrules.
The v-groove blocks align the fibers on 260 $\mu$m centers; the v-groove blocks are then glued to the slitheads.
The v-groove blocks were machined in the Physics Machine Shop at the University of Washington.
The stainless steel ferrules were manufactured by Swiss Screw Products\footnote{Swiss Screw Products, Inc., http://www.swissscrew.com}  After the fiber was inserted into the ferrule and
glued using a low-shrinkage adhesive,  the end of the fiber was  polished, removing a small amount of material and maintaining strict
tolerances on planarity.  The ferrule end was designed for insertion into the holes drilled in the plug-plates, where it is held by friction.
As with the SDSS configuration, a jacket of nylon tubing on the ferrule end protects the fibers from breakage during repeated plugging and unplugging operations.

\subsubsection{Fiber Performance}

\begin{figure}[htbp]
\begin{center}
\epsscale{1.0}
\plotone{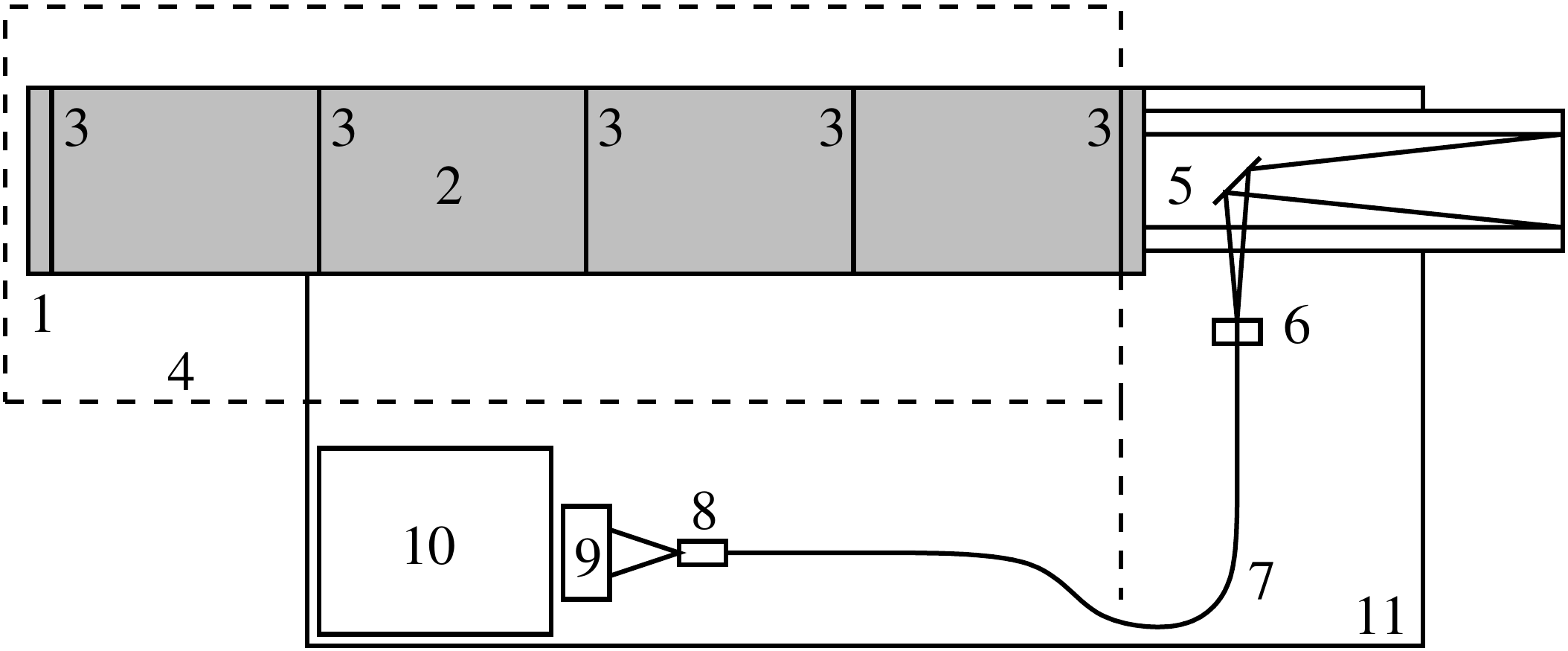}
\caption{{\bf BOSS fiber focal-ratio degradation test setup diagram:
(1) White LED array; (2) Diffuser tube; (3) Diffusing screens; (4)
Light shroud; (5) Newtonian reflector telescope stopped to $f/5$
with 27\% central areal obscuration; (6) Fiber input bracket; (7) BOSS optical
fiber; (8) V-groove block with fiber termination; (9) SBIG ST-8300M
CCD camera; (10) Data-acquisition computer; (11) Optical breadboard
(24$\times$48\,in). System shown approximately to scale.}}
\label{fig:fibertest}
\end{center}
\end{figure}

\begin{figure}[htbp]
\begin{center}
\epsscale{1.15}
\plotone{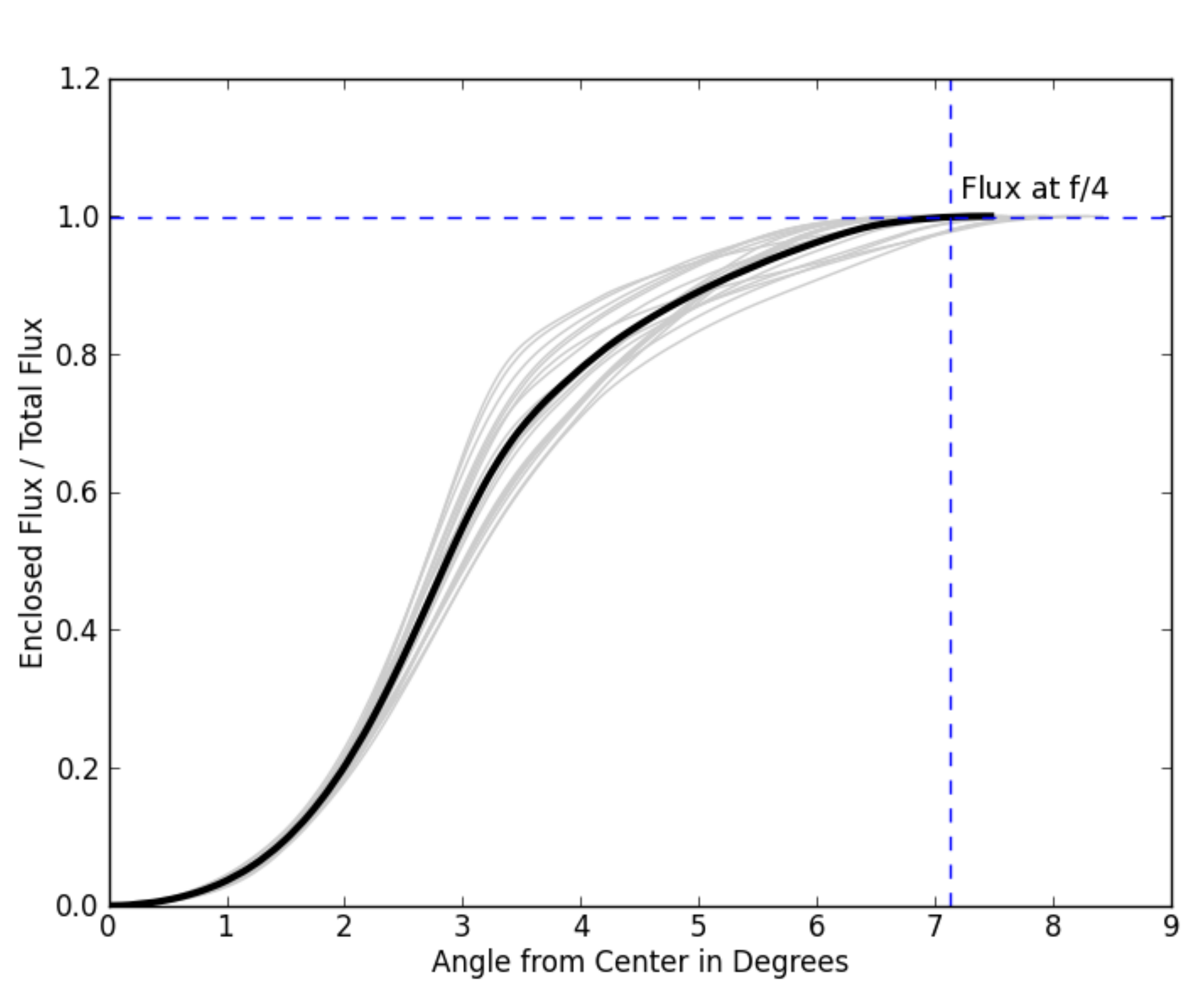}
\caption{{\bf Focal-ratio degradation curves for 18 tested BOSS fibers (thin gray),
with average curve (thick black).}}
\label{fig:FRDcurves}
\end{center}
\end{figure}

To measure the FRD characteristics of BOSS fibers as configured
for science observations (as well as to characterize the
effective exit pupil of the telescope--spectrograph system),
the testing setup diagrammed in Figure~\ref{fig:fibertest}
was constructed.   This system was used to illuminate
18 fibers, one at a time, within a spare
BOSS harness with a white-light input beam matched to the $f/5$ SDSS 2.5-m telescope
beam, \textit{including the 27\% areal obscuration that removes
the central rays from the cone of light incident on the
telescope focal plane.}
The ferrule of each fiber under investigation was held in place using
an aluminum plate drilled with a decommissioned SDSS-III drill bit.
A tip-tilt bracket was used to adjust the fibers for telecentricity
with respect to the input beam.  The diverging output beam
was imaged using an SBIG ST-8300M CCD camera\footnote{SBIG Astronomical Instruments, http://www.sbig.com} with 3326$\times$2504
pixels of 5.4\,$\mu$m pitch, with the detector plane located a distance
of $4.7 \pm 0.1$\,cm from the output termination of the fibers
in the v-groove block.  The resulting beam-profile images
show a characteristic ``fuzzy donut''
of the telescope pupil convolved with the FRD within the fiber.
The beam patterns were flux-centroided in the images,
and the fractional flux enclosed as a function of distance
from beam center was computed for each of the 18 fibers.
The resulting curves are shown in Figure~\ref{fig:FRDcurves},
with respect to an abscissa of beam half-angle in degrees.
The mean curve for all 18 fibers is shown in thick black.
We confirm that with an $f/4$ spectrograph collimator,
essentially all of the emergent flux is captured by the optical system.

\subsubsection{Slithead}
The BOSS slitheads are nearly identical in design to the original SDSS slitheads.  They consist of a thin, stiff aluminum slitplate with a curved edge of radius 640 mm.
There are two slitheads per cartridge, each having 500 fibers.  The 120 $\mu$m diameter fibers are mounted in groups of
twenty in v-groove blocks, with 25 v-groove blocks being glued to each slithead.
The center-to-center spacing between fibers on adjacent v-groove blocks is 624 $\mu$m, compared to 260 $\mu$m between fibers within a v-groove block.
The total length of the arc is 138.6 mm from outside edge to outside edge of the first and last fibers, slightly taller than the original SDSS design.

\subsection{Optical Upgrades}
\label{sec:BOSSoptic}

\subsubsection{Optical Design Overview}

The optical design of the BOSS spectrograph is nearly identical to the original SDSS design. For BOSS the bandpass is extended in both channels to cover $3560 < \lambda < 10,400$ \AA. 
Additionally the replicated surface relief gratings used in the SDSS design were replaced with volume phase holographic (VPH) gratings, and modern CCDs with higher quantum efficiency and smaller pixels were installed; the combined effect being an impressive 70\% peak instrumental efficiency, a significant improvement over the 45\% peak of the SDSS design.

\begin{figure*}[htbp]
\begin{center}
\epsscale{1.0}
\plotone{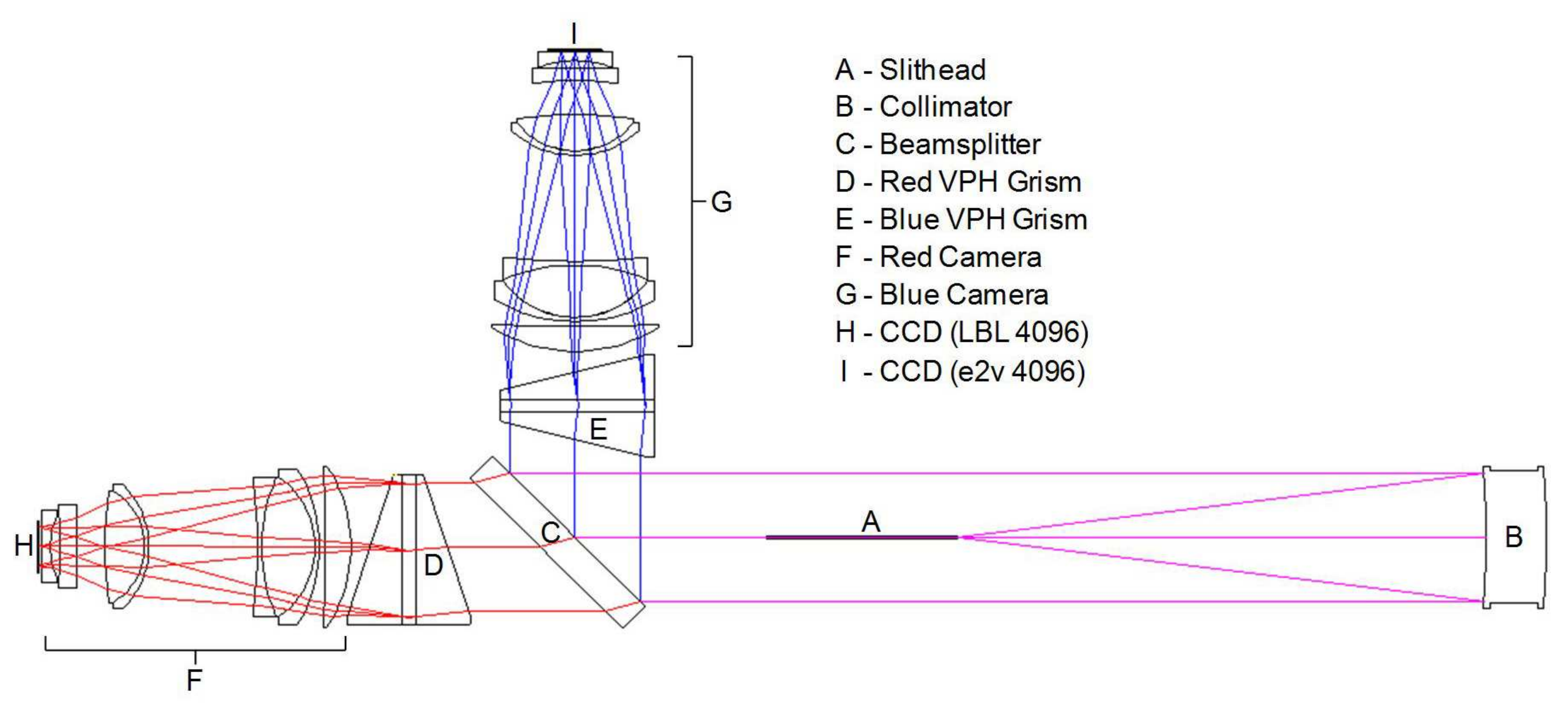}
\caption{{\bf Optical layout of the BOSS spectrographs. Light enters each spectrograph through 500, 120 $\mu$m diameter fibers, which terminate at a curved slit plate mounted inside the slithead.  The slit plate, which is integral to the slithead, positions the fiber ends on a radius concentric with the spherical collimating mirror, which  operates at f/4 and produces a 160 mm diameter beam.  The 45 degree dichroic beamsplitter reflects the blue portion of the bandpass ($\lambda < 6050$ \AA) and transmits the red wavelengths ($\lambda > 6050$ \AA).  Immediately after the beamsplitter in each channel is a VPH grism, consisting of a VPH grating sandwiched between a pair of prisms (the grating is tilted 3.5$^\circ$ to steer the Littrow ghost off the detector). The dispersed light exits the grisms and enters all-refractive, eight-element, f/1.5 cameras.  Each camera contains a 4k $\times$ 4k CCD with 15 $\mu$m pixels.  The camera demagnification from f/4 to f/1.5 produces fiber images that are just 3 pixels in diameter, resulting in 3 pixel tall spectra on the detector.}}
\label{BOSSoptical}
\end{center}
\end{figure*}

Figure~\ref{BOSSoptical} shows the BOSS optical layout. Light enters the spectrograph through 120 $\mu$m diameter fibers,
which terminate at a curved slithead in the same manner as the original SDSS slithead. The collimator, unchanged from the original design, collimates the f/4 beam producing a 160 mm diameter beam, reflecting it back toward the dichroic beamsplitter. Light blueward of 640 nm is reflected into the blue channel, with the remaining light being transmitted into the red channel. A VPH grism in each channel disperses the light, which is then imaged by upgraded cameras, each containing a single 4k $\times$ 4k CCD with 15 $\mu$m pixels. The use of smaller fibers and correspondingly smaller pixels, along with the same collimator/camera demagnification preserves the scale of the spectra on the detector, thus yielding 3 pixel wide spectral profiles separated by 6 pixels center-to-center; identical to the SDSS configuration. 
 
\subsubsection{Collimator}

The BOSS survey started in 2009 with the same collimators used for SDSS. At that time protected silver coatings offering significantly higher reflectivity at the blue end of the spectrum had been realized
on small optics. Given the obvious benefit of such a coating,
the BOSS project set out to re-coat the collimators, beginning with the spare collimator originally produced for SDSS. However, realization of such a coating on a large surface such as the collimator was challenging due to the process control required to achieve very uniform layer depositions over a large surface area. Attempts by one vendor ultimately failed after more than a year of trial and error. A successful coating was achieved by Infinite Optics\footnote{Infinite Optics, http://www.infiniteoptics.com}  in late 2011, and
newly coated collimators have been installed in both spectrographs.

\subsubsection{Beamsplitter}

The dichroic beamsplitter in each spectrograph was replaced for BOSS given the expanded bandpass of the instrument.  For BOSS, wavelengths shorter than 605 nm are reflected by the new coating, a 15 nm shift toward the red as compared to SDSS.  Although the BOSS bandpass is extended much more into the red than the blue, the crossover wavelength was shifted only a small amount.
In the process the entire beamsplitter, i.e., substrate and coating, were replaced; a logical consequence of the desire to keep the instrument operational while the new coatings were applied, and to mitigate the risk of potentially taking the instrument offline for an extended period should stripping the original coating lead to damage of the substrate.

An additional, and significant, consideration in replacing the beamsplitters
was the desire to obtain a denser coating using ion-assisted deposition,
which would be more immune to wavelength shifts with variations in humidity.
The SDSS beamsplitters suffered from throughput variations with changes
in humidity inside the instrument, which occurs due to the periodic insertion
of new cartridges. Each time a cartridge is installed the slithead door
is opened, allowing moist air to displace the dry air injected into the
bench to protect the calcium fluoride singlet in the camera. Once the door
closes the humidity level slowly reduces back to its original near-zero level.
The effect on the data are throughput variations as large as 5\% in
$\sim 50$ \AA\ wide ripples in the 3800-4800 \AA\ range, and shifts of
the dichroic transition region.
A more detailed description of this problem and its removal from the data
can be found in Section 8.1 of \citet{abazajian09a}. There is no evidence of throughput variations with the new coating.

\subsubsection{Gratings}

As in the SDSS spectrographs, a grism immediately follows the beamsplitter in each channel. However, the significant increase in efficiency offered by VPH gratings convinced us to abandon the SDSS replicated surface relief grism design in favor of a VPH grism. The grism consists of a VPH transmission grating sandwiched between two prisms; the use of prisms is necessary to preserve the straight-through beam layout of the SDSS design.

The BOSS design utilizes a ruling density of 400 l/mm in the red channel and 520 l/mm in the blue channel. The grisms were designed independently at Johns Hopkins University (using the Zemax raytrace program) and at Kaiser Optical Systems, Inc.\footnote{Kaiser Optical Systems, Inc., http://www.kosi.com} (KOSI), who built the gratings.  Both the VPH grating substrates and the prisms are made from BK7 or equivalent glass.  The blue channel components were made from Ohara Corporation's BSL7Y glass, a BK7 equivalent with enhanced UV transmission.  The prism apex angles (total for the two-prism assembly) are 28$^\circ$ and 35.6$^\circ$ for the blue and red grisms, respectively.

A phenomenon now well-known to builders of VPH grating-based spectrographs, is that of the Littrow ghost, or recombination ghost as it is sometimes called \citep[i.e., Littrow ghosts;][]{burgh07a}.  This is an undispersed (white-light) ghost image of the entrance slit or fibers of the instrument, and appears in the image plane at the location of the blaze or Bragg wavelength.  Because the ghost is undispersed, its intensity can be quite bright relative to the nearby features of interest in the spectrum.  This particular type of ghost is present in any spectrograph operating in the Littrow configuration ($\alpha$ = $\beta$) but plagues in particular instruments with VPH gratings as they are so often used at or near the Littrow mode in order to obtain the highest efficiency possible.

\begin{figure}[htbp]
\begin{center}
\epsscale{1.15}
\plotone{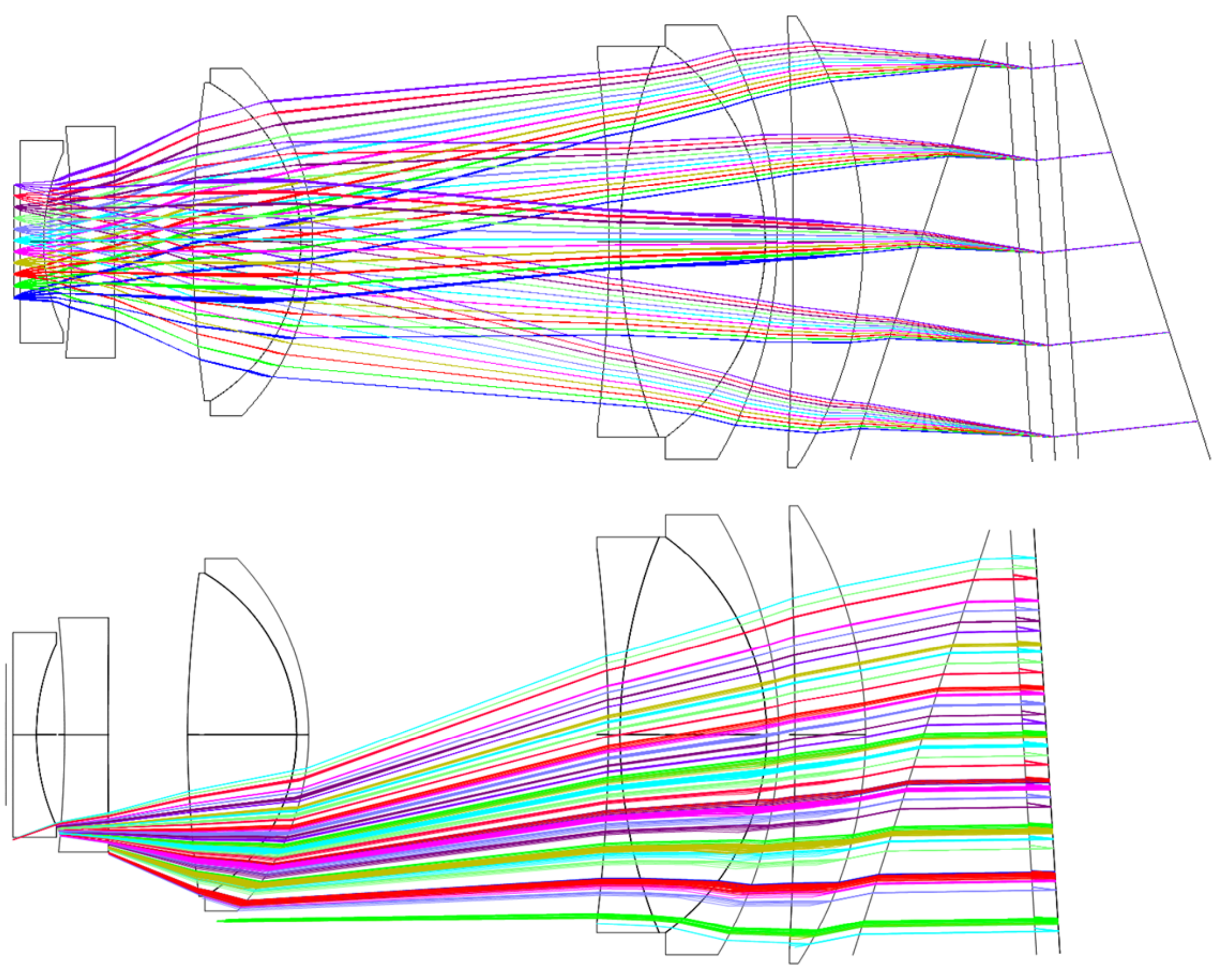}
\caption{{\bf Raytrace layouts illustrating the light path for the recombination (Littrow) ghost.  In these layouts the CCDs are on the left, the VPH grism assembly (or part thereof) is on the right.  The desired, first order diffracted light (top) travels right to left through the grism and camera lenses to the CCD.  The recombination ghost arising within the grating layer (bottom) is initiated by a reflection off the CCD, which sends some small percentage of the light back through the camera in a left-to-right direction.  The camera recollimates this light, which then encounters the grating for a second time. Some of this light is reflected and recombined by the grating and travels back through the camera to the CCD.  It is this last, right-to-left pass that is shown in the layout, the other passes are omitted for clarity.  Note how the grating is rotated by 3.5 degrees with the prism assembly in order to direct the recombination ghost entirely off the CCD.}}
\label{BOSS_VPH_Ghosts}
\end{center}
\end{figure}

To better understand the implications of the recombination ghost for the BOSS spectrographs, we undertook a rather extensive raytrace analysis of the possible ghost paths initiated by a reflection off the detector.  In addition, we analyzed the paths of the zero-order light and the undesired diffracted orders with any appreciable efficiency predicted by KOSI's rigorous coupled-wave analysis of the grating designs.  In order to move the recombination ghost away from the useful part of the spectrum, the grating must be moved away from the Littrow configuration.  One way to do this with a VPH grating is by rotating the fringes within the gelatin layer during exposure, to be off-normal with respect to the plane of the grating itself.  The grating is rotated by the same angle in use such that the fringe orientation is returned to the Bragg condition where efficiency is highest, but the grating is no longer at Littrow.  Our analysis indicated a rotation of three degrees would be sufficient to move the ghost completely off the detector, and we settled on 3.5 degrees to provide a bit of margin.  To accommodate the tilted grating each prism deviates from a right angle by the same 3.5 degree tilt, with one prism having an apex angle that is 3.5 degrees larger and the other 3.5 degrees smaller than the right angle design of an untilted grating.  We chose this approach, rather than using right angle prisms and rotating the entire grism assembly as a whole, due to extreme space constraints within the upgraded central optics assembly.  Figure~\ref{BOSS_VPH_Ghosts} shows the nominal, first order light path in the upper panel, and in the lower panel the recombination ghost path arising within the grating layer.

For BOSS, the VPH red grating peak efficiency is 82\% as compared to 64\% for SDSS, and the blue grating peak efficiency is 80\% as compared to 50\% for SDSS.

\subsubsection{Cameras}

Both red and blue camera optical designs were reoptimized for BOSS. The revised designs use the original SDSS prescriptions for the first six elements with the same multiplet configuration: a singlet, followed by a contact triplet, and then a contact doublet. As in the original design this sequence of elements is followed by two field flatteners, which reside just in front of the detector. For BOSS the spacings between these groupings were modified slightly, and the field flatteners were completely redesigned for improved imaging performance and larger focal plane arrays. These changes were motivated in part by the expanded wavelength range in each channel, and in part by new, larger (4k $\times$ 4k as compared to 2k $\times$ 2k), detectors having smaller pixels (15 $\mu$m for BOSS as compared to 24 $\mu$m for the SDSS design) and working over a slightly larger field. Hence imaging performance requirements for BOSS are more stringent than for SDSS. Nonetheless the BOSS design uses the same basic eight-element configuration as the original SDSS cameras.  In fact, the singlet and doublet elements are re-used from SDSS.  For the triplet, new lenses were fabricated for the first and third elements to take advantage of improved coatings that are properly matched to the BOSS bandpasses in each channel.; the central CaF$_{2}$ triplet, which is not coated, was re-used from SDSS.  For the field flatteners, new lenses were fabricated.

\subsubsection{Optical Prescription}

Table~\ref{BOSSPrescript} shows the optical prescription for the BOSS spectrograph.
The surface descriptions are the same as in Table~\ref{OpticalPrescript}.
As with the SDSS spectrographs, all materials are from Ohara Glass except for CaF$_{2}$ and the lens couplant, Dow-Corning Q2-3067.

\begin{table*}
\begin{center}
\caption{Optical Prescription for the BOSS Spectrographs\label{BOSSPrescript}}
\begin{tabular}{lrlrlrl}
\tableline\
Surface & \multicolumn{2}{c}{Radius (mm)} & \multicolumn{2}{c}{Thickness (mm)} & \multicolumn{2}{c}{Material} \\
 & Blue & Red & Blue & Red & Blue & Red  \\
\tableline
slithead            & -640.080               & -640.080               & 630.123    & 630.123    & air        & air\\
collimator mirror   & -1263.904              & -1263.904              & -1087.831  & -1087.831  & air        & air\\
prism               & plano($-14.0^{\circ}$) & plano($-17.8^{\circ}$) & -33.604    & -37.414    & BSL7Y      & BK7\\
cover plate         & plano($-3.5^{\circ}$)  & plano($-3.5^{\circ}$)  & -8.000         & -8.000         & BSL7Y      & BK7\\
grating             & plano($-3.5^{\circ}$)  & plano($-3.5^{\circ}$)  & -8.000         & -8.000         & BSL7Y      & BK7\\
prism               & plano($14.0^{\circ}$)  & plano($17.8^{\circ}$)  & -33.376    & -40.005    & BSL7Y      & BK7\\
singlet,front       & -182.804               & -182.804               & -30.531    & -30.531    & CaF$_{2}$  & CaF$_{2}$\\
singlet,back        & -1813.560              & -1813.560              & -6.604     & -6.960     & air        & air\\
triplet,1st,front   & -185.522               & -185.522               & -5.156     & -5.156     & LAL7       & LAL7\\
triplet,1st,back    & -105.867               & -105.867               & -0.076     & -0.076     & Q2-3067    & Q2-3067\\
triplet,2nd,front   & -105.867               & -105.867               & -63.525    & -63.525    & CaF$_{2}$  & CaF$_{2}$\\
triplet,2nd,back    & 226.136                & 226.136                & -0.076     & -0.076     & Q2-3067    & Q2-3067\\
triplet,3rd,front   & 226.136                & 226.136                & -5.105     & -5.105     & BSM2       & BSM2\\
triplet,3rd,back    & -661.416               & -661.416               & -131.750   & -129.718   & air        & air\\
doublet,1st,front   & -110.490               & -109.423               & -5.080     & -5.080     & BAL35Y     & BAL35Y\\
doublet,1st,back    & -87.274                & -81.331                & -0.076     & -0.076     & Q2-3067    & Q2-3067\\
doublet,2nd,front   & -87.274                & -81.331                & -44.120    & -46.990    & FPL51Y     & FPL51Y\\
doublet,2nd,back   & 429.006\footnote{Aspheric surface coefficients: $\alpha_4 = 7.517856E-04$, $\alpha_6 = -5.900703E-05$, $\alpha_8 = 2.644114E-06$}           & 432.892\footnote{Aspheric surface coefficients: $\alpha_4 = -7.880071E-04$, $\alpha_6 = 6.800388E-05$, $\alpha_8 = -2.676036E-06$}           & -44.094    & -34.823    & air        & air\\
flattener,1st,front & 294.716                & 3362.350\footnote{Aspheric surface coefficients: $\alpha_4 = 6.490621E-05 $, $\alpha_6 = 5.825222E-05 $, $\alpha_8 = -4.033623E-05$}          & -14.961    & -18.796    & BSM51Y     & CaF$_{2}$\\
flattener,1st,back  & plano                  & -503.453               & -9.398     & -12.141    & air        & air\\
flattener,2nd,front & 99.187                 & 94.971                 & -6.096     & -10.185    & BSM51Y     & S-LAH59\\
flattener,2nd,back  & plano                  & plano                  & -3.175     & -3.048     & air        & air\\
CCD                 & plano                  & plano                  & 0          & 0          & silicon    & silicon\\
\tableline
\end{tabular}
\end{center}
\end{table*}

\subsubsection{Predicted Optical Performance}

The following sections discuss the predicted optical performance of
the BOSS spectrographs, including image quality, spectral resolution,
and throughput. Measured performance is discussed in Section \ref{sec:Perf}.

\paragraph{Image Quality}

\begin{figure}[htbp]
\centering
\vspace{1mm}
\includegraphics[scale=.47]{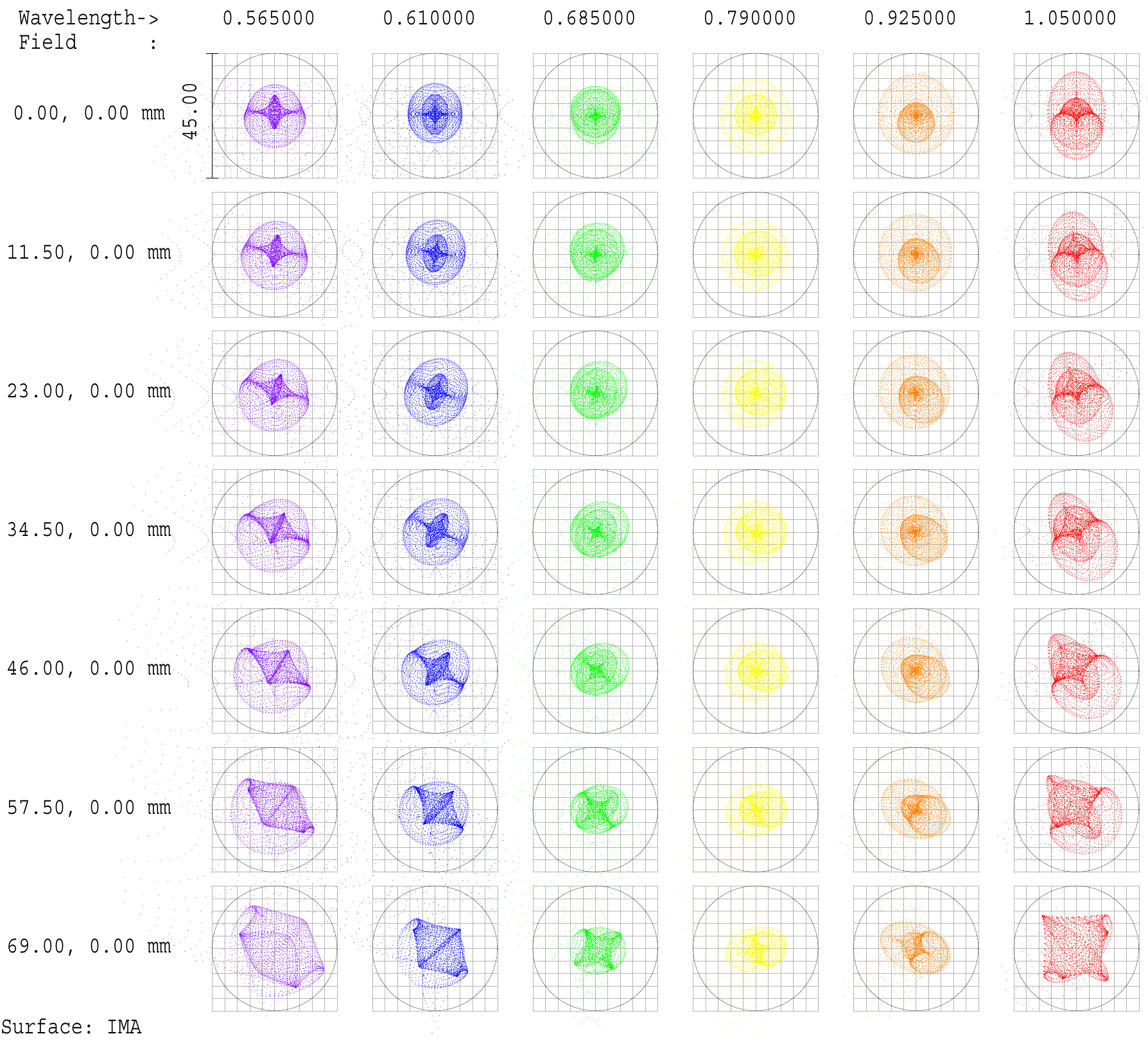}
\caption{{\bf Spot diagram for the BOSS red channel. The average RMS spot diameter is 18.3 $\mu$m and the maximum RMS diameter is 27.5 $\mu$m}}
\label{red_spots_BOSS}
\end{figure}

\begin{figure}[htbp]
\centering
\vspace{1mm}
\includegraphics[scale=.47]{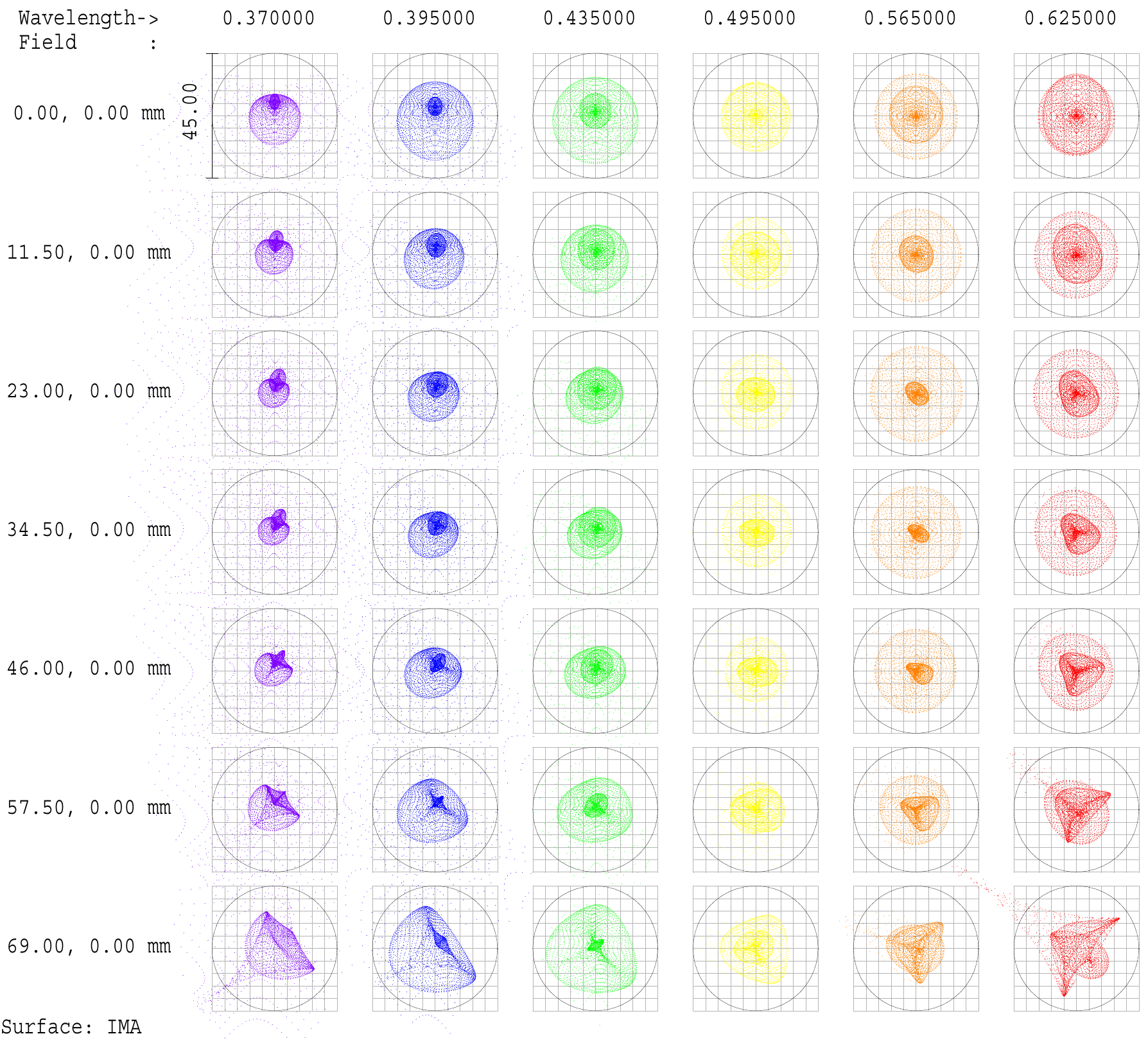}
\caption{{\bf Spot diagram for the BOSS blue channel. The average RMS diameter is 19.6 $\mu$m and the maximum RMS diameter is 30.9 $\mu$m.}}
\label{blue_spots_BOSS}
\end{figure}

Spot diagrams for the red and blue channels are shown in Figure~\ref{red_spots_BOSS} and Figure~\ref{blue_spots_BOSS}, respectively. 
The spots are shown within a 45 $\mu$m diameter circle, representing the
diameter ($2\2pr$) of the imaged BOSS fiber on the detector.  Each diagram covers
the full respective bandpass of the channel, and field points cover the
full length of the slithead.  The average RMS spot diameter for the red
channel is 18.3 $\mu$m and the maximum RMS diameter is 27.5 $\mu$m.  For the
blue channel the average RMS diameter is 19.6 $\mu$m and the maximum RMS
diameter is 30.9 $\mu$m.  Compared to the SDSS image quality, the BOSS
design is slightly better on average, and this is over an expanded bandpass and larger field of view (taller slit).  However, the fiber diameter is smaller for BOSS as are the detector pixels.  Nonetheless, when the
geometric aberrations are convolved with the fiber diameter the resulting
FWHM is still very near 45 $\mu$m, which matches three rows of pixels on the
detector.

\paragraph{Spectral Resolving Power}

\begin{figure}[htbp]
\begin{center}
\epsscale{1.18}
\plotone{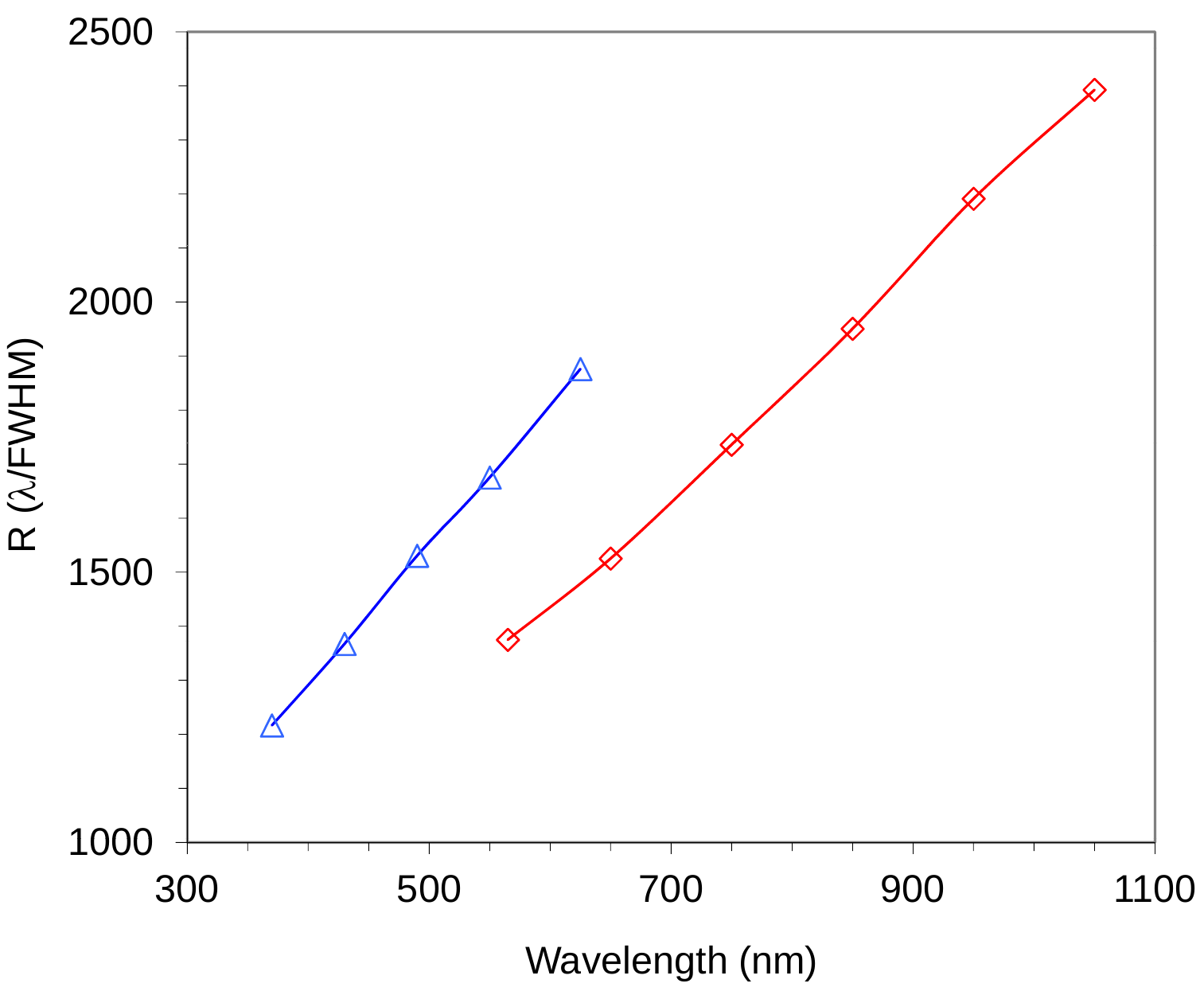}
\caption{{\bf Predicted resolving power, R,
as a function of wavelength for the BOSS spectrographs. As with the SDSS design, predictions are derived from the FWHM of simulated, uncollapsed images computed using Zemax. In reality, the resolving power of the instrument is determined using collapsed
spectra, so the prediction here is on the low side. However,
the Zemax analysis assumes perfect optics and alignment. For this
reason the predictions have been based on un-collapsed spectra, which
is the conservative approach. The expectation is that the actual resolving power
will be somewhat better than shown here.}}
\label{BOSS_blue_resolving_power}
\end{center}
\end{figure}

Spectral resolving power predictions for the BOSS spectrograph are shown
in Figure~\ref{BOSS_blue_resolving_power}. These predictions were derived in the same manner as
for the SDSS spectrograph resolving power discussed in Section \ref{sec:SDSSoptic}. Comparing
the BOSS and SDSS predictions, the blue channel is slightly lower than the SDSS design, while
in the red channel the resolving power is slightly higher. The improved resolving power
in the red channel is the result of a larger detector with more pixels, more than offsetting the expanded wavelength coverage. In the blue channel the grating dispersion was deliberately chosen to underfill the detector in the spectral direction, thus maximizing the signal-to-noise ratio while providing sufficient resolving power to carry out the desired science.

\begin{figure}[htbp]
\centering
\includegraphics[scale=0.250]{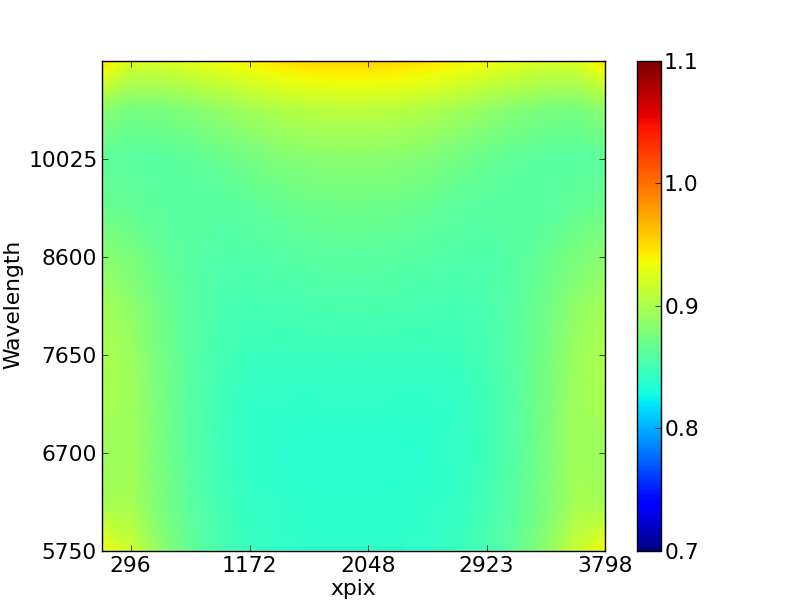}
\includegraphics[scale=0.250]{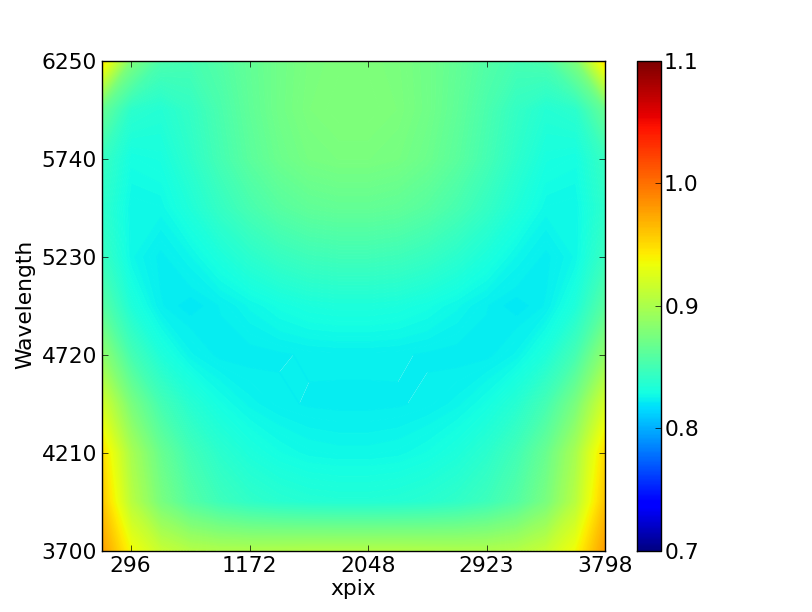}
\caption{{\bf Contour image
 of the BOSS CCDs showing the RMS width $\sigma_p$ of the line-spread function in native pixels 
to demonstrate the predicted variation in resolution over the focal plane.  The FWHM = $2.35 \times \sigma_p$; red CCD (top) and the blue CCD (bottom).  The wavelength decreases from top to bottom.}}
\label{fig:BOSS-spot-diagram}
\hspace{0.25cm}
\end{figure}

We also present a two-dimensional contour image,
Figure~\ref{fig:BOSS-spot-diagram}, of the BOSS CCDs showing the RMS width $\sigma_p$ of the line-spread function in native pixels 
to demonstrate the predicted variation in resolution over the focal plane.  The FWHM = $2.35 \times \sigma_p$.  Note that the linewidths are moderately degraded at the shortest wavelengths of the blue channel near the ends of the slit.

\paragraph{Throughput}

\begin{figure*}[htbp]
\begin{center}
\epsscale{0.8}
\plotone{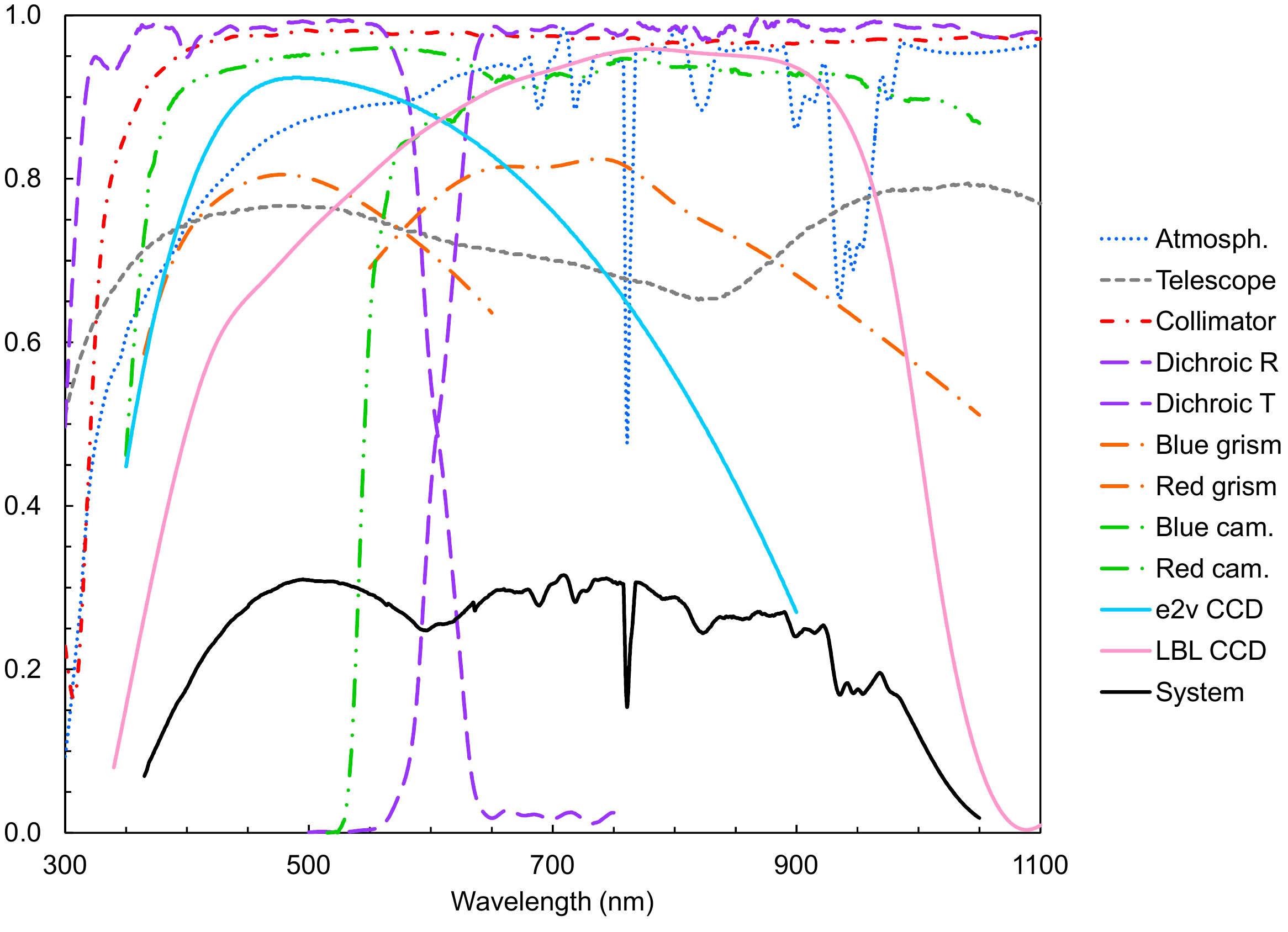}
\caption{{\bf Expected throughput for the BOSS spectrographs.  The plot shows all contributors to the throughput model having a wavelength dependence.  Not shown are those contributors with an  essentially flat response across the bandpass: average fiber transmission including Fresnel losses at the two faces (0.85), focal ratio degradation overfilling the collimator stop (0.96), and ``slit'' losses for 1$\2pr$ FWHM seeing conditions modeled with a double Gaussian PSF (0.83).  Not included in the model are losses due to centering and guiding errors.  Overall system throughput, shown by the solid black curve, is expected to peak at about 31\% in both the blue and red channels, a significant gain over the original SDSS spectrographs.}}
\label{BOSS_throughput}
\end{center}
\end{figure*}

As was the case for the SDSS spectrographs, the total throughput, on-sky, for the BOSS spectrographs was predicted from an end-to-end component model as a function of wavelength.  Included in the model were: atmospheric extinction, seeing (slit) losses, telescope, fibers, collimator,
dichroic, grism, camera, and CCD quantum efficiency (QE).  Atmospheric extinction was modeled at one airmass as before. Seeing losses were modeled using a double Gaussian PSF with a FWHM of 1$\2pr$ centered in a 2$\2pr$ aperture, resulting in a 17.5\% throughput loss assumed flat across the bandpass. The telescope efficiency is based on measured data for CO$_2$-cleaned bare aluminum mirrors (Wilson,RTO II), along with simulated anti-reflection coating curves to match the specifications of the two wide field corrector lenses (a small overall effect).  Measured curves were used for the dichroic, grism, camera coatings, and CCD QE.  The manufacturer's curve
for Denton FSS99 silver was used for the collimator coating.
The new silvered coating applied
to the collimators in 2012 by Infinite Optics, led to a system throughput somewhat better than what is
described in this paper, particularly at short wavelengths.
Internal transmission curves for the camera glasses were obtained from the manufacturer's data sheets.  The fiber efficiency (85\%) is based on lab measurements, representing an average value for the fibers measured and assumed flat across the bandpass.  An additional 3.6\% loss is included for focal ratio degradation beyond the f/4 beam the collimator was designed to accept.  Figure~\ref{BOSS_throughput} shows the individual component efficiencies used for this model along with the total expected system throughput.  Peak efficiencies in the blue and red channels are similar at about 31\%.  This represents a major gain relative to the SDSS spectrographs, where a similar component model resulted in expected peak efficiencies of 17\% and 22\% for the blue and red channels, respectively.

\subsection{Mechanical Upgrades}
\label{sec:BOSSmech}

\subsubsection{Overview}

\begin{figure*}[htbp]
\begin{center}
\epsscale{0.8}
\plotone{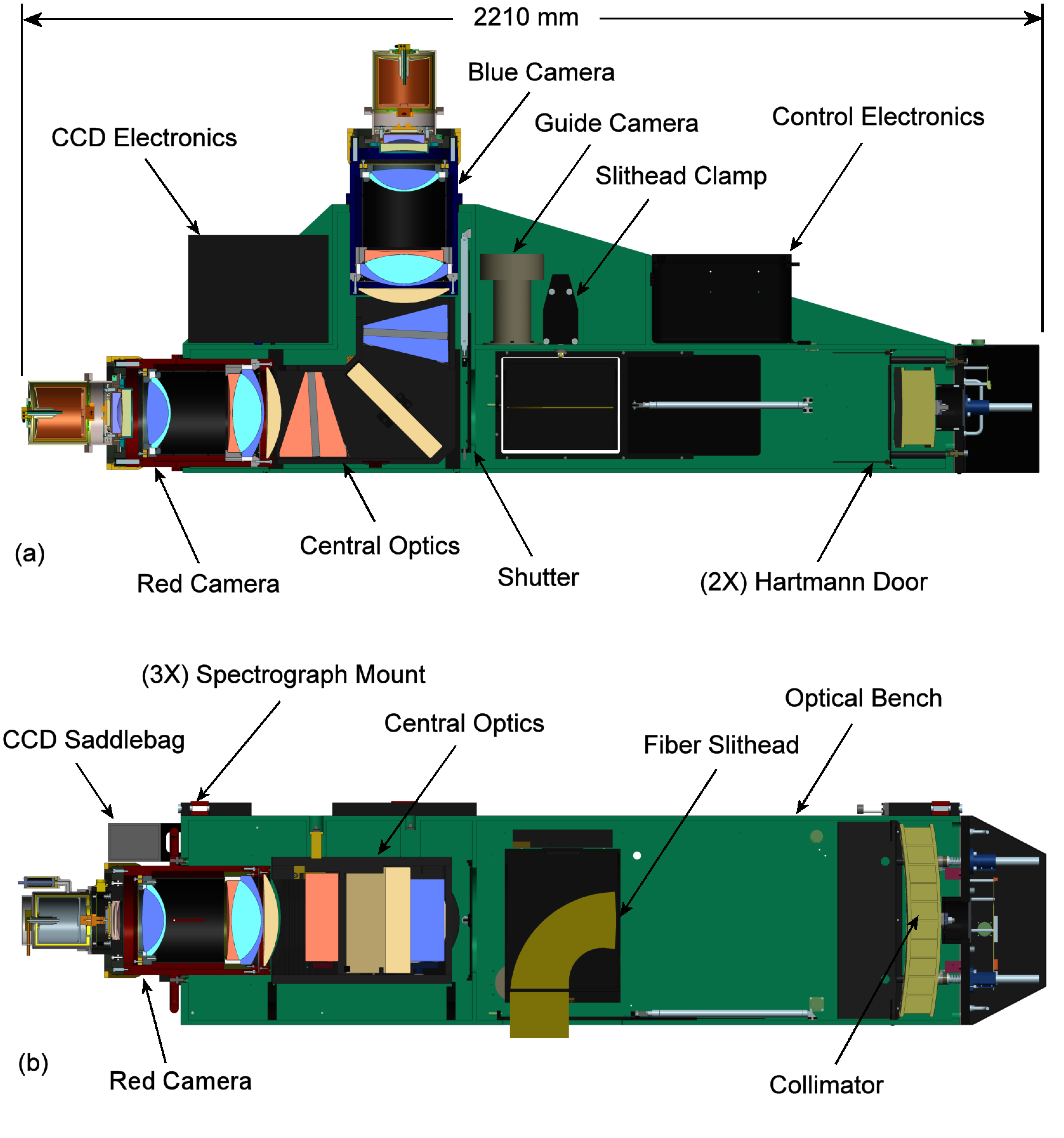}
\caption{{\bf Cross sections showing the BOSS spectrograph mechanical layout. The layout is identical to the SDSS design, however a revised central optics assembly to accommodate the VPH grisms is implemented here. In addition, there are subtle modifications to the lens group placements within the cameras, and a revised dewar design to accommodate new CCDs.}}
\label{BOSS_Spec_Section}
\end{center}
\end{figure*}

Figure~\ref{BOSS_Spec_Section} shows the mechanical layout for the BOSS spectrograph. The revised design is nearly identical to that for SDSS.
In fact, most of the hardware in the SDSS spectrograph is reused for BOSS. Upgrades include new fiber cartridges and slitheads, a new central optics assembly to house new beamsplitters and VPH grisms, modified lens mounts and hardware to accommodate the revised optical design, and redesigned dewars for the new CCDs. The fiber system upgrades were discussed in Section~\ref{sec:BOSSfiber}. In this section we describe the details of the remaining opto-mechanical upgrades.

\subsubsection{Collimator}

The collimator opto-mechanics are the same for BOSS as they were for SDSS.
However, a subtle, yet important, modification was made by inserting a small mask in front of the collimator.
A 3/16 inch wide vertical strip was placed down the middle of the collimator, aligned parallel to the slithead,
to prevent reflections off the tips of the fibers and back into the optical system.

\subsubsection{Central Optics Assembly}

As described in Section~\ref{sec:BOSSoptic}, the right angle grisms used in SDSS were
replaced with VPH grisms for BOSS. The new grisms consist of a plane,
parallel, VPH grating sandwiched between two prisms. Given the new
grism form factor, the central optics assembly was redesigned. The revised configuration can be seen in
Figure~\ref{BOSS_Spec_Section}. The new design utilizes the same
basic mounting schemes as its predecessor. Each grism is kinematically
located by six reference points and registered against those points
using spring plungers. Retention of the dichroic is identical to the
SDSS design with simple spring plungers to preload the optic against
six reference points. As in the original design, all but one of the
eighteen locating references (six for each of the three optics) are
machined into the base of the assembly, which ensures accurate placement.
A mask was added in this design just in front
of each grism to prevent undispersed light from sneaking past the grism; something that was discovered during commissioning. Overall the assembly is slightly larger and more
massive than the SDSS design. The mass of the BOSS version is 44 kg
as compared to 39 kg for the SDSS configuration. Fortunately the opening
in the side of the optical bench was large enough to accept the larger-sized assembly. However, a minor adjustment to the guide rails inside
the bench was required to allow for a slight shift in the location of the assembly relative to the opening in the bench.

\subsubsection{Camera Opto-Mechanics}
\label{sec:BOSS_Cameras}

\begin{figure}[htbp]
\begin{center}
\epsscale{1.18}
\plotone{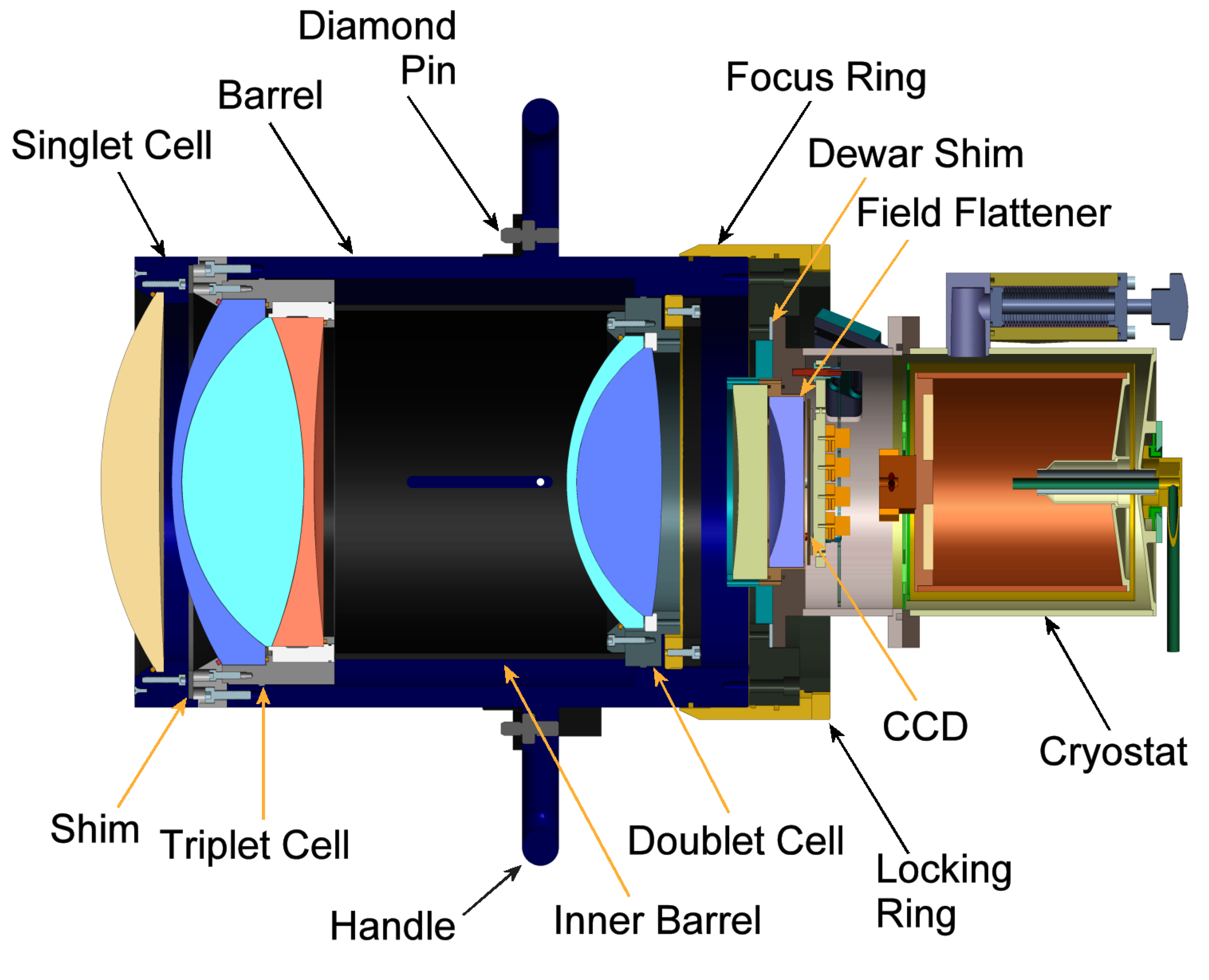}
\caption{{\bf  Cross section of the BOSS blue camera.  As with the original SDSS design, the red camera opto-mechanical design is identical except for subtle differences in the location of the lens groups.}}
\label{BOSS_Blue_Camera}
\end{center}
\end{figure}

Exclusive of the dewars, which are discussed below, only minor
opto-mechanical adjustments to the cameras were required to accommodate
the BOSS optical design.  The BOSS cameras (both red and blue) have new field-flattening lenses and revised singlet-to-triplet
and triplet-to-doublet spacings.  To achieve the increased spacing
between the singlet and the triplet the singlet cells were remade, the
new cells being thicker than the original.  The triplet-to-doublet
spacings were adjusted by fabricating new inner barrels with the appropriate length. 
Figure~\ref{BOSS_Blue_Camera} shows a cross-section of the BOSS blue
camera. 

While the SDSS cameras performed well for many years, the grease
coupling layers in the triplet degraded slowly over time, the result
being a random web-like pattern of voids, which reduced transmission and increased scattered light.  Considerable effort was invested in resolving this
problem.  Tests using the grease originally used, Dow Corning Q2-3067,
showed that these voids begin to appear after just one thermal cycle. 
Hence re-coupling the lenses with fresh grease as a routine maintenance
effort was not a viable solution.  After much trial and error using
alternative optical couplants such as Sylgard 184,\footnote{Produced by Dow Corning, http://www.dowcorning.com} and an optical gel,
Lightspan LS6943 from NuSil,\footnote{NuSil Technology, http://www.nusil.com} both of which failed during thermal
testing, it was determined that optical coupling fluid (i.e. an optically transparent oil) was the only reliable
solution. 
 
\begin{figure}[htbp]
\begin{center}
\epsscale{1.15}
\plotone{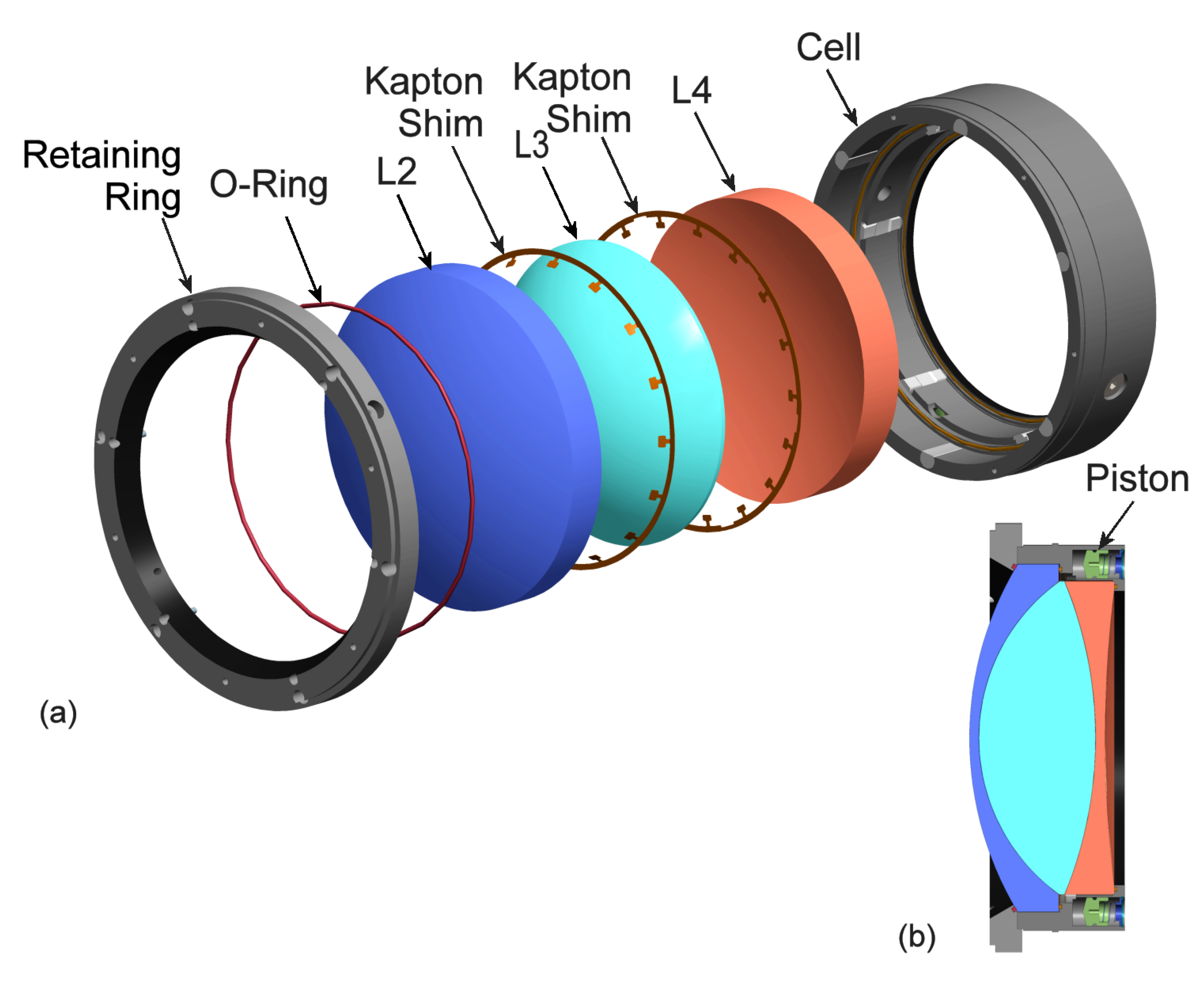}
\caption{{\bf (a) Exploded view of the BOSS camera triplet, and (b) a cross-section
of the BOSS triplet. In this revised design, the lenses are coupled by Cargille Laser Liquid 1074.  Lenses are spaced using 0.25 mm Kapton shims with fingers that extend radially inward near the lens outer diameters.  Gaps between the fingers allow fluid to flow into the gap between the lenses.  O-rings seal the liquid volume at the rear face of the last element and at the rear face of the front element.}}
\label{tripletcell}
\end{center}
\end{figure}

The triplet cell was redesigned with the lenses
being coupled by Cargille Laser Liquid 1074\footnote{Cargille Labs, http://www.cargille.com}.  The revised design is
shown in Figure~\ref{tripletcell}.  In the new design, the 0.25 mm gap
between the lenses is set by a kapton shim with fingers that extend
radially inward near the lens outer diameters.  The space between the
fingers allows the laser liquid to penetrate into the gap. Silicone O-rings seal
the liquid volume at the rear face of the last element and at the rear
face of the front element.  Two reservoirs with pistons were integrated
as well to accommodate differences in volumetric expansion between the
laser liquid and the space it occupies.  A drain and fill port, $180^\circ$
opposed, penetrate the outer diameter of the cell to facilitate
filling and draining the liquid. 

Some consideration was given to the choice of O-ring material given that the Robert Stobie Spectrograph (RSS) \citep{kobulnicky03a} suffered UV transmission loss when the coupling fluid became contaminated from interaction with the polyurethane material used for the expansion bladder in their liquid-coupled cell.  At the time the BOSS triplet cell was designed we were unable to conduct our own material compatibility testing, other than simple qualitative tests; the schedule did not permit it.  Instead we relied on data obtained from studies done for the Binospec instrument (Dan Fabricant, private communication), and from discussions with Cargille Labs.  Based on both sources of data, silicone was chosen for the large O-rings that seal against the front and rear elements, and Viton was used to seal the pistons and drain plugs. Silicone was used for the main seals because it was available in a lower durometer (i.e. a softer material).  Data published in 2010 on the RSS experience \citep{nordsieck10a} confirmed that both silicone and Viton have acceptably low reactivity with Laser Liquid, however, in general, silicone appears to be better for wavelengths $<$ 380 $n$m. 

In addition to the changes made to accommodate the revised optical design, the focus locking mechanism was also modified.  
It had been determined that while the locking mechanism was sufficient to lock the dewar to the focus ring, the focus ring itself was not properly locked to the barrel. As a result, the blue camera in Spectrograph 1 would suddenly drift out of focus within a predictable, narrow range of motion on the rotator; see Section~\ref{sec:SDSS_BOSS_Flexure} for a more detailed discussion of this issue.
To solve the problem three large locking screws were installed in the dewar flange, which press against the back face of the barrel.  Tightening the screws preloads the connection between the threaded focus ring and the barrel in a very positive way.  The screws are loosened when focus adjustments are made, and then tightened to secure the adjustment.  
Each screw is equipped with a large knob for ease of use.

\newpage

\subsubsection{Dewars}

Larger CCDs, larger field flatteners, and new readout electronics necessitated a redesign of the dewars for BOSS. The new design borrows considerably from SDSS with the front half of the dewar containing the field flatteners, the detector, and the pre-amp board, and the rear half of the dewar containing the  LN$_{2}$ reservoir. However, unlike the SDSS configuration, the red and blue detector packages are physically quite different, both from each other and from the original CCDs. The blue channel CCD has a thick silicon carbide package while the fully depleted device used in the red channel is mounted to an invar substrate. This led to a redesign of the detector mounting and adjustment scheme, the red and blue designs being unique.

The design of the  LN$_{2}$ reservoir is nearly identical to the SDSS design. The dewar volume is increased slightly with the usable volume being 0.4 liters. The heat load is approximately 4.5 W, slightly larger than in the original design. With the increased reservoir volume, the hold time is approximately four hours.  The dewar is filled automatically at 2 hour intervals.

\subsubsection{Dewar Autofill System}
\label{sec:BOSSAutofill}

BOSS uses the same autofill system implemented for SDSS, with one minor exception.  About a year after the start of BOSS, once the imaging camera was decommissioned, the 180-liter tank used to cool the imager was re-purposed for cooling one of the spectrographs.  Hence, the current autofill configuration uses two 180-liter supply dewars, one dedicated to each spectrograph.

\subsection{Flexure}
\label{sec:BOSS_Flexure}

Given the measured flexure for SDSS 
(see Section~\ref{sec:SDSS_BOSS_Flexure}), smaller fibers, 
and smaller CCD pixels, the flexure expected for BOSS is 0.5 pixels 
for $15^\circ$ on the sky.  This exceeds the 0.3 pixel goal 
set for SDSS by a modest amount.  However, 
the acceptable tolerances (spectral and spatial) for mechanical flexure
are, in practical terms, determined by the extent to which
they can be compensated for in the data reduction.

Based on the method developed for SDSS (see Section~\ref{sec:SDSS_BOSS_Pipeline}), we considered
0.5 pixels of flexure to be acceptable. 
If we consider a maximal 15$^{\circ}$ rotation per hour,
then we have a fiducial shift of 0.13 pixels in the spectral direction
and 0.13 pixels in the spatial direction
over the course of a single 15-minute exposure.
The magnitude of the blurring introduced
by this flexure within an exposure
will be given by the total shift scaled by $1/\sqrt{12}$, which
corresponds to the RMS of a tophat function.
When added in quadrature to the $\sim$~1-pixel dispersion
of the spectrograph PSF, this is
a fractional effect of $7 \times 10^{-4}$,
an insignificant effect on the resolution of the spectra.
It also does not degrade the quality of
near-infrared sky subtraction,
since the sky-modeling algorithm does not rely on any
calibrated estimate of the spectral resolution for sky lines,
but rather implicitly fits for the
spectral resolution as a low-order
function of fiber number simultaneously with the fit
for the sky spectrum in each exposure.
In the spatial
direction, this introduces only a negligible mismatch with the
cross-sectional profile measured from flat-field exposures
that is used to extract the science spectra.

\subsection{Detectors, Electronics and Data Acquisition Upgrades}
\label{sec:BOSSelex}

\subsubsection{Detectors}
Modern CCDs with higher quantum efficiency and smaller pixels were installed for BOSS.  
On the blue side of the spectrograph, e2v\footnote{e2v Technologies, http://www.e2v.com} CCDs were installed for improved quantum efficiency
in the far blue while fully depleted devices from Lawrence Berkeley National Laboratory (LBNL)
were installed for improved quantum efficiency near 10,000 \AA\ \citep{holland06a}.
With some improvements in the optics as described previously, the 15 $\mu$m pixel size provides
optimal 3-pixel sampling of the PSF and the larger format allows for
larger gaps between v-groove blocks to sample the wings of the PSF as well as additional
pixels in the dispersion direction to accommodate the larger range in wavelength.
Equally important, as evidenced in a comparison of Figure~\ref{SDSS_throughput}
to Figure~\ref{BOSS_throughput},
the upgrade to the new devices provides substantially improved quantum efficiency in
the regions 3560 \AA $< \lambda < $5000 \AA\ and 7000 \AA $< \lambda < $10,400 \AA.

\subsubsection{Readout Electronics} The readout electronics are almost
unchanged from the earlier system, with some additions and modifications
required for driving the p-channel LBNL CCDs and handling the larger
format of both new detectors.  A new dewar board incorporating the
preamp (now four channels, one for each quadrant of the new detectors) and
clock drivers was designed and built, one version each for the n-channel
blue CCDs and the p-channel red CCDs.  The Burr-Brown OPA627s were replaced with
Analog Devices AD8610s,\footnote{Analog Devices, http://www.analog.com} which are slightly noisier but faster and have
lower power dissipation; the gain of the new devices is so much higher
than that of the earlier CCDs that the slightly higher noise is of no
consequence, and the signal chain adds less than a third of an electron
in quadrature to the system noise.  New and more modern analog switches
were incorporated as well, partly to better incorporate the new faster
readout (88 kilopixels per second), which is twice as fast as the one used previously.
Some of the SDSS camera devices had only one
good amplifier and had to be read at this rate, so the signal chain was
designed to handle it from the beginning, but better and faster switches
make it work better.  With the 88 kHz rate, the larger format devices with four
channels are read in the same time as the earlier SDSS devices with two.  

The dewars are equipped with two hermetic micro-D connectors, one
with 51 pins for DC and video, and another with 37 pins for digital
clock signals.  Cables lead from connectors to saddlebags similar to the
originals, but with more boards.  The original bus receivers,
power-distribution, and signal-chain/bias boards are used with a few
gain-setting component changes, but now two sig/bias boards are used,
since each was built to handle only two outputs. For the ``upside down''
LBNL p-channel
device, a new board which inverts the bias voltages was designed and
built. This board also handles the control of the large substrate bias
voltage which places a large DC potential across the device to minimize
charge diffusion.

The data and control between the saddlebags and the controller are handled
on two cables, a standard 68-pin SCSI cable carrying the RS485 clock signals
and a few slow CMOS signals, and a 50-pin MDM cable carrying the digitized
video, ADC clocks, and the new voltages.

The controllers are substantially modified. A new power supply to 
generate the large negative voltages (-30 V) for VDD and VRD and the very high
positive substrate bias (+100V) for the LBNL devices was added. The
single-micro topology was changed to the camera's two-micro architecture
to accommodate the more complex operating environment using two rather
different kinds of devices. The clocks are generated synchronously for
both red and blue CCDs, but the two sets of clocks themselves are completely
different for the two CCD types. The architectures of the detectors are
different in detail but also grossly--the e2v blue devices are 
4-phase in the parallels, but the LBNL devices are three-phase, and the
way charge is passed to the serial register is rather different.

The FOXIs have been replaced by a fast ethernet-interfaced micro, the
Netburner, described in the next subsection.

\subsubsection{Data Acquisition}
The data acquisition (DAQ) system is responsible for data transfer from the spectrograph to the instrument control computer (ICC), located approximately 300 feet from the telescope. The DAQ receives pixel data from the spectrograph electronics, performs some processing including simple data integrity tests, formats the data according to the FITS standard, and saves the images on a disk array.  These operations have to be performed in real time to maximize throughput and overall survey efficiency.  The architecture of the BOSS DAQ is similar to the proven design of the original SDSS DAQ, described in Section~\ref{sec:SDSSelex}.  The main change was the use of standard commodity computing components to replace the original, aging hardware modules, many of them custom-built and no longer supported. An additional benefit was the replacement of licensed software packages with free, open-source products. 

\begin{figure}[htbp]
\begin{center}
\epsscale{1.17}
\plotone{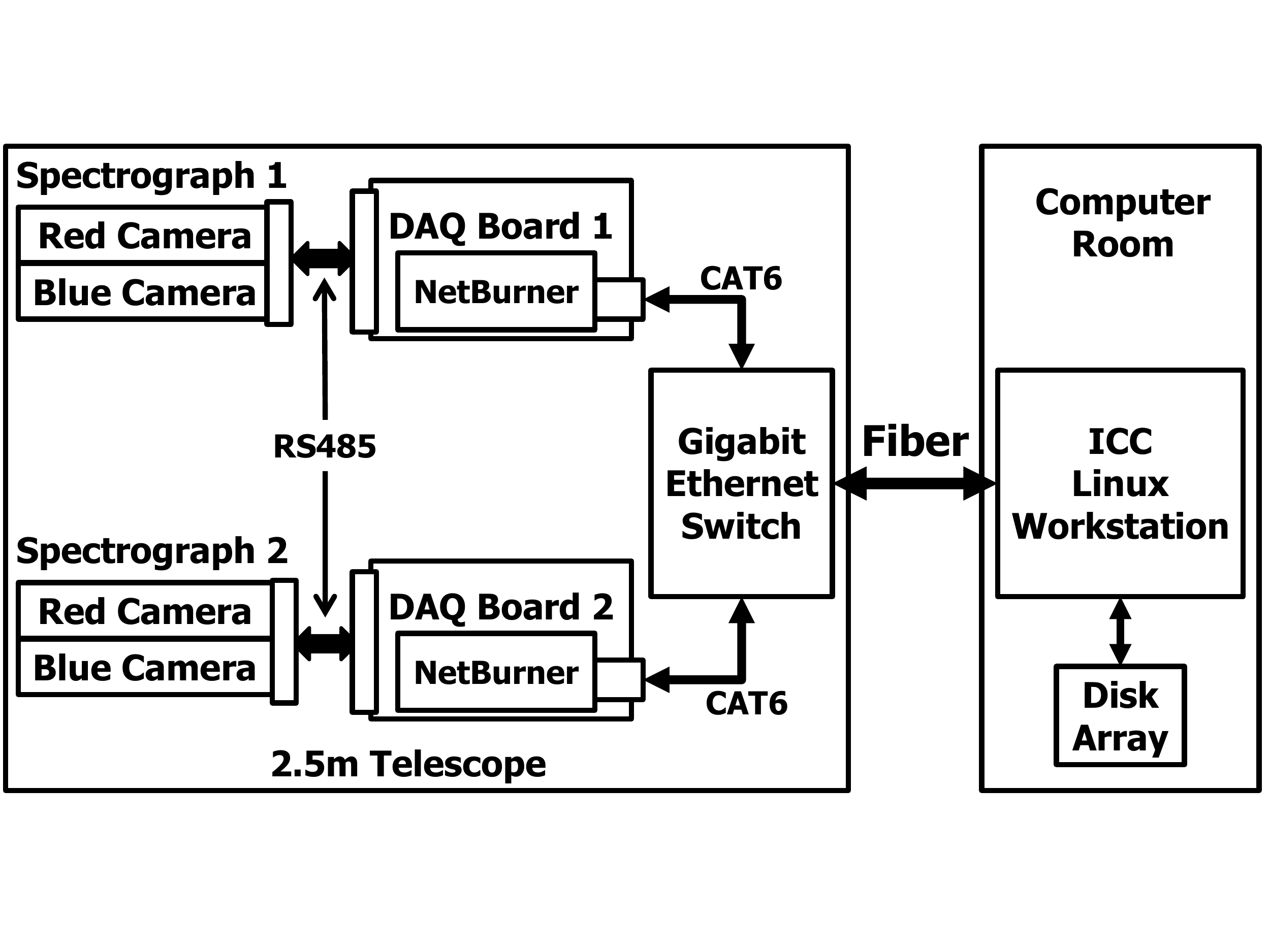}
\caption{{\bf  Block diagram of the upgraded BOSS DAQ system.}}
\label{BOSSDAQtopblock}
\end{center}
\end{figure}

A block diagram of the BOSS DAQ system is shown in Figure~\ref{BOSSDAQtopblock}. Separate DAQ modules connect to the digitizer boards of the two cameras in each of the spectrographs. Copper Ethernet cables connect the DAQ modules  to a network switch mounted on the telescope, where the data is converted to optical signals and travels over optical fibers to the computer housed in a separate building.

\begin{figure}[htbp]
\begin{center}
\epsscale{1.2}
\plotone{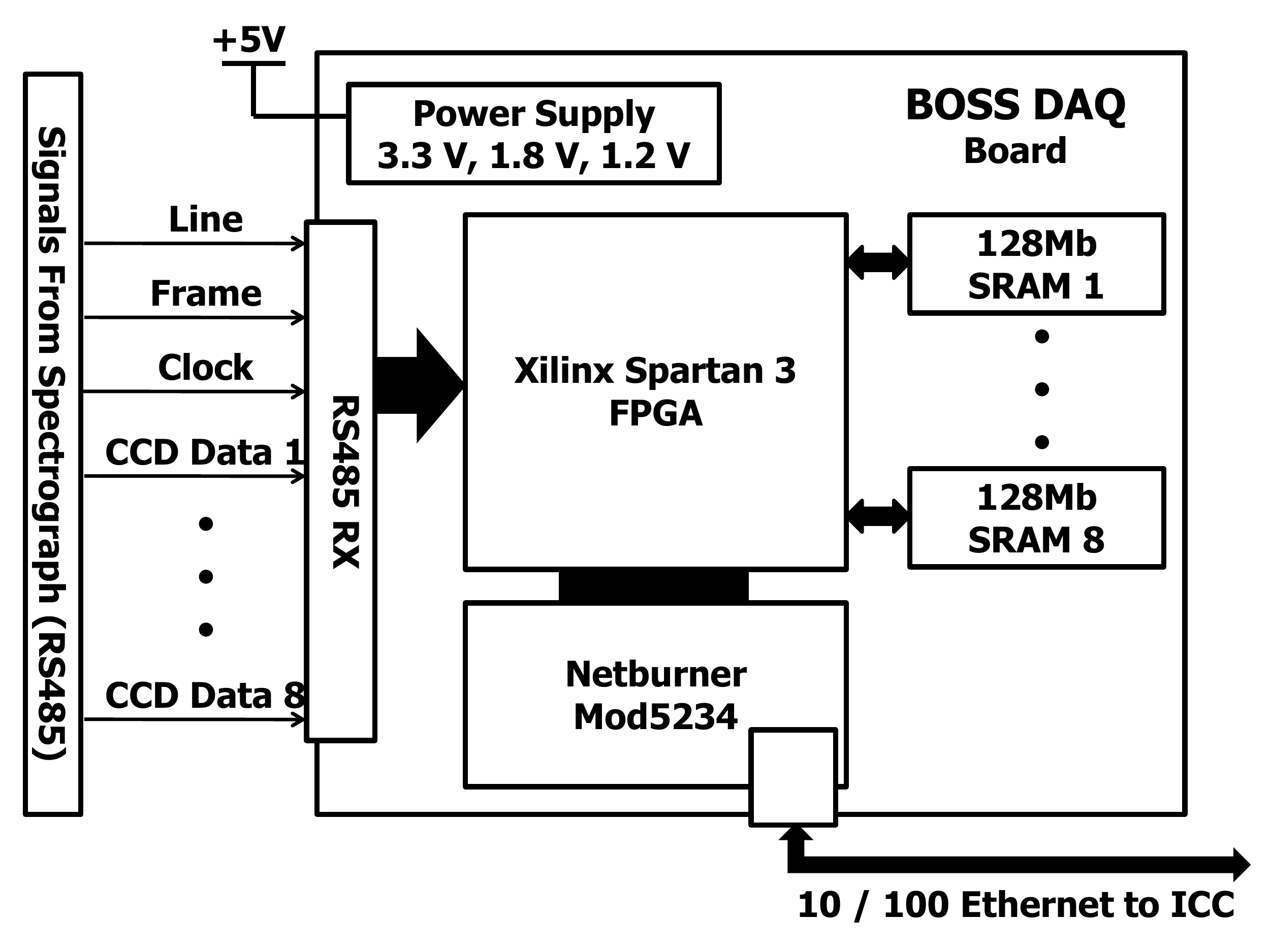}
\caption{{\bf  Block diagram of the BOSS DAQ card.}}
\label{BOSSDAQblock}
\end{center}
\end{figure}

\paragraph{DAQ Hardware}
To efficiently collect, package, monitor, and transmit data from the BOSS spectrographs, a custom printed circuit card was designed.  The card is called the DAQ board and a block diagram of its contents is shown in Figure~\ref{BOSSDAQblock}.  One DAQ board is required for each BOSS spectrograph. During normal operation, the DAQ board receives 8 synchronous streams of pixel data, coming from the two cameras on each spectrograph, each camera having one CCD with four outputs.  The incoming data, in RS-485 format, are stored in a frame buffer to decouple input and output operations, reformatted  and transferred  via a 100 Mbit Ethernet link to the ICC over optical fiber to the computer room. The input stage of the DAQ board operates at the pixel clock rate set by the front-end electronics. The buffer memory is of sufficient size to store an entire exposure, but under normal circumstances the network interface is fast enough to keep up with the input rate. In order to fit in the limited space available on the spectrograph, the DAQ card was designed to be compact, $65 \times 115$ mm, and to operate from a single 5 V supply. The design of the DAQ card is based on three main components; a Xilinx FPGA, a NetBurner Ethernet Core Module, and eight 8M$\times$16 pseudo-static ram chips (PSRAM).  An overview of these components and their functions is given below.  

The Xilinx FPGA, a Spartan XC3S1400AN,\footnote{Xilinx, Inc., http://www.xilinx.com} controls the input stage of the DAQ module and the frame buffer memory. The bus interface to the NetBurner module is also implemented in the FPGA. The pixel data streams are de-serialized and recorded in the frame buffer which is organized as a First-In/First-Out (FiFo) memory. The FPGA is configured to write to all eight memory modules simultaneously, thus maximizing the data recording speed.  At the FPGA level data integrity checks are performed by keeping pixel and line counts. The acquisition software running on the NetBurner module can access this information and, if necessary, raise appropriate flags in the ICC.   For testing and debugging purposes the FPGA also contains logic to emulate the spectrograph signals and load test images to the frame buffer, which in turn can be read and sent to the ICC.  With this feature, problems upstream from the DAQ card may be located and the ICC and analysis code can be debugged easily using well-defined images.  In addition, the FPGA firmware includes a watchdog timer that will raise a flag if the camera data is not loaded into the DAQ card within a reasonable time limit.

The NetBurner Ethernet Core Module, model 5234,\footnote{NetBurner, Inc., http://www.netburner.com} coordinates the communication between the DAQ card and the ICC.  The software for the NetBurner is written in C/C$^{++}$ and will be discussed in detail in the following section. The NetBurner contains a Freescale ColdFire 5234 microprocessor, 2 MB Flash memory, 8 MBytes of SDRAM, and 10/100 Ethernet interface. The ColdFire microprocessor has a full 32-bit architecture providing more than 140 million instructions per second  of processing power. The NetBurner module is interfaced to the FPGA via a 16-bit external data bus which provides a simple and fast communication link.  

The 8M$\times$16 pseudo-static DRAM ram chips used to buffer and store the spectrograph images are Micron MT45W8MW16BGX.\footnote{Micron Technology, Inc., http://www.micron.com}  Pseudo-static PSRAM is a memory technology that features an SRAM-like architecture using a DRAM cell with internal refresh operations that do not interfere with user read or write cycles.  This approach significantly reduces the complexity of the DAQ board while still allowing for sufficient memory to buffer an entire exposure. Should a problem arise with the communication link to the ICC, this feature allows the data to be resent repeatedly until it is received correctly. The PSRAM type chosen is limited to a read/write timing of 70 ns, but consumes little power; less than 50mA, during a read or write cycle.

\paragraph{DAQ Software}
The DAQ software consists of two major pieces. The acquisition part, written in C, runs on the NetBurner modules on each of the DAQ boards. The second software component, a Python module, provides image formatting as well as an interface to the BOSS online system. This software is executed on the instrument control computer in the computer room. The acquisition program takes advantage of the soft real time functionality provided by the NetBurner software tools. One task implements a command interface and support for the emulation mode used for hardware and software debugging. A second task is responsible for reading out the data  from the frame buffer on the DAQ module and for data transmission to the ICC using the TCP/IP protocol. Additional functionality provided by the application includes a separate reset connection to restart the software in case of problems, as well as telnet support to allow remote connections to the NetBurner low-level debugger. Furthermore, the NetBurner allows for software updates over TCP/IP, permitting remote modification and testing in the event that the acquisition software developer is not on site.

Both BOSS spectrographs operate independently at the DAQ level and are seen by the online system as independent devices. A Python module was developed to allow the online system to interact with the DAQ controllers. Responsibilities for the data transfer, connection management and data integrity checks, such as the correct number of pixels, lie with this module. The Python module also rearranges the pixel streams from the four amplifiers for each of the two CCDs, generating FITS formatted files with contiguous data sections and with the overscan regions moved to the edges of the array. Header keywords are generated and added, and for each camera a separate FITS file is written to disk. 
\paragraph{DAQ Performance}
In almost two years of survey operation the BOSS DAQ system has performed efficiently and reliably. The readout time has been measured to be 55.6 seconds. Only 1.6 seconds of this period is added by the DAQ system, while the bulk of the readout time is used for CCD digitization and is set by the clocking speed, which is optimized to minimize read noise. 

\subsection{BOSS Calibration}
\label{sec:BOSScal}

The calibration system described in Section~\ref{sec:SDSScal} was largely
preserved for calibration in BOSS with the exception of two minor changes.
The neon lamp was replaced by a carefully constructed neon-argon
tube\footnote{Custom made by Absolutely Neon, Inc., http://www.absolutelyneon.com} to provide argon lines farther to the red than the neon spectrum reaches.
Pixel-to-pixel variation was mapped using a ``lossy-fiber''
that was placed in the position of the usual fibers along the slithead of an engineering cartridge.
This fiber was acquired from MeshTel (model\# GF-400-1-SMA)\footnote{MeshTel INTELITE, Inc., http://www.meshtel.com} and is 
illuminated by a Deuterium Tungsten Halogen light source provided by Ocean Optics (model\# DT-Mini-2-GS).\footnote{Ocean Optics, Inc., http://www.oceanoptics.com}
It is designed with a diffusive cladding and transmittive coating so that it glows uniformly along its length.
While the usual flat fields are illuminated through discrete fibers,
the lossy-fiber spectra is aligned with the slit so that it produces a similar spectral dependence but with
nearly uniform illumination in the spatial direction.

\subsection{Guider Upgrade}
\label{sec:BOSSguider}

\subsubsection{Overview}

As mentioned in Section \ref{sec:SDSSguider}, 
the original Roper Scientific camera used for the
guider in SDSS-I and -II was unsatisfactory on several counts, and it was
decided at the outset of SDSS-III to replace it with a cooled camera with
a better detector. In addition, although the pointing of the telescope is
usually accurate enough that guide stars are found in two large ($11\2pr$)
guide fibers, the efficiency of the
survey would be enhanced if we had larger acquisition fibers, and we
implemented two of these per cartridge in SDSS-III.

With the original SDSS system, focus was achieved by
trial and error, adjusting focus of the telescope and tracking changes
in the FWHM of the stars in the guide fibers.
The focus system for the SDSS {\it imaging} camera worked quite well.
In this system, stars were simultaneously imaged onto one focal plane
that was set some distance behind the scan data and another set of
stars were imaged onto a focal plane
that was set some distance in front of the scan data.
From the images mapped in these three focal planes, it was possible to
derive a focus error signal, and it was decided to implement a similar
design in the new spectroscopic guide system for SDSS-III.
In a similar fashion to the imaging system,
a set of guide fibers was placed in the same focal plane
as the science fibers for the spectrograph, a second set was slightly displaced
outside of focus, and a third set inside.
In addition to the two large acquisition fibers, the number of guide fibers was increased
from 11 to 14 to account for any potential decrease in sensitivity in the unfocused fibers
to perform the other guider functions.

\subsubsection{Guider Camera}

The new guider camera is an Alta E47
from Apogee Imaging Systems.\footnote{Apogee Imaging Systems, http://www.ccd.com}
The camera employs a 100 Mb/s ethernet interface, uses a 
thermoelectrically cooled e2v CCD47-10 AIMO 1024 $\times$ 1024 back-illuminated 
CCD with 13.5 $\mu$m pixels, normally binned 2 $\times$ 2. This results
in 27 $\mu$m, $0.45\2pr$ pixels. The images are nearly always well-sampled
for the seeing conditions at APO. The camera is much more sensitive
and stable than the old one, and allows the use of much fainter guide stars.

The camera was supplied with a liquid cooling head, which
we replaced with a similar part made of copper.
The replacement is intended to alleviate prior
bad experiences in the glycol system with aluminum parts.
Otherwise, the camera is used as supplied. 
The reimaging system from the output fiber block to the CCD and the
broad visual (BG38) filter from the old system was retained for the new guider.

\subsubsection{Guide Fibers}

The new system uses 14, $8\2pr$ (500 micron) Sumitomo (model\# IGN-05/10) coherent fibers 
of the same kind used in the old system for guiding. Eight of these are
in ferrules which place them at the nominal focus of the plate, and 
three each are 200 $\mu$m inside and outside of nominal focus. The
focus offset adds $0.7\2pr$ FWHM in quadrature to the seeing of the images, which
is not yet enough to make them unsatisfactory for guiding, but still
provides a sensitive measure of focus.

In addition to the 14 guide fibers, there are two much larger $24\2pr$ (1.5 mm)
acquisition fibers, which are inexpensive coherent plastic fiber bundles manufactured by NanOptics (model\# B29060J-175),\footnote{NanOptics, Inc., http://www.nanoptics.com}
used for guide star acquisition only.
Though not used for guiding, their quality seems to be
almost as good as the glass guide fibers.

\subsubsection{Software, Control, and Performance}

The software for the new control system is an adaptation of
the old guider code, written mostly in C.  There are a few enhancements over the
old system, including a fast acquisition mode that uses the acquisition
fibers only, a mode in which
successive guider images can be accumulated to improve
stability when using bright guide stars and therefore short exposures,
and a slow servo loop to actively and automatically control the
scale of the telescope (in the SDSS-I/II surveys the scale of the telescope was adjusted manually).
The upgraded BOSS guider automatically tracks in RA, Dec, rotation, and scale.
All of the fits used to generate
guider offsets are simple least-square deviation fits; it is possible
that we could do
better if a refraction model using the actual distribution of the
guide stars on the plate were used, but this has never been implemented.
The present system is satisfactory if the guide stars are reasonably
uniformly distributed over the plate; we strive
to achieve this distribution in the process of designing each plate.

The display code was completely rewritten for the new system but is
functionally identical to the old one.
A `raw' CCD image is displayed to the observers as is an
image in which the fiber images are distributed approximately as they are on the sky.
This display makes qualitative evaluation of the guider behavior
easy. In addition, the guider errors are displayed for each fiber for
each image and are recorded in the guider frame headers.

\section{SDSS and BOSS Spectrograph Performance}
\label{sec:Perf}

Here we describe the methods for determining the instrument performance for SDSS and BOSS
in relation to the predictions outlined earlier in the paper.
A summary of the design predictions and measured performance is presented in Table~\ref{table:reqperf}.
Instrument characterization is presented in three parts:
the spectroscopic pipeline, the spectrograph performance, and CCD performance.  
Data found in DR8 \citep{aihara11a} are used in the characterization of the SDSS spectrograph and
data in DR9 \citep{ahn12a} are used in the characterization of BOSS.

\begin{table*}
\begin{center}
\caption{Design Predictions and Measured Performance for SDSS and BOSS}
\label{table:reqperf}
\begin{tabular}{lcccc}
  \hline
 & \multicolumn{2}{c}  {(SDSS)} & \multicolumn{2}{c} {(BOSS)} \\
Quantity   & Design  & Measured  & Design  & Measured  \\
\hline
\multicolumn{5}{l} {Spectrograph 1} \\
\hline
Wavelength Range & 3900--9100  &  3800--9220 \AA\ & 3560--10,400 & 3560--10,400 \AA\  \\
Resolving Power ($\lambda<3800$ \AA) & N/A & N/A & 1000 & 1220--1480\\
Resolving Power ($3800<\lambda<4900$ \AA) & 1500 & 1500--1900 & 1400 & 1200--1900\\
Resolving Power ($\lambda>4900$ \AA) & 1500 & 1900--2600 & 1000 & 1550--2550\\
Total Throughput (4000 \AA) & 0.11 & 0.08 & 0.18 & 0.15 \\
Total Throughput (6000 \AA) & 0.20 & 0.17 &  0.25 & 0.21 \\
Total Throughput (8000 \AA) & 0.14  & 0.13 & 0.29 & 0.25 \\
Total Throughput (10,000 \AA) & N/A & N/A & 0.12 & 0.10 \\
Flexure: Blue Channel, Spectral (pixels per $15^\circ$ on the sky) & 0.3  & 0.33  & 0.5 & 0.40 \\
Flexure: Blue Channel, Spatial (pixels per $15^\circ$ on the sky)  & 0.3 & 0.18 & 0.5 & 0.11  \\
Flexure: Red Channel, Spectral (pixels per $15^\circ$ on the sky) & 0.3 & 0.30 & 0.5 & 0.41  \\
Flexure: Red Channel, Spatial (pixels per $15^\circ$ on the sky) & 0.3  & 0.21 & 0.5 & 0.18  \\
\hline
\multicolumn{5}{l} {Spectrograph 2} \\
\hline
Wavelength Range & 3900--9100 &  3800--9220 \AA\ & 3560--10,400 & 3560--10,400 \AA\  \\
Resolving Power ($\lambda<3800$ \AA) & N/A & N/A & 1000 & 1300\\
Resolving Power ($3800<\lambda<4900$ \AA) & 1000 & 1700--2200 & 1400 & 1350--1950\\
Resolving Power ($\lambda>4900$ \AA) & 1000 & 1850--2650 & 1000 & 1700--2600\\
Total Throughput (4000 \AA) & 0.11 & 0.09 &  0.18 & 0.14 \\
Total Throughput (6000 \AA) &  0.20 & 0.18 &0.25  & 0.24 \\
Total Throughput (8000 \AA) & 0.14 & 0.12 & 0.29 & 0.25 \\
Total Throughput (10,000 \AA) & N/A & N/A & 0.12 & 0.10 \\
Flexure: Blue Channel, Spectral (pixels per $15^\circ$ on the sky) & 0.3  & 0.30  & 0.5 & 0.43  \\
Flexure: Blue Channel, Spatial (pixels per $15^\circ$ on the sky)  & 0.3 & 0.31 & 0.5 & 0.07  \\
Flexure: Red Channel, Spectral (pixels per $15^\circ$ on the sky) & 0.3 & 0.27 & 0.5 & 0.44  \\
Flexure: Red Channel, Spatial (pixels per $15^\circ$ on the sky) & 0.3  & 0.25 & 0.5 & 0.15  \\
\end{tabular}
\end{center}
\end{table*}

\subsection{Spectroscopic Pipeline}
\label{sec:SDSS_BOSS_Pipeline}

The spectroscopic redshifts and classification for SDSS and BOSS are based on the
data reduction pipeline {\bf idlspec2d}.\footnote{Current and development
versions of the software are found at http://www.sdss3.org/dr9/software/products.php}
A brief explanation of the routines for SDSS is
found in \citet{adelman-mccarthy08a} and \citet{aihara11a}.  
SDSS data were processed using idlspec2d v5\_3\_12, which is the same version of the software
used in DR8.
The pipeline was modified
for BOSS to account for the increase in wavelength coverage, number of fibers, and other changes
discussed in Section \ref{sec:Upgrade}.
For a brief description of the BOSS software reductions, see \citet{dawson13a} and
for a full description of the automated classification, see \citet{bolton12a}.
BOSS data were processed using v5\_4\_31, a somewhat earlier
version of the v5\_4\_45 software used for DR9 \citep{ahn12a}.
There are minor changes between these two versions, but the differences do not
change the conclusions of the work presented here.
The intermediate data products produced by the idlspec2d pipeline are used to characterize
the instrument performance; we include a brief explanation of the relevant functions here.

SDSS and BOSS observations of each plate follow the sequence:
mount cartridge, slew to field, focus spectrograph, take exposure with calibration arc lamps
and flat field lamps, and obtain a series of 15 minute science exposures.
In the first step of the data reduction, the read noise and bias level from the raw images is
determined from the overscan of the image.
Next, raw images are bias-subtracted and counts are converted into electrons using CCD amplifier
gains specific to each quadrant of each detector.
Cosmic rays are then identified using a filter to discriminate between the sharp edges
of cosmic ray trails and the smoother profile of optical sources.
The noise of each pixel is determined by taking the Poisson noise from the electron counts
in quadrature with the measured readnoise.  Finally, a 
flat-field image is used to correct the pixel-to-pixel response in the CCD.

Each fiber projects a spectrum onto the CCD with an RMS in the spatial
direction of roughly one pixel and a spectral dispersion of roughly one pixel
RMS in the direction of the parallel clocking.
The data from the flat field exposures are used to model the profile of each fiber on the CCD and
also to normalize fiber-to-fiber throughput variations.
The data from the arc exposures are used to determine the wavelength solution for each fiber.

\begin{figure}[htbp]
\centering
\vspace{2mm}
\includegraphics[scale=0.55]{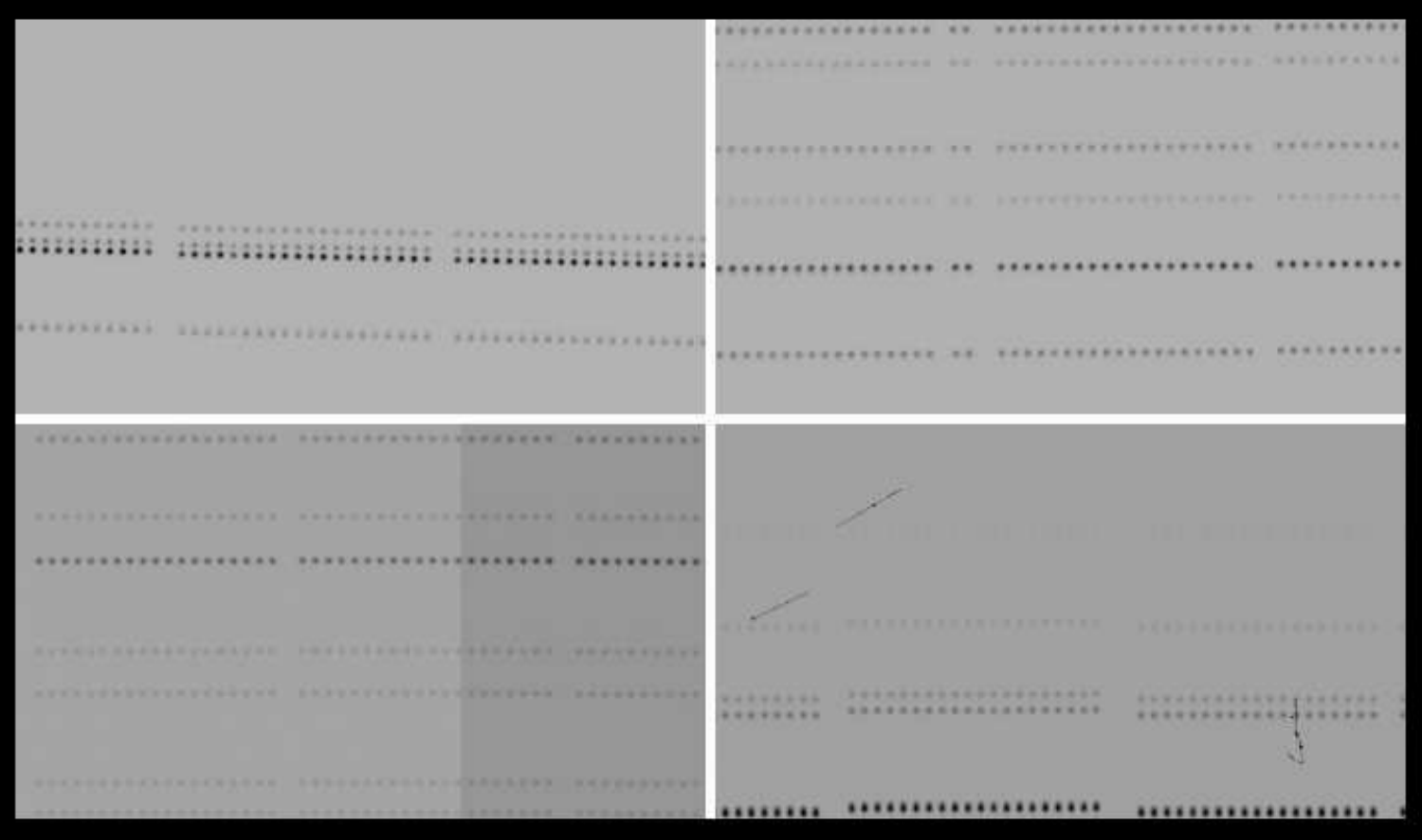}
\caption{{\bf Raw arc image taken with the BOSS spectrograph.
Top left panel:  3650 -- 3850 \AA\ on blue CCD;
Top right panel:  5050 -- 5250 \AA\ on blue CCD;
Bottom left panel:  6000 -- 6250 \AA\ on red CCD;
Bottom right panel:  8850 -- 9100 \AA\ on  red CCD.
Each panel covers roughly 50 spectra.}
}
\label{fig:BOSS-arc-image}
\end{figure}

Figure~\ref{fig:BOSS-arc-image} shows an example of the arc lines obtained in a BOSS exposure.
The top left panel is taken near the edge of the blue CCD
and shows a slight tilt in the wavelength direction due to optical distortion.
This image also demonstrates the bundling of 20 fibers separated by a gap on either side.
The top right panel demonstrates a dropped fiber near the right edge of the
first bundle.
The bottom left image from a red camera is taken near the center of the CCD and shows
the amplifier gain between two quadrants at a different background count.
The lower right panel shows several cosmic rays
interacting with the CCD on a red camera.
The bottom right image also shows an offset in the wavelength direction between the central bundle
and the surrounding bundles due to imperfections in the slithead assembly.
The top two images are more representative of the SDSS CCDs
because the red BOSS CCDs are much thicker and lead to more pronounced cosmic ray trails.
We use the wavelength solution and line spread function derived from the arc exposures
to determine wavelength coverage and resolution.

Each two dimensional science image is collapsed into a series of one-dimensional spectra
using the fitted profile and wavelength solution from the calibration exposures for each fiber.
At this point, the ``SKY'' fibers (fibers assigned to areas with no detected
objects) are used to model the background as a function of position on
the detector.  The model sky background is subtracted from the one-dimensional spectra.
These one-dimensional spectra contain all of the information required to characterize
the instrument performance.  For science applications, the
spectrum from each fiber is resampled to a pixel size of 69 km/s and combined with
the spectra from the same fiber for each exposure in the sequence to generate a high signal-to-noise spectrum.
Redshifts and object classification are determined from these co-added spectra as is described
in \citet{bolton12a}.

The data reduction pipeline also compensates for spectral shifts due to mechanical flexure.  
Shifts between exposures  in the spatial direction are
compensated by iteratively re-centering
the traces of all fibers starting
from the positions determined from flat-field exposures,
and adopting the median spatial shift as a global value
for each exposure.  For this process to converge properly,
the accumulated spatial shift between flat and
science frames must be significantly less than
one-half the typical fiber-to-fiber spatial separation,
a condition that is satisfied for any reasonable
number of science exposures in sequence.
Shifts between exposures in the spectral direction
are compensated by measuring the positions of
known bright sky emission lines and fitting a
shift-plus-scale for each fiber and each exposure
relative to the wavelength solution determined
from arc-lamp calibration frames.
The ultimate validation of the combined hardware--software
system for flexure control is the $\sim$~1\% level
of systematic residual errors in near-infrared sky
subtraction achieved for BOSS \citep{bolton12a}.

\subsection{Spectrograph Performance}
\subsubsection{Wavelength Coverage}

The SDSS spectrographs have excellent wavelength calibration between
the Hg I arc line at 3901.87 \AA\ and the Ne I arc line at 9148.67 \AA.
The BOSS spectrographs have excellent wavelength calibration between
the Cd I arc line at 3610.51 \AA\ and the Hg arc line at 10140 \AA.
In between these detected arc lines for SDSS (3902 \AA\ -- 9149 \AA) and BOSS
(3611 \AA\ -- 10140 \AA), the wavelength solutions are typically accurate
to 3 km/s.  Beyond those wavelengths the solutions are extrapolated
and the accuracy has not been evaluated.

The scientific rationale behind the wavelength coverage
for the SDSS and BOSS surveys was described in Section \ref{sec:SDSSreq} and Section \ref{sec:BOSSreq}, respectively.
Including the extrapolation beyond the prominent arc lines,
the wavelength range included in data reductions for the SDSS spectrographs is 3800--9220 \AA.
This is slightly broader than the design wavelength range of 3900--9100 \AA.
The BOSS spectrographs have useful data in the 3560 to 10,400 \AA
wavelength range that are included in all data releases.
Although the detectors see light beyond this wavelength
range, it is discarded in the reductions since the throughput is very low and the longer wavelengths suffer from second-order light.

Both the SDSS and BOSS blue limits ensure the CaII H and K lines are detected at zero redshift.
The BOSS blue limit provides extended coverage of the Lyman-$\alpha$ forest for quasars at $z>2.2$.
The BOSS red limit measures the
Mg b 5175 \AA\ and the Na D 5893 \AA\ absorption features for galaxies
redshifted to $z < 0.72$.
The wavelength coverage for SDSS and BOSS leads to detection of at least 31 common
astrophysical emission lines, as detailed in Table 5 of \citet{bolton12a}.

\subsubsection{Spectral Resolution}

The spectral resolution is measured from calibration arc images taken before each set of science exposures.
The one-dimensional arc image is
first masked to include only pixels within 12 pixels of the center of each arc line on each fiber.
A Gaussian of width $\sigma_\lambda$ is fit to each spectral profile using the 25 unmasked pixels.
A fourth order Legendre polynomial model is fit to the derived $\sigma_\lambda$
as a function of wavelength to model the dispersion over the full wavelength range.
The resolving power for SDSS and BOSS is then formally defined as
$R=\frac{\lambda}{2.35 \times \sigma_\lambda}$. The resolution is
the FWHM of the Gaussian, $2.35 \times \sigma_\lambda$.

\begin{figure}[htbp]
\centering
\includegraphics[scale=0.53]{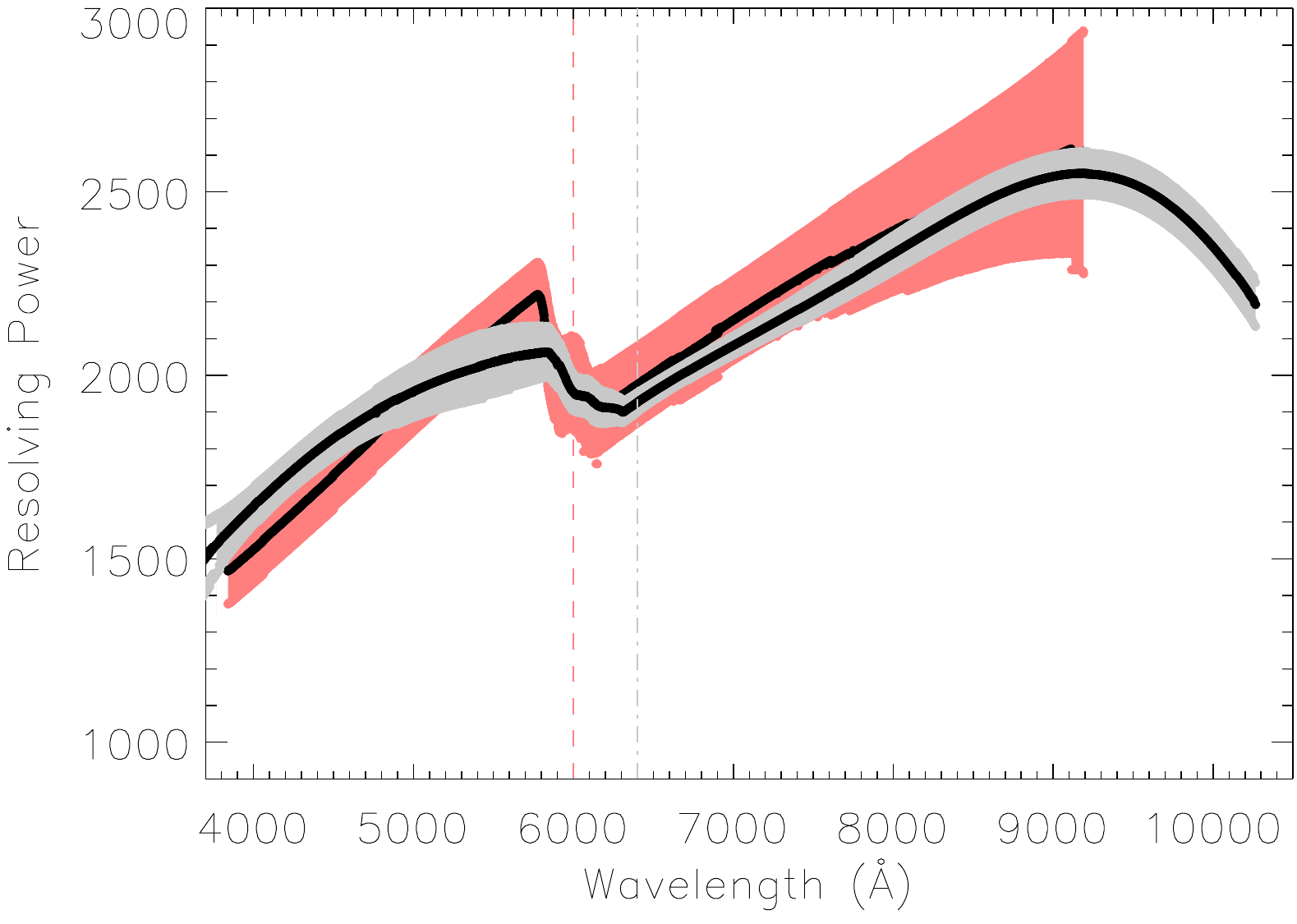}
\vspace{3mm}
\includegraphics[scale=0.53]{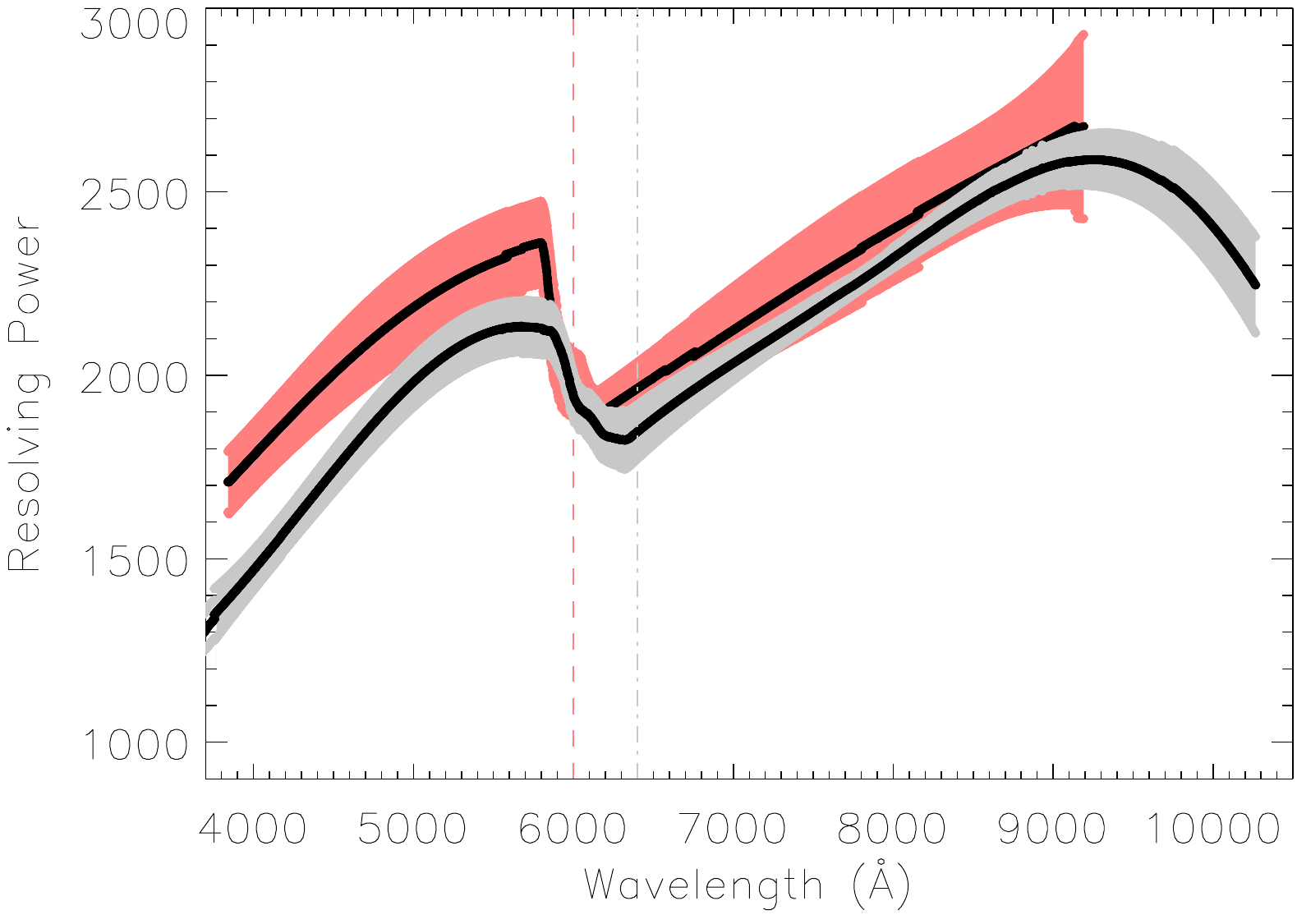}
\vspace{2.5mm}
\caption{{\bf The resolving power for BOSS (gray) and SDSS (red).  The shaded
regions correspond to the regions that contain 68\% of the plates
in the measurement.  The top panel shows the results for Spectrograph
1 and the bottom panel shows the results for Spectrograph 2.}
}
\label{fig:resolution-compare}
\hspace{0.25cm}
\end{figure}

We measured the resolving power as a function of wavelength from a sample
of 100 SDSS plates and 100 BOSS plates.
For each plate and each camera, we computed the mean $R$ as a function of wavelength.
Figure~\ref{fig:resolution-compare} shows a comparison of the SDSS resolving power to that for BOSS.

The extended coverage at long wavelengths from the new BOSS CCDs is evident
redward of 9200 \AA\ while
the extended coverage at short wavelengths is more subtle, but also shown in the figure.
The presence of a dichroic for SDSS and BOSS is also revealed near 6000 \AA\ where the resolving power no longer
increases monotonically with wavelength.
The apparent transition around the dichroic is due to the fact that both the red and blue
channels contribute to this region, each with different dispersive elements.
In the regions where we count flux from both cameras, we simply determine
the resolving power by a mean weighted by the relative throughput of the two cameras.

\begin{figure}[htbp]
\centering
\includegraphics[scale=0.53]{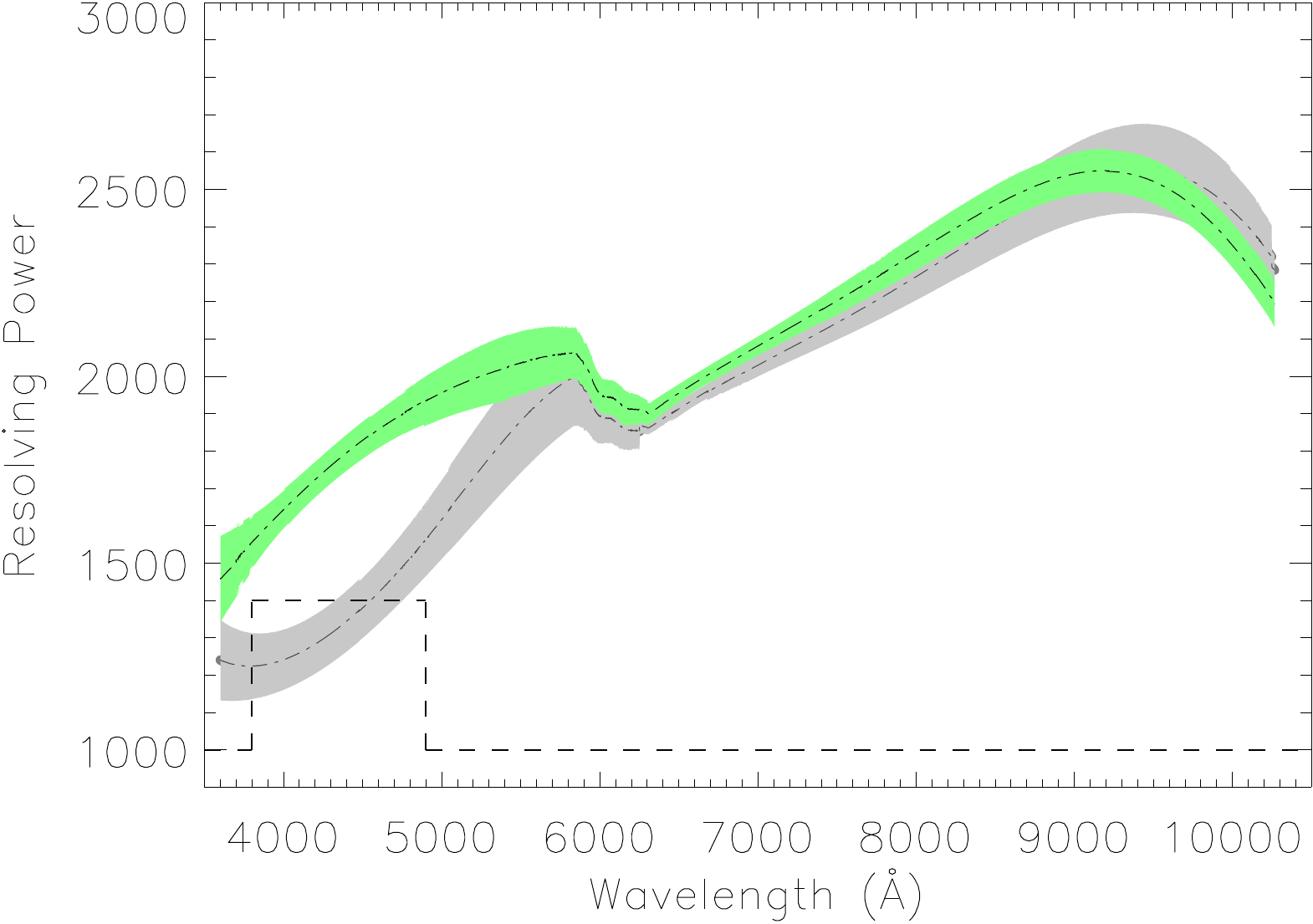}
\vspace{3mm}
\includegraphics[scale=0.53]{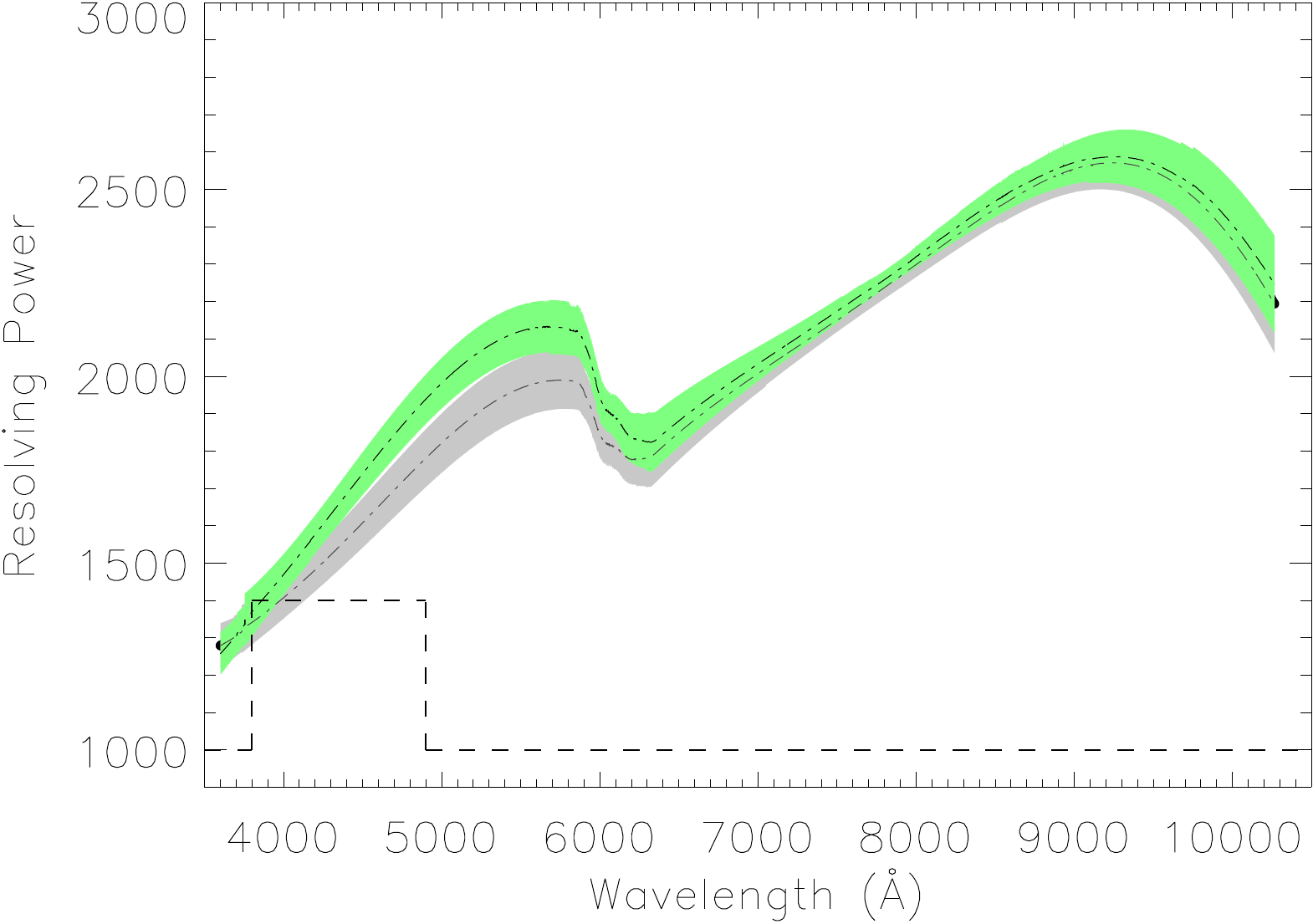}
\vspace{2.5mm}
\caption{{\bf The requirements and the measured resolving power for the two BOSS spectrographs. The results for Spectrograph 1 are shown in the top panel. 
Results for Spectrograph 2 are shown in the bottom panel. 
The dashed black curve is the requirement for the resolving power.
The green curve is the 68\% confidence limit about the mean of the resolving power for
the central fiber and is representative of $\sim$80\% of the fibers.
The gray curve is the 68\% confidence limit about the mean of the resolving power for
a representative fiber near the edge of the spectrograph slit.}
}
\label{fig:BOSS-resolution}
\hspace{0.25cm}
\end{figure}

The BOSS requirements and measured resolving power are presented
in Figure~\ref{fig:BOSS-resolution}.
For BOSS, the position of the fiber on the CCD impacts the resolution due to optical distortion
over the larger detectors.
The resolving power of the middle fiber is representative of $\sim$80\% of the BOSS fibers, 100\% of the SDSS fibers,
and clearly exceeds specifications at all wavelengths for the first spectrograph.
However, the fibers near the edge of the first spectrograph have a resolving power that is degraded
by up to 20--25\% relative to the central fibers.
Also shown in the figure, the two blue channels show quite different patterns of image quality,
with most fibers in the second spectrograph missing the requirement for resolving power around 380 nm.
The variations in resolution around 4000--5000 \AA\ are likely due to 
slight differences in the alignment of the detectors or optics,
affects that tend to be amplified at blue wavelengths.
A fraction ($\lesssim 25$\%) of the BOSS fibers therefore do not meet specifications
for spectral resolving power in the wavelength range 3800 $< \lambda <$ 4900 \AA.

\begin{figure}[htbp]
\centering
\includegraphics[scale=0.25]{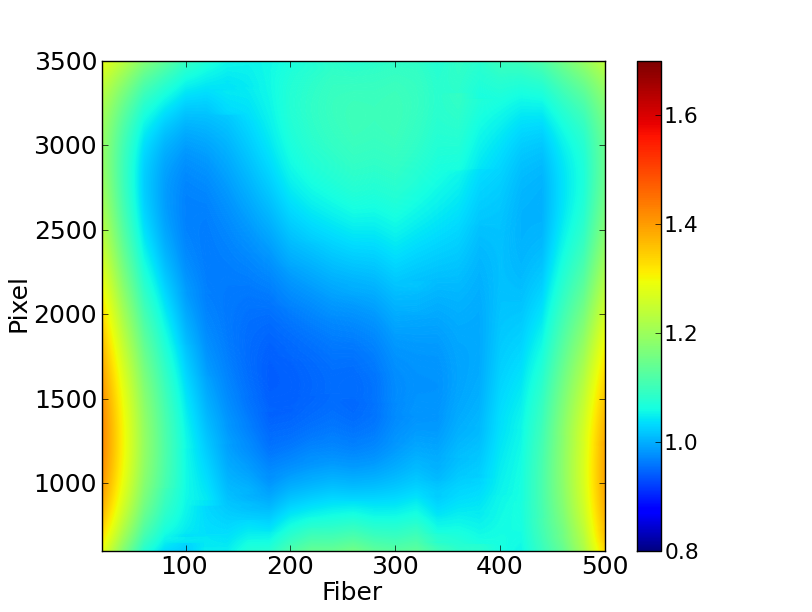}
\includegraphics[scale=0.25]{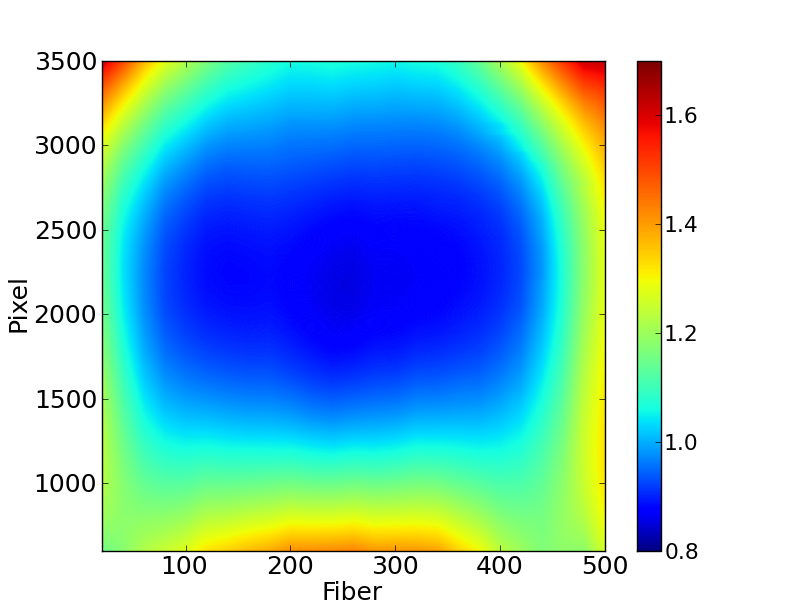}
\includegraphics[scale=0.25]{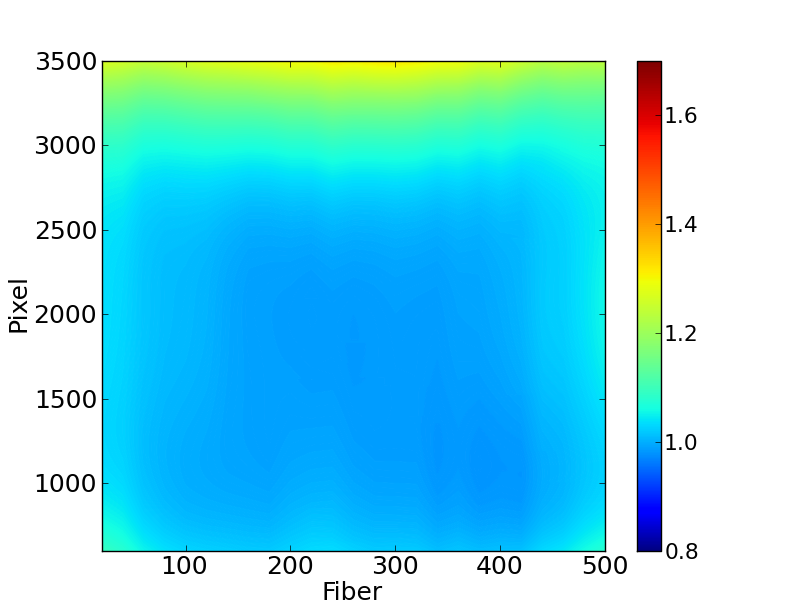}
\includegraphics[scale=0.25]{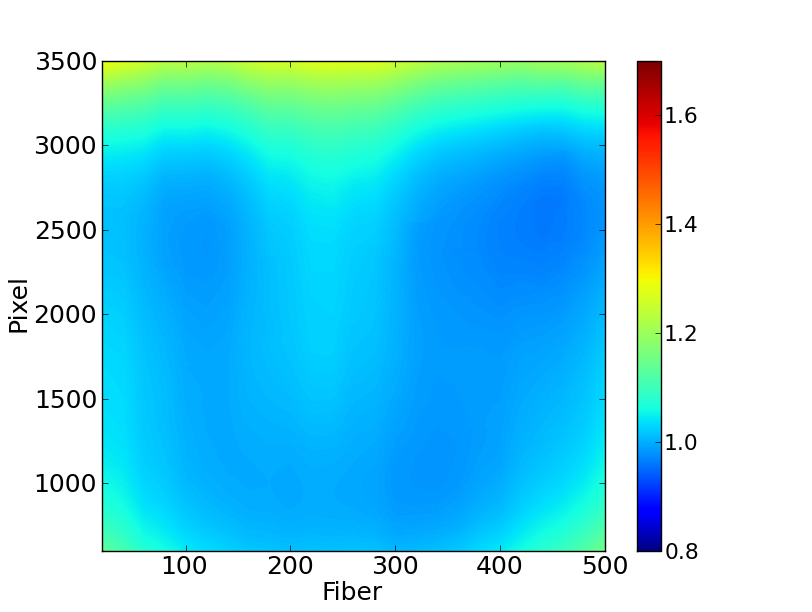}
\caption{{\bf The RMS width in pixels in the wavelength direction for each of the four BOSS CCDs.
The top panels represent the two blue cameras while the bottom panels represent the two
red cameras.  In all cases, the wavelength increases from the bottom of the image to the top
of the image.  The SDSS images have similar characteristics and are not shown here.}
}
\label{fig:BOSS-resolution-contour}
\hspace{0.25cm}
\end{figure}

In addition to the resolving power as a function of wavelength, we have also determined
the RMS width $\sigma_p$ of the line-spread function in native pixels 
to demonstrate the spatially varying resolution over the focal plane.
Figure~\ref{fig:BOSS-resolution-contour} shows the measured RMS pixel resolution of the illuminated region
for each of the four BOSS CCDs.
For the blue cameras, the spectra near the edges of the detector have a higher $\sigma_p$, reflecting the
lower optical quality and deviations from flatness of the focal plane at large field angles.
The optics project a saddle-shaped focal plane which contributes to 
this poorer resolution, an effect that is broadly consistent with predictions in Section \ref{sec:BOSSoptic}..
The two red cameras display increased RMS width between 9500 $< \lambda <$ 10,200 \AA,
mostly as a consequence of the increased path length of longer wavelength light
in the silicon.

\subsubsection{Throughput}

As discussed in Section \ref{sec:Upgrade}, the primary motivation for rebuilding the spectrographs for BOSS was to
increase the instrument throughput.
We define the throughput as the ratio of the measured flux for a point source
relative to the incoming flux outside the atmosphere.
A higher throughput is important for the fainter objects targeted by BOSS, particularly
the highest redshift galaxies and the quasars near the magnitude limit of the survey.
The throughput described here is the throughput of the entire instrument;
the predicted throughput of the individual components was discussed
in Section \ref{sec:SDSSoptic} and Section \ref{sec:BOSSoptic}.

The throughput is measured from the spectrophotometric standard stars on the plate during science exposures.
For each standard star, raw photon counts are measured using a 3-pixel radius boxcar extraction on the calibrated,
two-dimensional science frame before flux calibration.  The aperture is centered
on the spectral trace using the fiber flat-fields from the calibration sequence.
The photon counts from the source are estimated at each pixel using the observed
magnitudes from the SDSS imaging survey and the synthetic spectral template derived from spectrum.
The synthetic spectrum is converted to photons per pixel using the wavelength solution,
the effective collecting area of
the obstructed primary mirror, and the time of each exposure.
The throughput at each pixel is calculated by taking the ratio of the measured raw counts to 
the number of photons per pixel as predicted by the model.  

The throughput for SDSS was measured by averaging the throughput from 84 different standard stars
observed under photometric conditions at an airmass of $\sim$1.0 and seeing $\le 1.15\2pr$.
Because BOSS has smaller fibers, the measured throughput on the BOSS spectrographs is more susceptible 
to guiding errors.  We attempt to mitigate this effect by averaging the throughput over the four
stars that produce the highest throughput on each camera.
The throughput for BOSS was measured by averaging the throughput of 24 stars over three plates,
split evenly between the two spectrographs.
As with SDSS, we selected standard stars
observed under photometric conditions at an airmass of $\sim$1.0 and seeing $\le 1.15\2pr$.
In both SDSS and BOSS measurements, we correct small differences in seeing by normalizing the observed flux
to a double Gaussian PSF (as described in Section~\ref{subsec:sdssopticalperformance})
with FWHM of $1\2pr$ integrated over an aperture that corresponds to the area of the $2\2pr$ or $3\2pr$ fiber.
The throughput curves have been flat-field corrected to account for fiber-to-fiber variations,
thereby normalizing each fiber to the median for each plate. 
 
\begin{figure}[ht]
\centering
\vspace{3mm}
\includegraphics[scale=0.52]{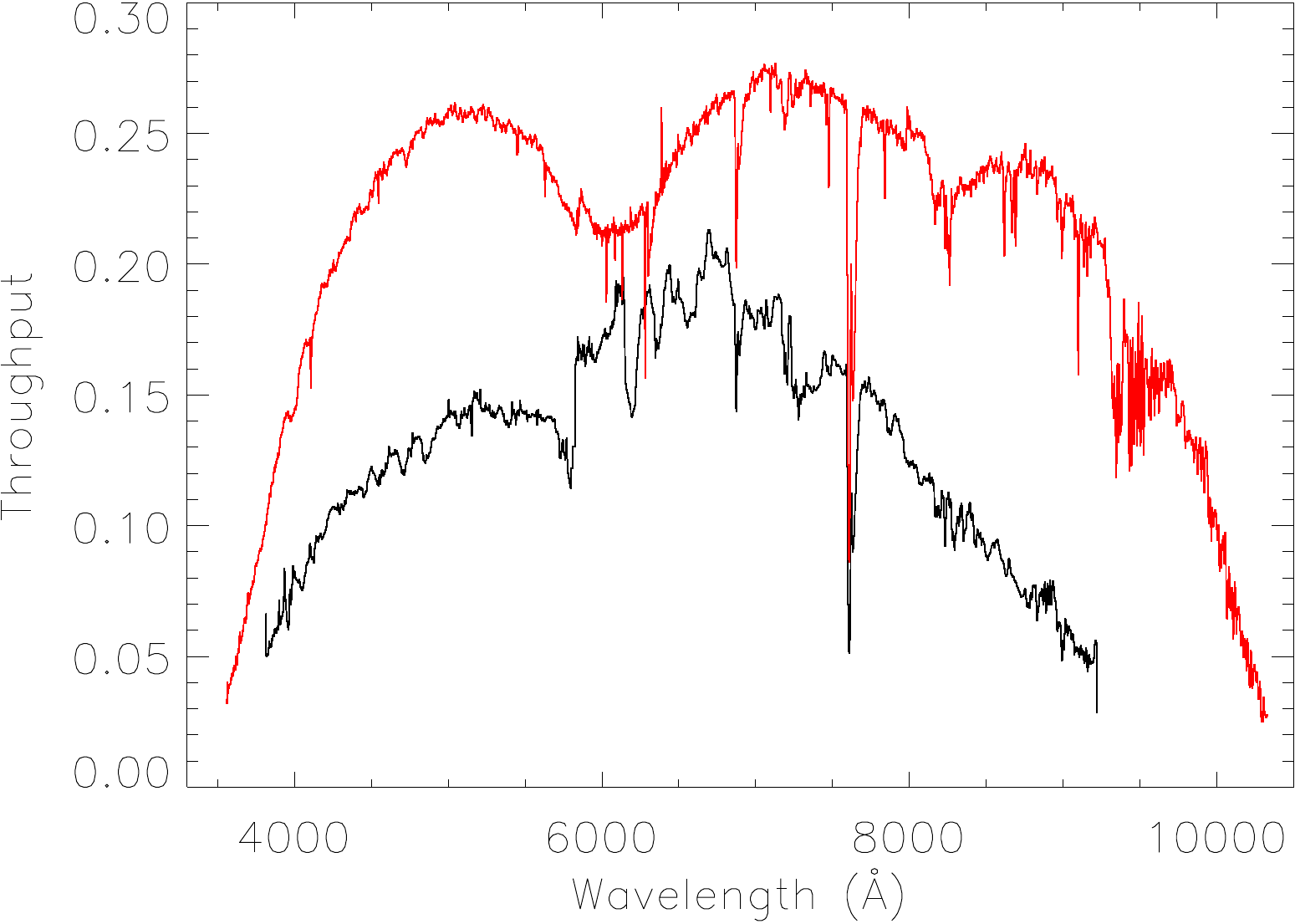}
\vspace{3mm}
\includegraphics[scale=0.52]{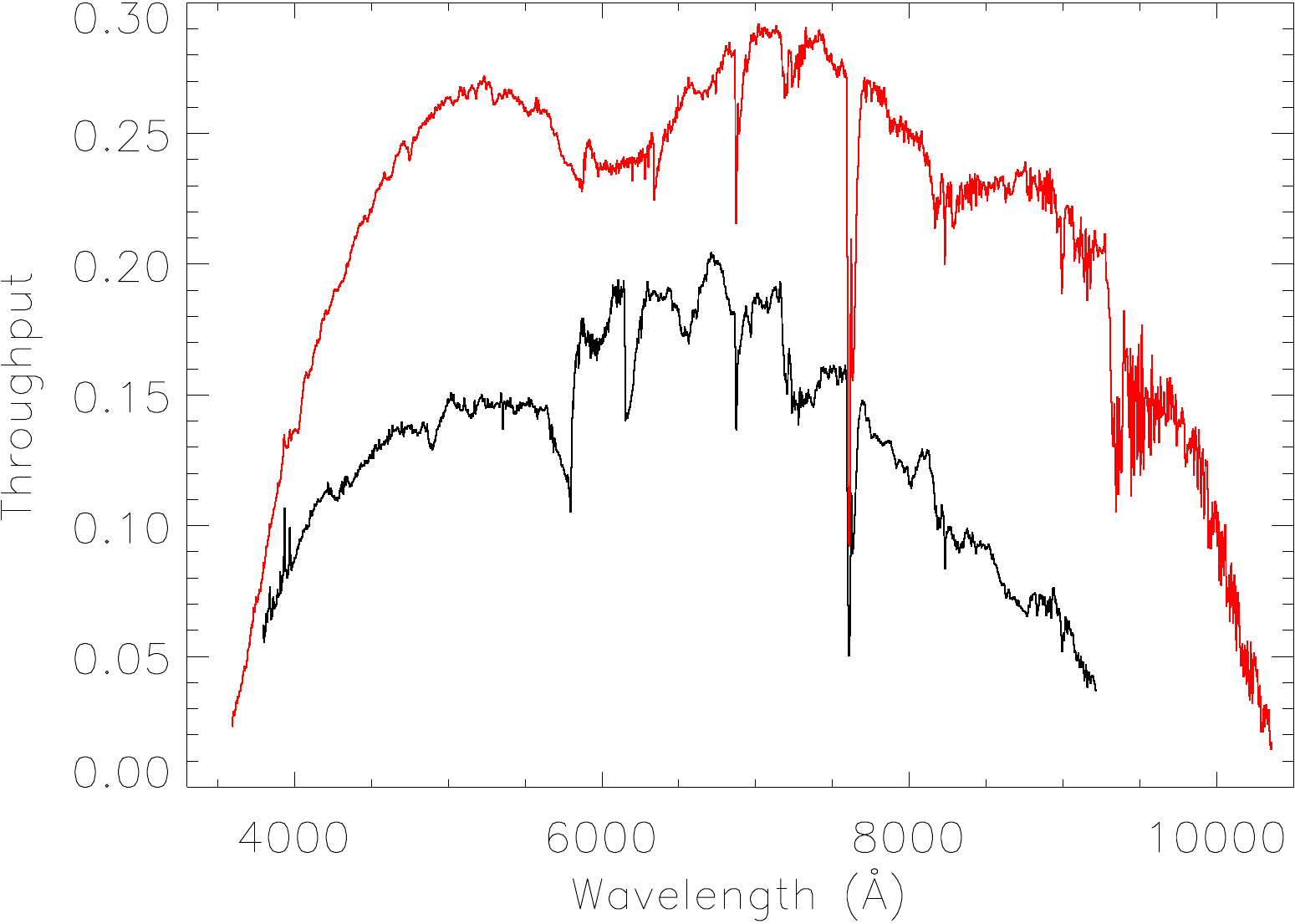}
\vspace{2.5mm}
\caption{{\bf Throughput curves for SDSS (black) and BOSS (red).  The results for
Spectrograph 1 are displayed in the top panel.  The results for Spectrograph
2 in the bottom panel.  Note the telluric absorption features. }
}
\label{fig:throughput-compare}
\end{figure}

\begin{figure}[htbp]
\centering
\includegraphics[scale=0.54]{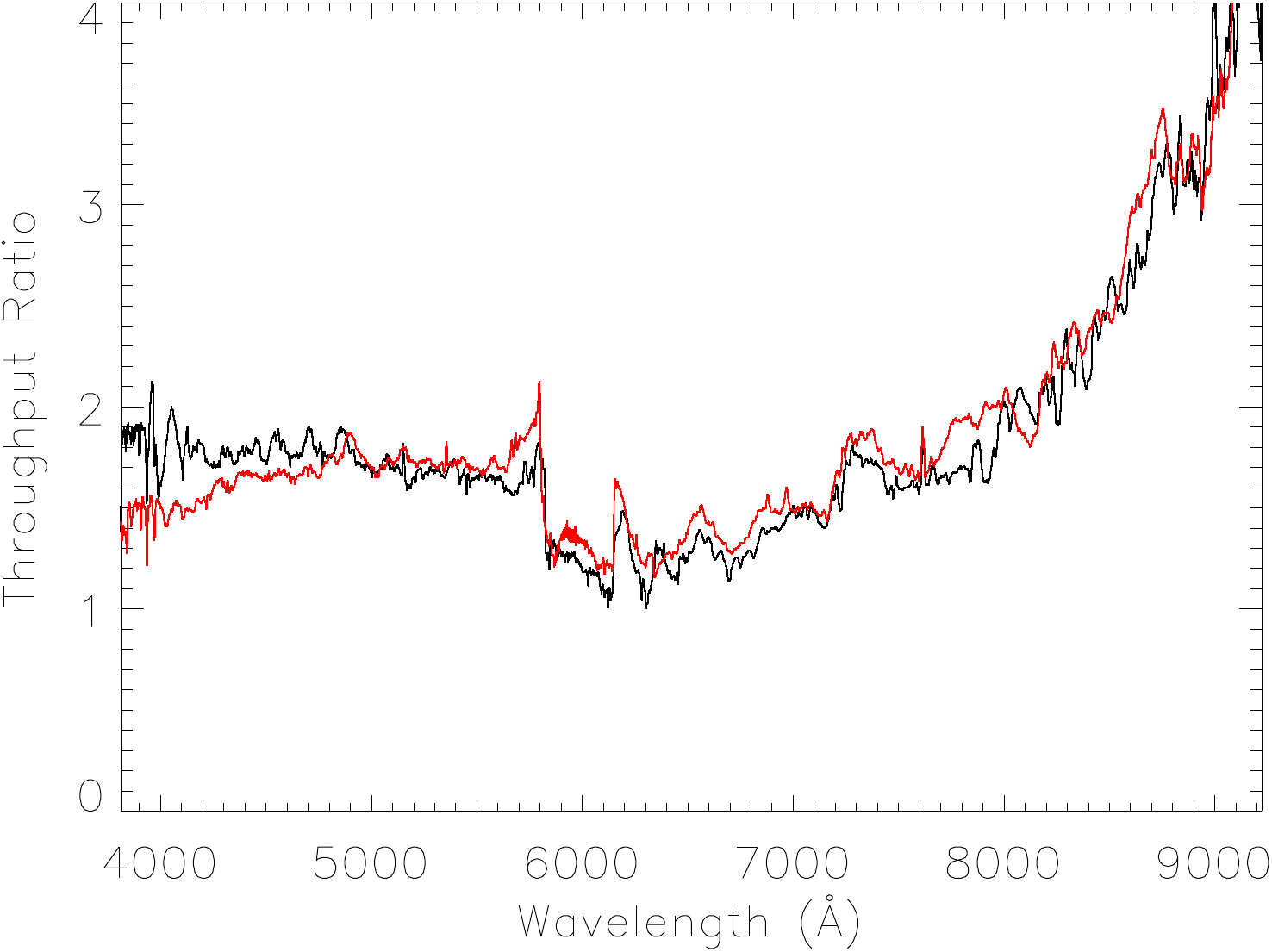}
\vspace{2.5mm}
\caption{{\bf Ratio of throughput BOSS/SDSS after applying a median smoothing kernel
of width 100 pixels.  The figure displays results for Spectrograph 1 (black)
and Spectrograph 2 (red).}
}
\label{fig:throughput-ratio}
\end{figure}

As shown in Figure~\ref{fig:throughput-compare},
the throughput after the BOSS upgrades has been significantly improved.
The ratio of BOSS to SDSS throughput is also shown in Figure~\ref{fig:throughput-ratio}.
Not demonstrated here is a fiber dependent variation in the throughput.
As in the resolution measurements, fibers near the edge of the BOSS detectors
have a slightly lower throughput than fibers near the middle of each spectrograph.
For BOSS, there is an additional $\sim$10 \% RMS variation on the average throughput curve (particularly at
short wavelengths) between standard stars.
This variation in BOSS is likely caused by scale changes in the telescope over the
course of an observation and possible guiding errors (BOSS is more sensitive to guiding because of the smaller fibers).  The effect is probably present
in the SDSS data as well, but is amplified
in BOSS due to the decrease in fiber diameter.

Overall, the SDSS spectrographs delivered on-sky throughput that was consistent between the two instruments to within approximately 10\% across all wavelengths, fell short of design predictions by roughly 10--25\% depending on wavelength, and met the requirements set forth in Section~\ref{sec:SDSS_Throughput_S/N} except at the bluest wavelengths where the 8\% and 9\% delivered throughputs at 4000 \AA, for Spectrographs 1 and 2, respectively, is compared to the requirement of 10\%.   The throughput predictions were based on a combination of measured, published, and modeled efficiency curves for a large number of system components, and the agreement with measured throughput to 10--15\% over the majority of the bandpass is satisfying, as is the agreement in overall spectral shape. The largest uncertainty in the throughput prediction is the telescope, which is not well characterized and continually changes with exposure to the environment and periodic cleaning of optical surfaces. The telescope contains two skyward looking surfaces, the primary mirror and the first surface of the dual element wide field corrector. These surfaces, in addition to the secondary mirror, are periodically cleaned with CO2 snow, which deals with much of the dust but not {\it stickier} contaminants such as pollen. This contamination undoubtedly degrades the UV throughput of the telescope more than the longer wavelengths but again, is not well characterized.  In addition, the efficiency of several other system components in the UV is rolling off precipitously; thus, the wider discrepancy between predicted and measured throughput in the UV is not unexpected.

The BOSS spectrographs exhibit somewhat more variation in throughput between the two instruments, but not to any degree of surprise or concern; the greatest variation is about 15\% in the dichroic crossover region near 6000 \AA, while the difference in peak throughput is only 5\%.  The discrepancy between predicted and measured throughput is flatter across the bandpass than was the case for SDSS, with an average shortfall in measured throughput of roughly 15\%.  The discrepancy is highest at both ends of the bandpass, but not dramatically so and this may exonerate the telescope throughput model somewhat as a major suspect in the UV discrepancy for SDSS.  Again, many component efficiencies are rolling off in the UV and at the other end for BOSS, at 10,000 \AA the CCD efficiency is falling steeply.

The requirement on BOSS throughput from Section~\ref{BOSS_Throughput_S/N} is a factor of two increase in peak throughput relative to SDSS.  While the actual peak throughput was increased by only about 40\%, from 21\% to 29\%, the throughput was nearly doubled over much of the blue bandpass, and was more than doubled longward of 8000 \AA.

\subsection{Flexure}
\label{sec:SDSS_BOSS_Flexure}

\begin{figure}[htbp]
\centering
\includegraphics[scale=0.38]{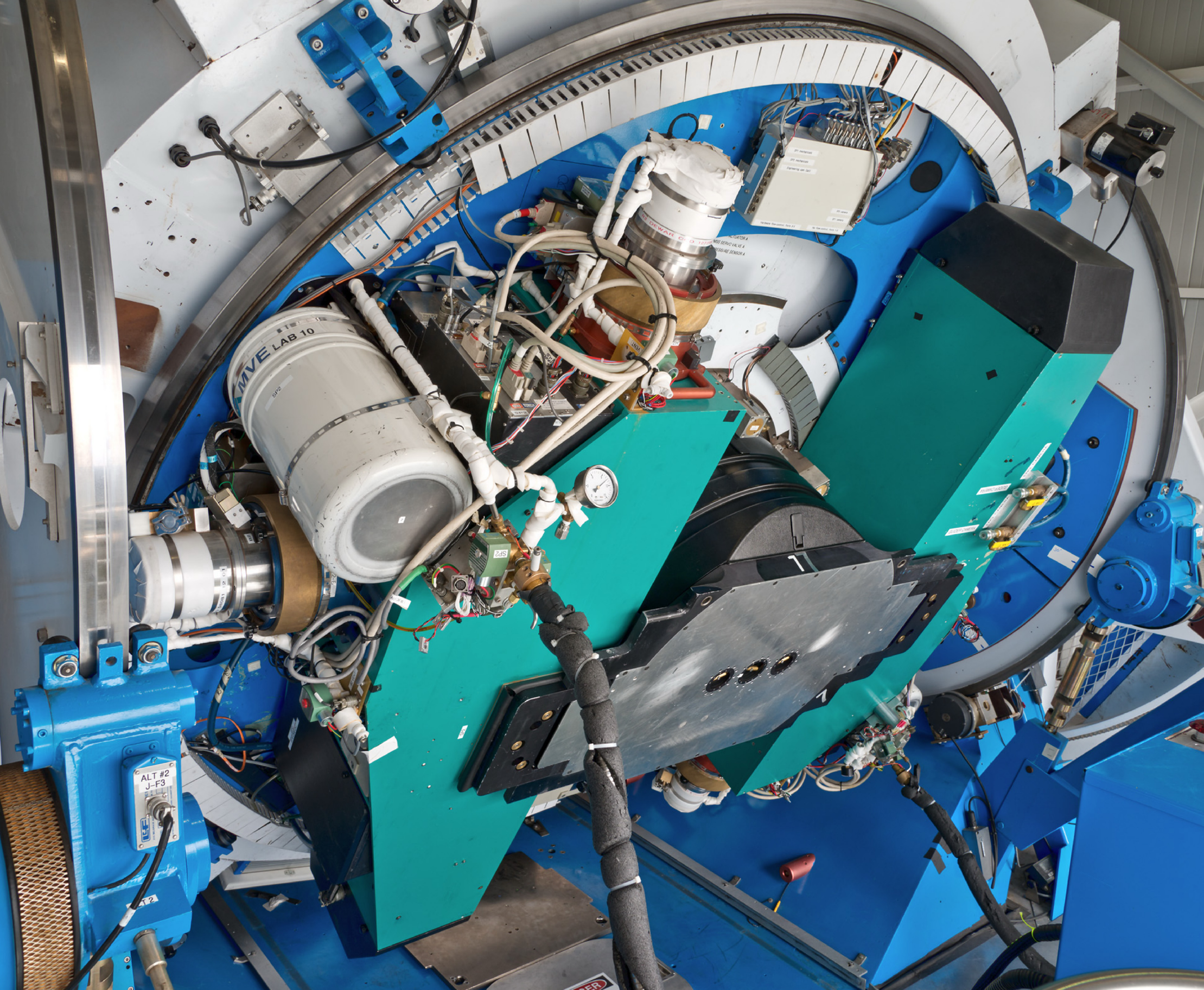}
\caption{{\bf Photograph of the spectrographs mounted to the Cassegrain rotator. Spectrograph 1 is on the right.  Spectrograph 2 is on the left.  In this orientation the rotator is at zero degrees.  Positive rotation is defined as counterclockwise motion as viewed from behind the rotator.   Here, the telescope is at a zenith angle of $60^\circ$. }}
\label{fig:Spectrographs_Photo}
\end{figure}

Flexure was characterized for both SDSS spectrographs in 1999.
Measurements were taken with the telescope near the horizon at a zenith angle of $72^\circ$.
A sparsely populated plate was illuminated with Ne and HgCd arc lamps to provide an array of spots at mostly
random positions on the CCDs.
Data were taken approximately every $45^\circ$ over the entire $360^\circ$ range of the rotator.
The sign convention is such that positive rotator angles represent a counterclockwise rotation of the rotator
as viewed from behind the rotator.
Positive values in the spectral direction represent motion of the spots toward longer wavelengths,
and positive values in the spatial direction represent motion of the spectra toward higher column number on the CCD. For reference, a photograph 
of the spectrographs on the back of the rotator is shown in Figure~\ref{fig:Spectrographs_Photo}. As shown, the rotator is at the zero degree orientation, and the telescope is at a zenith angle of $60^\circ$.

\begin{figure}[htbp]
\begin{center}
\epsscale{1.23}
\plotone{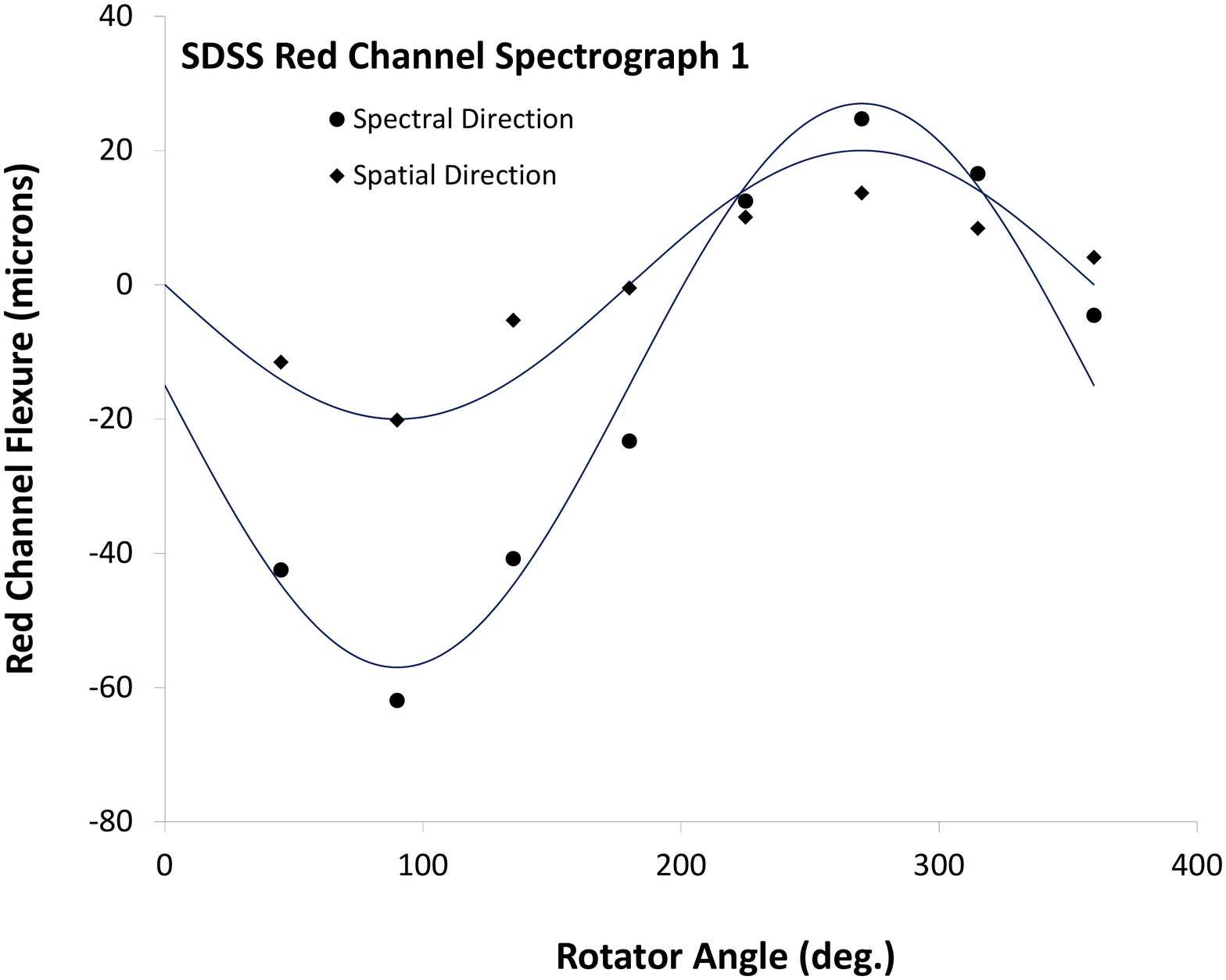}
\plotone{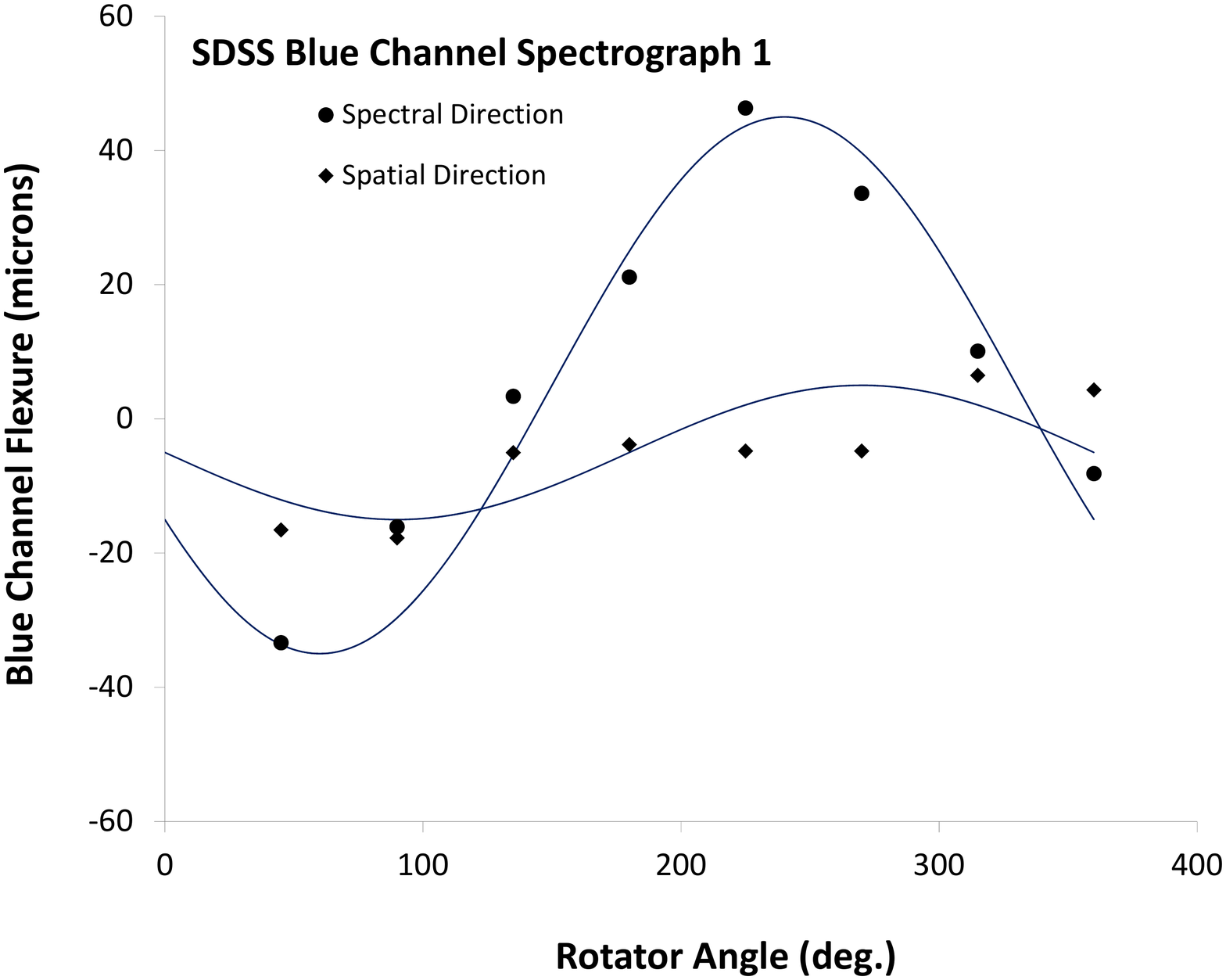}
\caption{{\bf Gravity induced image motion in SDSS Spectrograph 1 as a function of rotator angle for a zenith angle of 72$^{\circ}$.  Twenty-four microns represents one pixel. The top and bottom plots show flexure in the red and blue channels, respectively.  Flexure in the spectral direction is greater than in the spatial direction due to compliance in the collimator mount; a natural consequence of a tall mirror. }}
\label{fig:SDSS_flexure}
\end{center}
\end{figure}

Figure~\ref{fig:SDSS_flexure} shows image motion as a function of rotator angle for both the spectral and spatial directions
in Spectrograph 1.
Results for Spectrograph 2 are comparable and a summary can be found in Table~\ref{table:reqperf}.
At 24 $\mu$m per pixel, flexure in the spectral direction was measured to be $\sim 0.3$ pixels in the worst
case in the red channel and $\sim 0.33$ pixels in the worst case in the blue channel over a $15^\circ$ rotation.
Total flexure in the spectral direction, over the $360^\circ$ rotator range, was measured to be $\sim 3.6$ pixels
peak-to-valley in the red channel and $\sim 3.3$ pixels peak-to-valley in the blue channel.
In the spatial direction the flexure was lower, about 0.2 pixels in both channels over $15^\circ$.
Total flexure in the spatial direction was measured to be $\sim 1.4$ pixels and $\sim 1.1$ pixels peak-to-valley in the red
and blue channels, respectively.
It should be noted that these results are near worst case since most fields observed are much further from the horizon,
and flexure is at a maximum at the horizon.  Flexure scales as the sine of the zenith angle for instruments on a Cassegrain rotator.

These results are in-line with the performance goal of 0.3 pixels for $15^\circ$ on the sky that
was discussed in Section~\ref{sec:SDSS_Flexure}.
A noteworthy observation from the results is that the flexure is much worse in the spectral direction,
a surprise that should have been anticipated given the collimator mount design.
The tall rectangular collimator is supported by a thin membrane fastened to the rear face of the collimator.
Having the support behind the center of gravity, combined with the fact that the tip/tilt/piston actuators are not
infinitely rigid, allows the mirror to tilt slightly as the gravity vector changes.
Considering the contact points of the actuators, the collimator tilts considerably more in the narrow
(spectral) direction than in the tall (spatial) direction. 

Not long into SDSS operations a subtle problem appeared in the blue channel of one of the spectrographs.
It was discovered that the blue camera on Spectrograph 1 (b1), would go out of focus more often than the other cameras.
This sudden defocus would tend to happen over a particular range of rotator angle ($90^\circ$ - $135^\circ$),
sometimes requiring a refocus of the collimator.  
Measurements in 2004 indicated the CCD tilted approximately 35 $\mu$m (corner-to-corner) over a small range of rotator motion
but the location of the spectra did not change.
In the early phases of BOSS, the problem suddenly got worse,
probably due to the increased mass of the BOSS dewars.
It was discovered that the focus locking mechanism, by design, did not lock-out the backlash in the focus ring threads.
The locking mechanism was redesigned as discussed in Section~\ref{sec:BOSS_Cameras} and the problem was finally solved. 

\begin{figure}[htbp]
\begin{center}
\epsscale{1.23}
\plotone{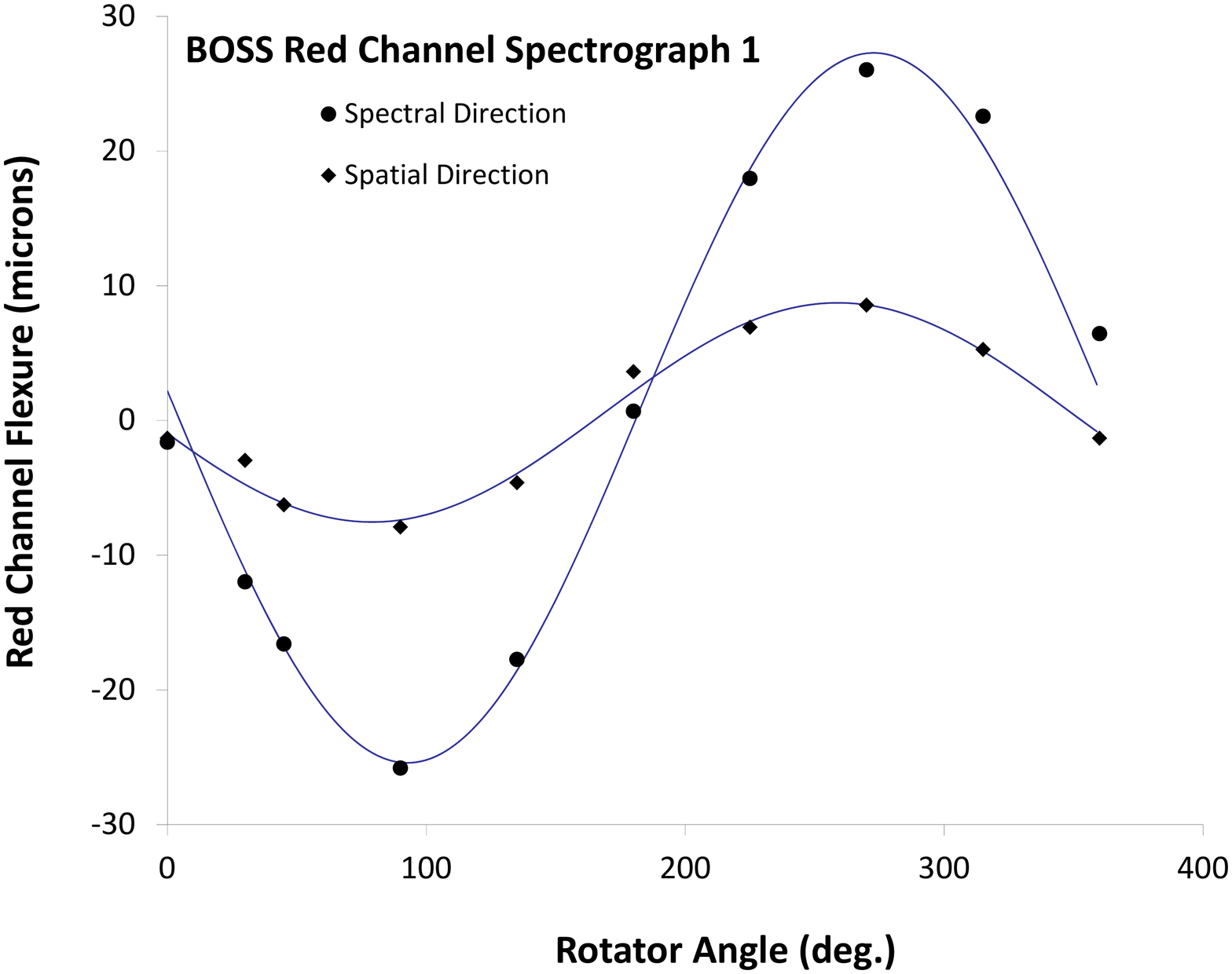}
\plotone{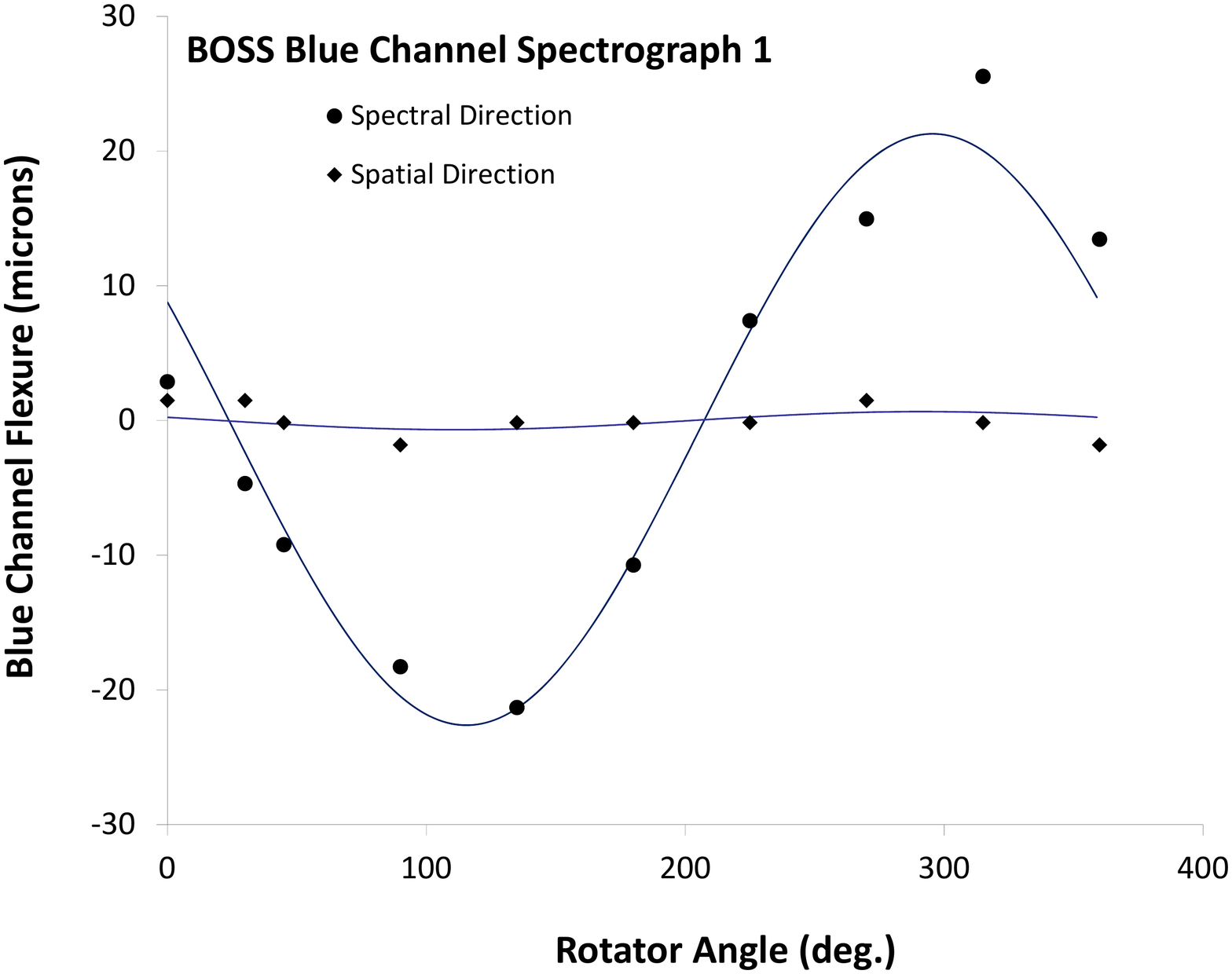}
\caption{{\bf Gravity induced image motion in BOSS Spectrograph 1 as a function of rotator angle for a zenith angle of 72$^{\circ}$.  Fifteen microns represents one pixel. The top and bottom plots show flexure in the red and blue channels, respectively.}}
\label{fig:BOSS_flexure}
\end{center}
\end{figure}

With the implementation of upgraded cameras and a redesigned focus locking mechanism, flexure was re-characterized for BOSS.
Figure~\ref{fig:BOSS_flexure} shows image motion as a function of rotator angle for both channels of Spectrograph 1.
Measurements were taken with the telescope zenith angle at 60$^{\circ}$, then scaled to a zenith angle of 72$^{\circ}$ for
comparison with the results for the SDSS spectrographs. 
From Figures~\ref{fig:SDSS_flexure} and~\ref{fig:BOSS_flexure} it is clear that the new focus locking mechanism reduced flexure overall.
Total flexure in the spectral direction, in microns, was reduced from $\sim 85$ $\mu$m peak-to-valley in the red channel of
SDSS Spectrograph 1 to $\sim 52$ $\mu$m peak-to-valley in BOSS.
Similarly, flexure in the blue channel was reduced from $\sim 80$ $\mu$m peak-to-valley to $\sim 47$ $\mu$m peak-to-valley
in the spectral direction.
However, given the smaller, 15 $\mu$m pixel CCDs used in BOSS, the results are comparable to SDSS in units of pixels.
For 15$^{\circ}$ on the sky, the spectral shift is 0.41 pixels (worst case) in both channels of BOSS Spectrograph 1,
modestly lower than the design expectation of 0.5 pixels;
Spectrograph 2 is 0.44 pixels worst case. 
In the spatial direction, the worst case flexure is 0.18 pixels in Spectrograph 1,
and 0.15 pixels in Spectrograph 2. Again, see Table~\ref{table:reqperf} for a summary of the results.

\subsection{CCD and Electronics Performance}

\begin{table*}
\begin{center}
\caption{CCD Performance for SDSS and BOSS}
\label{table:CCD}
\begin{tabular}{  l  l  l  l  r l  }
  \hline
   & read noise  & dark current & gain & number of  & fraction of  \\
 & ($e^-$) &  ($e^-$/pix/15 min) & ($e^-/$ADU) & bad columns & bad pixels\\ \hline
\multicolumn{ 5}{l}  {SDSS} \\  \hline
   b1 & 2.8--3.8 & 1.05--1.10 & 1.05--1.10 & 4 & \\
   b2 & 3.2--3.9 & 1.23--1.26 & 1.23--1.26 & 4 &   \\
   r1 & 3.5--4.1 & 1.00--1.09 & 1.00--1.09 & 14 &  \\
   r2 & 3.6--4.4 & 1.04--1.05 & 1.05 & 10 & \\
\multicolumn{ 5}{l}  {Start of BOSS} \\  \hline
   b1 & 1.79--1.98 & 0.64--0.68 & 1.01--1.05 & 0 & 2.0e-05 \\
   b2 & 1.74--2.04 & 0.61--0.65 & 0.99--1.04 & 2 & 2.1e-04 \\
   r1 & 2.36--2.72 & 0.88--1.10 & 1.54--1.97 & 5 & 1.4e-04 \\
   r2 & 2.42--2.99 & 0.95--1.97 & 1.54--1.96 & 3 &1.8e-04 \\
\multicolumn{ 5}{l}  {MJD 55300 (r2 replaced)} \\  \hline
  b1 & 1.83--2.01 & 0.51--0.53 & 1.01--1.05 & 0 & 2.1e-05 \\
  b2 & 1.87--2.03 & 0.53--0.56 & 0.99--1.04 & 2 & 2.1e-04 \\
  r1 & 2.43--2.73 & 0.63--0.80 & 1.54--1.97 & 5 & 3.3e-04 \\
  r2 & 2.73--2.89 & 1.19--2.27 & 1.59--1.66 & 11 & 4.4e-04 \\
\multicolumn{ 5}{l}  {MJD 55800 (r1 replaced)} \\  \hline
  b1 & 1.77--2.02 & 0.46--0.49 & 1.01--1.05 & 0  & 2.1e-05 \\
  b2 & 1.86--2.01 & 0.56--0.59 & 0.99--1.04 & 2  & 2.1e-04 \\
  r1 & 2.45--2.82 & 0.57--0.82 & 1.47--1.93 &  1 & 3.3e-04 \\
  r2 & 2.85--2.88 & 1.30--1.56 & 1.59--1.66 & 11 & 2.5e-04 \\
\end{tabular}
\end{center}
\end{table*}

We measured the CCD and electronics performance
for SDSS and BOSS and report the performance in Table~\ref{table:CCD}.
Because the detectors are divided into multiple quadrants,
we report the range of each measurement between the lowest value
and highest value obtained in the quadrants of each detector.
While only a few aspects of detector performance were specified during the 
design of the surveys, we describe the results of the detector performance below.
When detector performance was specified, the CCDs exceeded specifications in all
cases except for small regions on the r2 camera that slightly exceed
the design goal of 2 $e^-$/pix/15 min of dark current.

First, the CCD images were checked for cosmetic defects such
as saturated hot pixels, pixels with a large deviation from neighbors,
blocked columns at low flux levels, and other bad pixels.
These bad pixels are masked during the data processing.
The hot pixels are recorded by identifying any pixel greater than 15,000 ADU in the
bias exposures.
Hot columns are identified in bias exposures as regions of connected
pixels that lie above the background, in the direction of parallel clock transfer, by an amount
equal to five times the readnoise.  For SDSS, pixel-to-pixel variation was mapped using the 
flat field dither technique described in Section~\ref{sec:SDSScal}.  For BOSS,
pixel-to-pixel variation is mapped using the images
taken with the lossy-fiber as described in Section~\ref{sec:BOSScal}.
Blocked columns and defective pixels are identified in these images as clustering
of pixels that lie 50\% below the mean background.
The flat fields from these tests are also applied at the same time as
the bad pixel masks in the data processing.  
Other bad pixels are found from a combination of flats,
biases, and dark frames.

In addition to cosmetic defects, the read noise, dark current and
gain of the CCDs were measured as shown in Table~\ref{table:CCD}.
The read noise was measured in a series of 
zero second exposures; bias frames that were obtained throughout the surveys.
A series of 15 minute exposures is averaged pixel-by-pixel to determine the dark current in each pixel.
We include any light source detected by the CCDs that does not originate from the fibers,
including light leaks, in our measurement of dark current.
The dark current in each quadrant of a CCD is averaged, and we report the minimum and maximum
values of dark current in the four quadrants.
Gain is determined for each quadrant from photon statistics in a series of flat field exposures.

For BOSS, we use measurements in the detector lab at LBNL to determine
full-well depth for the red CCDs.
We refer to documentation from e2v for these measurements on the blue CCDs.
The full-well depth exceeds the maximum number of electrons expected (65,000 $e^-$/pixel)
in a science exposure for all cameras.
The charge transfer efficiency (CTE) records the fraction of electrons that are successfully
transferred during each clock cycle during readout.
CTE for the red CCDs was measured using an $^{55}$Fe source \citep[as described in][]{bebek02a,dawson08a}
in the LBNL detector lab for the red CCDs and recorded from the e2v documentation for the blue CCDs.
All detectors exceed the CTE requirement of 0.99999 in both serial transfer and parallel transfer.
Finally, to meet our goal for overall observing efficiency, a maximum of 70 seconds can be used
for the CCD pre-exposure flush time and end-of-exposure readout time.
The readout time of the BOSS CCDs has been measured to be 55.6 seconds as described above.
This requirement means that not more than 10\% of the observing time is used for the
CCD flush and readout for two calibration exposures and four 900 second science exposures.  
Because the detectors have been replaced by newer technologies,
we do not report values for full-well depth, CTE, or readout time for the SDSS detectors.

\section{Conclusion}
\label{sec:conclusion}

The SDSS spectrographs were designed to be high-throughput, robust instruments,
to conduct an extragalactic survey producing a million redshifts over a five year lifetime.
These goals have clearly been met, with 1.6 million spectra of stars, galaxies,
and quasars over 2880 plates, primarily in the northern Galactic cap at high Galactic latitudes.
Spectra over 9274 deg$^2$ were collected in nine years of operation
between the start of SDSS-I in 1999 and the completion of the first of two extended
phases, SDSS-II, including SEGUE which
produced roughly 250,000 stellar spectra between 2004-2008 \citep{yanny09a}.
These data were released to the astronomy community nearly every year;
the full data set from SDSS-I and II is included in DR8 \citep{aihara11a}.
Aside from a few minor technical issues where the instruments
did not meet predicted design performance, the spectrographs proved to be quite reliable and
exceptionally productive instruments.
With improvements in CCD technology, the use of VPH gratings, and enhancements
to the optics (i.e. fluid coupled camera triplets, new dichroics,
and higher reflectivity collimator coatings) the instrumental throughput was
enhanced significantly, allowing the use of smaller diameter fibers and making
the BOSS survey possible.
Along with improvements to the guide camera, modifications to improve flexure,
and a host of minor upgrades, the BOSS spectrographs
have proven to be as productive as the original SDSS design and have many years of life remaining.

The spectroscopic data from SDSS have enabled studies across a broad range of astronomical disciplines including
the evolution and clustering of galaxies \citep[e.g.,][]{kauffmann04a,tegmark04a},
gravitational lensing \citep[e.g.,][]{bolton06a},
the properties of quasars \citep[e.g.,][]{vandenberk01a},
and stellar astrophysics \citep[e.g.,][]{west08a}.
One of the prominent scientific contributions
from SDSS and SDSS-II data is the discovery of acoustic oscillation
signatures in the clustering of galaxies \citep{eisenstein05a}, opening the door to a new
method of cosmological measurement.
SDSS imaging and spectroscopic data have been included in more than 3500 refereed papers.

As described in detail in \citet{dawson13a}, the BOSS survey will cover 10,000 deg$^{2}$ in five years,
obtaining spectra for 1.35 million luminous galaxies with redshifts $0.15<z<0.7$,
and approximately 160,000 quasars with redshifts between $2.15<z<3.5$.
The galaxies provide a highly biased tracer of matter, and at these densities over this area provide
percent-level distance scale measurements through studies of BAO.
The quasars provide a backlight to illuminate neutral hydrogen through Lyman-$\alpha$ absorption
along the line of sight, thereby mapping large-scale structure in the foreground of each quasar.
These distance measurements will represent a significant improvement on the accuracy of existing BAO measurements,
and will tighten the constraints on the dark energy equation of state, $w=p/\rho$, where $p$ is pressure
and $\rho$ is density, and the evolution of $w$ with time.  Detection of the BAO feature in quasar spectra would allow BOSS
to make the first characterization of dark energy at early times, when dark energy should be sub-dominant
according to the prevailing $\Lambda$CDM theory.

BOSS completed imaging of the SGC spectroscopic footprint in 2008 and 2009;
spectroscopic observations began December 5, 2009 after a three month commissioning phase.
The algorithm for selecting galaxy targets is outlined briefly in \citet{white11a} and \citet{eisenstein11a}.
The quasar target selection is described in \citet{ross12a} using photometry from the
SDSS imaging survey supplemented with data from programs at ultraviolet, infrared, and
radio wavelengths to enhance the detection efficiency.
These first two years of BOSS spectra include 324,198 unique galaxy targets
($z>0.43$) with a 98.7\% rate of successful classification, and
103,729 unique lower redshift galaxy targets ($0.15<z<0.43$) with a 99.9\%
successful rate of classification.
There were more than 85,000 quasars observed, 61,933 of which were at $z>2.15$.
The spectra of these objects are classified as described
in \citet{bolton12a} and each quasar is visually inspected
and recorded as explained in \citet{paris12a}.
Data from the first two years of the survey were included in
DR9 \citep{ahn12a} and include the spectra from 831 plates.

The higher redshift sample of galaxies from DR9 has already been used to
fit the position of the acoustic peak at an effective redshift $z=0.57$ \citep{anderson12a}.
With 1.7\% accuracy, this is the most precise distance constraint ever obtained from a galaxy survey at any redshift.
The potential for distance measurements at $z>2$ using Lyman-$\alpha$ forest absorption
from quasars is promising; the first detection of flux correlations across widely separated
sightlines was obtained in the first year of BOSS data to comoving separations of 60 $h^{-1}$Mpc \citep{slosar11a}.
More recently, the acoustic peak was measured in the Lyman-$\alpha$ forest, two complimentary
analyses produced distance measurements at $z>2$ with 3\% \citep{busca13a} and 2\% \citep{slosar13a} precision.
BOSS is on pace to complete the 10,000 deg$^2$ spectroscopic footprint by the middle of 2014, providing
new cosmology constraints and a wealth of data for studies of galaxy evolution and quasar physics.

The optical spectrographs at APO will have unique capabilities for widefield,
multiplexed spectroscopy well beyond the 2014 completion of the BOSS survey.
The SDSS collaboration has proposed an ``After SDSS III'' (AS3) program for 2014-2020
that will capitalize on this resource.
Four separate programs in AS3 will use the BOSS spectrograph,
likely with only minor modifications.
The extended Baryon Oscillation Spectroscopic Survey (eBOSS) will probe
dark energy and fundamental physics by making distance measurements with BAO
in the redshift range $0.6<z<2$.
The Time-Domain Spectroscopic Survey (TDSS) will obtain spectra of 50,000 -- 100,000 Galactic and
extragalactic variable sources with simultaneous, multi-epoch optical imaging data.
In a program to follow up x-ray sources in new 0.2–8 keV
data obtained from the extended ROentgen Survey with
an Imaging Telescope Array \citep[eROSITA:][]{predehl10a},
the SPectroscopic IDentification of eROSITA Sources (SPIDERS) survey will
follow up 50,000 -- 100,000 objects.
Finally, Mapping Nearby Galaxies at Apache Point Observatory (MaNGA) will perform
spatially resolved spectroscopy on approximately 10,000 nearby galaxies
using 15 integral field units integrated into new BOSS-like cartridges.
As with the original SDSS spectroscopic survey, these four surveys will
provide a premier data sample for astrophysical studies from Galactic to
cosmological scales.

\acknowledgments

Funding for SDSS-III has been provided by the Alfred P. Sloan Foundation, the Participating Institutions, the National Science Foundation, and the U.S. Department of Energy Office of Science. The SDSS-III web site is http://www.sdss3.org/.

SDSS-III is managed by the Astrophysical Research Consortium for the Participating Institutions of the SDSS-III Collaboration including the University of Arizona, the Brazilian Participation Group, Brookhaven National Laboratory, University of Cambridge, Carnegie Mellon University, University of Florida, the French Participation Group, the German Participation Group, Harvard University, the Instituto de Astrofisica de Canarias, the Michigan State/Notre Dame/JINA Participation Group, Johns Hopkins University, Lawrence Berkeley National Laboratory, Max Planck Institute for Astrophysics, Max Planck Institute for Extraterrestrial Physics, New Mexico State University, New York University, Ohio State University, Pennsylvania State University, University of Portsmouth, Princeton University, the Spanish Participation Group, University of Tokyo, University of Utah, Vanderbilt University, University of Virginia, University of Washington, and Yale University.

\bibliographystyle{apj}
\bibliography{spectro}

\end{document}